\documentclass[english,aip,jcp,reprint]{revtex4-2}
\usepackage[T1]{fontenc}
\usepackage[latin9]{inputenc}
\usepackage{babel}
\usepackage{amsmath}
\usepackage{graphicx}
\usepackage[unicode=true,pdfusetitle,
 bookmarks=true,bookmarksnumbered=false,bookmarksopen=false,
 breaklinks=false,pdfborder={0 0 1},backref=false,colorlinks=false]
 {hyperref}

\makeatletter

\providecommand{\tabularnewline}{\\}

\usepackage[version=4]{mhchem}
\usepackage{braket}
\usepackage{url}

\usepackage{cleveref}
\AtBeginDocument{%
\let\ref\cref
}

\@ifundefined{showcaptionsetup}{}{%
 \PassOptionsToPackage{caption=false}{subfig}}
\usepackage{subfig}
\makeatother

\begin{document}
\global\long\def\ERI#1#2{(#1|#2)}%
\global\long\def\bra#1{\Bra{#1}}%
\global\long\def\ket#1{\Ket{#1}}%
\global\long\def\braket#1{\Braket{#1}}%

\newcommand*\citeref[1]{ref. \citenum{#1}} 
\newcommand*\Citeref[1]{Ref. \citenum{#1}} 
\newcommand*\citerefs[1]{refs. \citenum{#1}} 
\newcommand\Erkale{\textsc{Erkale}}
\newcommand\Libxc{\textsc{Libxc}}
\newcommand\HelFEM{\textsc{HelFEM}}
\newcommand\PsiFour{\textsc{Psi4}}
\newcommand\xtwodhf{\textsc{x2dhf}}

\title{Importance profiles. Visualization of atomic basis set requirements}
\author{Susi Lehtola}
\affiliation{Department of Chemistry, University of Helsinki, P.O. Box 55 (A. I. Virtasen
aukio 1), FI-00014 University of Helsinki, Finland}
\email{susi.lehtola@alumni.helsinki.fi}

\begin{abstract}
Recent developments in fully numerical methods promise interesting
opportunities for new, compact atomic orbital (AO) basis sets that
maximize the overlap to fully numerical reference wave functions,
following the pioneering work of Richardson and coworkers from the
early 1960s. Motivated by this technique, we suggest a way to visualize
the importance of AO basis functions employing fully numerical wave
functions computed at the complete basis set (CBS) limit: the importance
of a normalized AO basis function $|\alpha\rangle$ centered on some
nucleus can be visualized by projecting $|\alpha\rangle$ on the set
of numerically represented occupied orbitals $|\psi_{i}\rangle$ as
$I_{0}(\alpha)=\sum_{i}\langle\alpha|\psi_{i}\rangle\langle\psi_{i}|\alpha\rangle$.
Choosing $\alpha$ to be a continuous parameter describing the orbital
basis, such as the exponent of a Gaussian-type orbital (GTO) or Slater-type
orbital (STO) basis function, one is then able to visualize the importance
of various functions. The proposed visualization $I_{0}(\alpha)$
has the important property $0\leq I_{0}(\alpha)\leq1$ which allows
unambiguous interpretation. We also propose a straightforward generalization
of the importance profile for polyatomic appliations $I(\alpha)$,
in which the importance of a test function $\ket{\alpha}$ is measured
as the increase in projection from the atomic minimal basis. We exemplify
the methods with importance profiles computed for atoms from the first
three rows, and for a set of chemically diverse diatomic molecules.
We find that the importance profile offers a way to visualize the
atomic basis set requirements for a given system in an \emph{a priori}
manner, provided that a fully numerical reference wave function is
available.
\end{abstract}
\maketitle

\section{Introduction}

Most quantum chemical calculations reported in the literature employ
the linear combination of atomic orbitals (LCAO) approach,\citep{Lehtola2019_IJQC_25968,Lehtola2020_M_1218}
in which the spin-$\sigma$ molecular orbitals (MOs) $\psi_{i\sigma}(\boldsymbol{r})$
are expanded in terms of atomic orbital (AO) basis functions $\chi_{\mu}(\boldsymbol{r})$
as 
\begin{equation}
\psi_{i\sigma}(\boldsymbol{r})=\sum_{\mu}C_{\mu i}^{\sigma}\chi_{\mu}(\boldsymbol{r}).\label{eq:lcao}
\end{equation}
The AO $\chi_{\mu}(\boldsymbol{r})$ centered on $\boldsymbol{R}_{\mu}$
is defined as a product of a radial function $R_{nl}(r)$ with a spherical
harmonic $Y_{lm}$ as

\begin{equation}
\chi_{\mu}(\boldsymbol{r})=R_{n_{\mu}l_{\mu}}(|\boldsymbol{r}-\boldsymbol{R}_{\mu}|)Y_{l_{\mu}m_{\mu}}(\widehat{\boldsymbol{r}-\boldsymbol{R}_{\mu}}).\label{eq:AO}
\end{equation}
Typically, LCAO approaches employ real-valued spherical harmonics;
however, in the case of linear molecules---which also trivially includes
the cases of atoms and diatomic molecules---complex spherical harmonics
$Y_{l}^{m}$ that afford the analytical solution with respect to the
angle $\phi$ around the bond axis\citep{Lehtola2019_IJQC_25968}
may also be employed in \ref{eq:AO}.

The accuracy of LCAO calculations is controlled by the AO basis set,
that is, the radial functions $R_{nl}(r)$ in \ref{eq:AO}. AO basis
sets are typically optimized to reproduce total energy differences
around the chemical equilibrium in order to facilitate cost-efficient
evaluation of reaction energies, for instance.\citep{Hill2013_IJQC_21,Jensen2013_WIRCMS_273}
Indeed, the reason for the popularity of LCAO calculations is that
they often offer reliable estimates of molecular properties, for instance,
because basis set truncation errors---the differences in energy between
the value predicted by the AO basis and the complete basis set (CBS)
limit---tend to be systematic across various geometries and electronic
states.\citep{Lehtola2019_IJQC_25968}

Unfortunately, while making \emph{a }basis set is straightforward,
making a \emph{good }basis set is terribly difficult because of the
conflicting requirements that define such a basis set. On the one
hand, the basis set should be as small as possible, because the larger
the atomic basis set is, the more costly it is to use in polyatomic
calculations. On the other hand, the basis set should also be transferable:
it should be similarly accurate across a wide variety of systems.
It is easy to make a basis set more transferable by adding more functions;
however, this is in opposition to the first criterion. Although it
is possible to formulate approaches to generate sequences of basis
sets that approach the CBS limit from first principles\citep{Lehtola2020_JCP_134108}---fully
numerical basis sets being an extreme example thereof---the issue
is that such benchmark quality basis sets are considered much too
large for routine calculations.

The tradeoff between the two aforementioned criteria is not always
simple. It historically led to the development of the pioneering Pople-type
$x$-$yz$G basis sets such as 3-21G,\citep{Binkley1980_JACS_939}
6-31G,\citep{Hehre1972_JCP_2257} and 6-311G,\citep{Krishnan1980_JCP_650}
and the zoo of their polarized counterparts. These basis sets have
become obsolete with the introduction new families of basis sets,
which afford an optimal balance of cost and accuracy. We would especially
like to point out here the problematic nature of the 6-311G family:
it is merely of valence double-$\zeta$ quality instead of the intended
valence triple-$\zeta$ quality,\citep{Grev1989_JCP_7305} and can
also lead to peculiar chemistries.\citep{Moran2006_JACS_9342} Unfortunately,
this is still not widely appreciated, as demonstrated by a recent
benchmark study.\citep{Pitman2023_JPCA_10295}

Modern basis set families have been designed to afford systematic
convergence towards the complete basis set (CBS) limit. The cost-accuracy
tradeoff is solved by the introduction of basis sets of prefixed size,
ranging from split-valence polarization or polarized double-$\zeta$
quality to polarized triple-$\zeta$ and higher basis sets; this also
greatly simplifies choosing the basis set, as one only needs to pick
a suitable rung, that is, the cardinal number of the basis set. Examples
of modern basis sets include the correlation consistent family,\citep{Dunning1989_JCP_1007}
the TURBOMOLE default basis sets,\citep{Weigend2003_JCP_12753} and
the polarization consistent family.\citep{Jensen2001_JCP_9113} 

We note in passing that such standard energy-optimized basis sets
are often suboptimal for modeling properties other than (differences
in) the total energy. Specially optimized basis sets that yield faster
convergence to the CBS limit have been reported for various properties
in the literature, such as magnetic properties\citep{Manninen2006_JCC_434,Lehtola2015_JCC_335,Jensen2006_JCTC_1360,Jensen2008_JCTC_719}
and electron momentum densities.\citep{Lehtola2012_JCP_104105,Lehtola2013_JCP_44109}

Extreme environments are an even more challenging case for standard
basis sets, as has been recently demonstrated for the case of strong
magnetic fields.\citep{Lehtola2020_MP_1597989,Aastroem2023__} The
magnetic fields that can be found in the atmospheres of white dwarfs
and neutron stars are strong enough to result in qualitative changes
in the electronic structures of atoms and molecules. As a result,
the basis set requirements for calculations at finite magnetic fields
are more stringent than for those at zero field, and novel types of
basis sets are required.\citep{Lehtola2020_MP_1597989,Aastroem2023__}

Having discussed various general challenges in the development of
AO basis sets, we can now comment on the practical aspects of basis
set development. Typically, the optimization of basis sets begins
by choosing a level of theory and a training set of atoms and molecules.
Sequences of basis sets with various numbers of functions are then
optimized with a given training set of systems in order to determine
the optimal composition of the basis set, which is usually chosen
with the notion of correlation\citep{Dunning1989_JCP_1007} or polarization\citep{Jensen2001_JCP_9113}
consistency; \citet{Shaw2023_JCP_44802} have recently reported a
Python package for performing such optimizations.

This procedure implies the need to carry out a large number of electronic
structure calculations with varying basis sets. If the training set
is changed by the addition or removal of some atoms or molecules,
the electronic structure calculations need to be repeated in full,
because the optimal basis set changes when the training set changes.

The question we now pose is: can we find a way to avoid having to
carry out such a large number of electronic structure calculations
when modifying the training database? The answer is yes: the maximal
overlap method of Richardson and coworkers\citep{Richardson1962_JCP_1057,Richardson1963_JCP_796}
offers a shortcut for basis set optimization: computing projections
onto a precomputed CBS limit wave function is much cheaper than carrying
out full, repeated electronic structure calculations of total energies
in the AO basis set under optimization. Thereby, optimization of the
projection is much faster than self-consistent energy optimization.
The attractiveness of the maximal overlap idea is obvious from the
number of times it has been described in the literature: projection
techniques have become an established technique in the literature
for forming compact basis sets. For instance, as discussed by \citet{Francisco1987_IJQC_279},
the later works by \citet{Kalman1971_JCP_1841} and \citet{Adamowicz1981_IJQC_545}
both describe approaches analogous to that of \citeauthor{Richardson1962_JCP_1057}. 

Fully numerical calculations on diatomic molecules were already possible
four decades ago,\citep{Lehtola2019_IJQC_25968} and already \citet{Adamowicz1983_IJQC_19}
applied an overlap maximization procedure to fully numerical references
in order to produce accurate AO basis sets. For completeness, we mention
in this context also the work on studying the deficiencies in Gaussian
basis calculations with fully numerical diatomic calculations of Kobus,
Moncrieff, and Wilson.\citep{Kobus1997_MP_1015,Kobus2001_MP_315} 

More recently, projections to fully numerical atomic wave functions
have been used by \citet{VanLenthe2003_JCC_56} to fit Slater-type
orbital basis sets. Projection techniques have likewise found use
in the solid state: reference calculations with plane waves have been
used to construct small numerical atomic basis sets.\citep{SanchezPortal1995_SSC_685,SanchezPortal1996_JPCM_3859,Chen2009_PRB_165121,Chen2010_JPCM_445501,Chen2011_JPCM_325501,Lange2011_PRB_85101,Lin2021_PRB_235131}
Although these techniques are likewise derivative of the maximal overlap
technique of Richardson and coworkers\citep{Richardson1962_JCP_1057,Richardson1963_JCP_796}
and its many later applications, this connection does not appear to
have been made before in the literature.

As already hinted above, fully numerical methods (see \citeref{Lehtola2019_IJQC_25968}
for a recent review) allow the direct determination of CBS limit total
energies and wave functions for Hartree--Fock, density functional
theory,\citep{Hohenberg1964_PR_864,Kohn1965_PR_1133} as well as for
complete active space self-consistent field calculations, and are
nowadays tractable for systems of appreciable size.\citep{Jensen2017_JPCL_1449,Brakestad2021_JCP_214302,Vaughn2021_JCP_110101,Valeev2023__}
All-electron calculations have recently become feasible with plane
waves, as well, through the use of a regularized nuclear Coulomb potential.\citep{Gygi2023_JCTC_1300,Lehtola2023_JCTC_4033}
These developments merit new attention on the maximal overlap method:
a database of fully numerical all-electron wave functions for a set
of chemically diverse systems would offer an excellent starting point
for fitting novel, systematic and error-balanced AO basis sets with
machine learning techniques, for instance. 

As a key step to building such a database, we point out that the projections
involved in the computation of the overlaps can also be used as a
visual tool. We will introduce \emph{importance profiles} $I(\alpha)$
that measure the projection of a AO test function with parameter $\alpha$
(here the exponent of a Gaussian or Slater type orbital) onto the
fully numerical orbitals of a given system. These importance profiles
thereby reveal the electronic structure of the system, and can also
be used to study what kind of AO basis functions should be used to
model the studied system.

The driving idea behind this work was to find out whether the importance
profile would offer an unambiguous way to characterize the electronic
structure of various molecules: would it be possible to see the differences
in polarization effects and optimal polarization exponents for atoms
in different molecules, for example H in \ce{H2} \emph{vs }H in HF?
Quantifying the dissimilarity of the basis set requirements of various
molecules would be of great help in constructing the training and
test sets of molecules that could be used to optimize new, compact
and efficient AO basis sets following the maximal overlap technique
of \citeauthor{Richardson1962_JCP_1057}.\citep{Richardson1962_JCP_1057,Richardson1963_JCP_796}

The layout of this work is the following. Next, in \ref{sec:Method},
the basis of the proposed visualization method is outlined: the employed
radial functions are discussed in \ref{subsec:Radial-functions},
the closely related completeness profiles of \citet{Chong1995_CJC_79}
are briefly reviewed in \ref{subsec:Completeness-profile}, and the
importance profiles are introduced in \ref{subsec:Importance-profile}.
Details of the implementation of the importance profiles are presented
in \ref{sec:Implementation}. Applications of the method on a set
of chemically diverse diatomic molecules are presented in \ref{sec:Results}.
We describe the atomic calculations used to build the minimal NAO
basis set in \ref{subsec:Atomic-calculations}. The molecular calculations
are discussed in \ref{subsec:Molecular-calculations}. \Cref{subsec:Spin-polarized-atomic-calculatio}
delves into the questions of non-orthogonality and exactness of the
NAO basis set with atomic calculations in the diatomic numerical basis.
The goodness of STOs \emph{vs} GTOs is compared in \ref{subsec:STO-analysis}.
Deficiencies in the minimal NAO basis are analyzed in \ref{subsec:Minimal-basis-analysis}.
The article concludes in a brief summary and discussion in \ref{sec:Summary-and-discussion}. 

\section{Method \label{sec:Method}}

\subsection{Radial Functions \label{subsec:Radial-functions}}

As recently reviewed in \citeref{Lehtola2019_IJQC_25968}, various
radial functions can be used in the AOs of \ref{eq:AO}. Gaussian-type
orbitals (GTOs) have a radial part defined by
\begin{equation}
R_{nl}^{\text{GTO}}(r)=\frac{2^{l+2}\alpha_{nl}^{\left(2l+3\right)/4}}{\left[\left(2\pi\right)^{1/4}\left(2l+1\right)!!\right]^{1/2}}r^{l}e^{-\alpha_{nl}r^{2}}.\label{eq:r-gto}
\end{equation}
GTOs are the pre-eminently employed basis set in quantum chemistry.
Although \ref{eq:r-gto} shows the primitive form, GTOs are typically
used in contracted form.\citep{Hill2013_IJQC_21,Jensen2013_WIRCMS_273}
However, the analysis of the projection onto the primitive GTOs of
\ref{eq:r-gto} offers a good starting point for the construction
of contracted GTO basis sets, as well.

Slater-type orbitals (STOs) whose radial part is given by

\begin{equation}
R_{nl}^{\text{STO}}(r)=\frac{\left(2\zeta_{nl}\right)^{l+3/2}}{\left[\left(2l+2\right)!\right]^{1/2}}r^{l}e^{-\zeta_{nl}r}\label{eq:r-sto}
\end{equation}
are a less commonly used option, as molecular integrals are more difficult
to evaluate in this basis set. It is commonly argued that STOs are
a better basis set than GTOs,\citep{Gueell2008_JPCA_91} because they
have the right asymptotic form to satisfy the Kato cusp condition
at the nucleus,\citep{Kato1957_CPAM_151} as well as to capture the
exponential decay of the Hartree--Fock and Kohn--Sham orbitals.\citep{Ahlrichs1972_CPL_609,Ahlrichs1973_CPL_521,Katriel1980_PNASUSA_4403,Ishida1992_TCA_355}
But, whether this actually holds in general systems is debatable:
as the asymptotic behavior far from the nucleus depends on the energy
of the highest occupied orbital which is system dependent, it is not
obvious that a basis set that has the right asymptotic form for fixed
values of $\zeta$ yields more accurate results than a GTO basis set,
for instance, because the asymptotic decay of the STO basis functions
will not match that of a general polyatomic system.

We note that numerical atomic orbitals (NAOs) are yet another option;
see \citerefs{Lehtola2019_IJQC_25968} and \citenum{Lin2023_WCMS_1687}
for reviews. NAOs are extremely powerful in principle, as they can
afford the exact solution to the non-interacting atom: not only do
NAOs have the right asymptotic behavior close to the nucleus and far
away from it like STOs, NAOs are also exact everywhere in-between
in the case of the non-interacting atom. Although the methods discussed
herein are also applicable to NAOs, the complication of NAOs is that
the form of $R_{nl}(r)$ is not restricted to a simple analytic form
with a single adjustable parameter like $\alpha$ in \ref{eq:r-gto}
or $\zeta$ in \ref{eq:r-sto}; thus, for simplicity, we will not
discuss NAOs and will explicitly focus on GTOs and STOs in this work.
Indeed, many NAO codes such as GPAW\citep{Larsen2009_PRB_195112}
and FHI-aims\citep{Blum2009_CPC_2175} employ GTOs or STOs as polarization
functions.

We also note the same drawback for NAOs as for GTOs and STOs, in that
NAOs are not exact for polyatomic systems. However, their flexibility
means that NAO basis sets can be more accurate than GTO or STO basis
sets.\citep{Jensen2017_JPCL_1449,Jensen2017_JPCA_6104,Feller2018_JPCA_2598}

\subsection{Completeness Profile \label{subsec:Completeness-profile}}

The completeness profile\citep{Chong1995_CJC_79} is a way to visualize
the completeness of AO basis sets, which can be quantified by the
goodness of satisfaction of the of the resolution of the identity
\begin{equation}
\sum_{\mu\nu}\ket{\chi_{\mu}}(\boldsymbol{S}^{-1})_{\mu\nu}\text{\ensuremath{\bra{\chi_{\nu}}}}\approx\boldsymbol{1}\label{eq:RI-basis}
\end{equation}
which is inherent in the LCAO expansion of \ref{eq:lcao}. The overlap
matrix, whose inverse is employed in \ref{eq:RI-basis}, has elements
$S_{\mu\nu}=\braket{\chi_{\mu}|\chi_{\nu}}$.

Studying how well the basis set can represent a normalized ($\braket{\alpha|\alpha}=1$)
primitive test function $\ket{\alpha}$ parametrized by $\alpha$,
which is typically an exponent, one obtains the completeness profile
\begin{equation}
Y(\alpha)=\sum_{\mu\nu}\braket{\alpha|\chi_{\mu}}\braket{\chi_{\mu}|\chi_{\nu}}^{-1}\text{\ensuremath{\braket{\chi_{\nu}|\alpha}}}.\label{eq:cpl}
\end{equation}
The completeness profile has been employed in completeness optimization,\citep{Manninen2006_JCC_434,Lehtola2015_JCC_335}
which has been succesfully applied to parametrization of basis sets
tuned for the reproduction of various properties.\citep{Manninen2006_JCC_434,Ikaelaeinen2008_JCP_124102,Ikaelaeinen2009_PCCP_14,Ikaelaeinen2010_PRL_153001,Ikaelaeinen2012_JCTC_91,Lantto2011_JPCA_23,Fu2013_JCP_204110,Vaara2013_JCP_104313,Abuzaid2013_MP_1390,Vaehaekangas2013_PCCP_41,Vaehaekangas2014_JPCC_23996,Lehtola2012_JCP_104105,Lehtola2013_JCP_44109,Lehtola2015_JCC_335,Rossi2015_JCP_94114,Hanni2017_PRA_32509}
A two-electron completeness profile for the assesment of suitability
for electron correlation effects has also been suggested.\citep{Auer2002_JCC_5}

\subsection{Importance Profile \label{subsec:Importance-profile}}

The completeness profile of \ref{eq:cpl} can be straightforwardly
applied to calculations with real-space basis sets with \cref{eq:RI-basis,eq:cpl}
as a way to visualize the flexibility of the real-space basis set.
A flexible real-space basis set is able to represent AO basis functions
for a wide range of exponents $\alpha$, which is demonstrated by
$Y(\alpha)\approx1$.

Alternative metrics can also be fashioned. A projection of the test
function onto the occupied orbitals computed at the CBS limit in the
real-space basis yields 
\begin{equation}
I_{0}(\alpha)=\sum_{i\text{ occ}}\braket{\alpha|\psi_{i}}\braket{\psi_{i}|\alpha}=\sum_{i\text{ occ}}\left|\braket{\psi_{i}|\alpha}\right|^{2},\label{eq:base-importance-profile}
\end{equation}
where the sum runs over the occupied orbitals $i$. As the metric
in \ref{eq:base-importance-profile} measures the weight of the test
function $\alpha$ in the electronic structure, we will call $I_{0}(\alpha)$
the \emph{free-atom importance profile}. Like the completeness profile,
this importance profile satisfies $0\leq I_{0}(\alpha)\leq1$.

The importance profile has an important connection to the maximal
overlap method. Inserting the resolution of the identity in the AO
basis, \ref{eq:RI-basis}, into the occupied-space projection
\begin{equation}
\sum_{i}\ket{\psi_{i}}\bra{\psi_{i}}=\boldsymbol{1},\label{eq:occ-proj}
\end{equation}
we obtain the occupied-orbital AO projection
\begin{equation}
P(\{|\mu\rangle\})=\sum_{\alpha\beta}\sum_{i\text{ occ}}\braket{\psi_{i}|\mu}(\boldsymbol{S}^{-1})_{\mu\nu}\braket{\nu|\psi_{i}}\label{eq:overlap}
\end{equation}
which is the quantity that is maximized in the maximum overlap method
of \citeauthor{Richardson1962_JCP_1057}\citep{Richardson1962_JCP_1057,Richardson1963_JCP_796},
also used by a variety of other authors in the literature (see Introduction
for discussion), by optimizing the parameters in the AO basis set. 

It is easy to see that in the case of an orthonormal AO basis set,
$S_{\mu\nu}=\delta_{\mu\nu}$, the sum of the basis functions' \emph{importances}
$\sum_{\mu}I_{0}(\alpha_{\mu})$ equals the overlap $P$ of the occupied
orbitals and the basis functions of \ref{eq:overlap}. $I_{0}(\alpha)$
therefore carries information on the overlap of the basis function
with parameter $\alpha$, and can be used to inspect basis function
requirements in atomic systems. Basis functions parametrized by $\alpha$
that have large $I_{0}(\alpha)$ should likely be included in the
maximal overlap AO basis due to the connection of \cref{eq:importance-profile,eq:overlap}. 

When applied to polyatomic systems, the importance profile given by
$I_{0}$ also carries information on the non-orthogonality of atomic
basis functions on different centers, which is the major headache
in the design and development of atomic basis sets. A complete basis
set can in principle be spanned by functions centered on a single
atom, and this is the physical interpretation of $I_{0}(\alpha)$.
Although $I_{0}(\alpha)$ can still be used to illustrate the non-orthonormality
of polyatomic AO basis sets, it does not afford good chemical insight.

In order to isolate the effects of chemistry---the aim being to dig
out the changes in the electronic structure from the free atoms---we
therefore need to build in the baseline of non-interacting atoms.
This is easily achieved by formulating an analogue of $I_{0}(\alpha)$
computed in the presence of a minimal NAO basis on each atom in the
system. Such an analysis should allow for the isolation of the effects
of breathing and polarization functions, the former describing changes
in the effective size of the atom in a molecule, and the latter describing
inhomogeneities in the electron density compared to the free atom;
minimal basis techniques are commonly used for various kinds of chemical
analyses.\citep{Lu2004_JCP_37,Lu2004_JCP_2638,Lu2004_PRB_41101,Knizia2013_JCTC_4834,Knizia2015_ACIE_5518,Janowski2014_JCTC_3085,Clement2021_JCTC_7406} 

A straightforward generalization of \ref{eq:base-importance-profile}
is afforded by the difference in overlap between that afforded by
a minimal NAO basis padded with the function $\ket{\alpha}$ and that
of the baseline of simply the minimal NAO basis on the atoms. The
\emph{importance profile}, which measures the importance of the test
function $\ket{\alpha}$ in the presence of a minimal NAO basis, can
therefore be computed as 
\begin{equation}
I(\alpha)=P(\text{\ensuremath{\ket{\text{minimal basis}}}}+\ket{\alpha})-P(\ket{\text{minimal basis}}),\label{eq:importance-profile}
\end{equation}
where $P$ is the occupied-orbital AO projection of \ref{eq:overlap}.
Note that in the absence of a minimal basis, \ref{eq:importance-profile}
reduces to \ref{eq:base-importance-profile}. We note that minimal
NAO basis sets are commonly used with GTO or STO polarization functions
e.g. in the FHI-aims program.\citep{Blum2009_CPC_2175}

\section{Implementation \label{sec:Implementation}}

We have recently described finite element implementations for all-electron
Hartree--Fock and Kohn--Sham density functional theory\citep{Hohenberg1964_PR_864,Kohn1965_PR_1133}
for atoms\citep{Lehtola2019_IJQC_25945,Lehtola2020_PRA_12516,Lehtola2023_JCTC_2502,Lehtola2023_JPCA_4180,Lehtola2023_JCTC_4033}
as well as diatomic molecules\citep{Lehtola2019_IJQC_25944} in the
\textsc{HelFEM} program.\citep{Lehtola2018__} The projection code
supports both GTOs and STOs, and either can be used to probe the completeness
of (i) the finite element basis set or (ii) the occupied orbital space.

The minimal NAO basis set is obtained from atomic calculations with
full spin-restriction and fractional orbital occupations,\citep{Lehtola2020_PRA_12516,Lehtola2023_JCTC_2502}
following the standard practice in the NAO literature.\citep{Sankey1989_PRB_3979,Porezag1995_PRB_12947,Horsfield1997_PRB_6594,SanchezPortal1997_IJQC_453,Kenny2000_PRB_4899,Junquera2001_PRB_235111,Anglada2002_PRB_205101,Ozaki2003_PRB_155108,Ozaki2004_JCP_10879,Ozaki2004_PRB_195113,Blum2009_CPC_2175,Larsen2009_PRB_195112,Shang2010_IRPC_665,Louwerse2012_PRB_35108,Corsetti2013_JPCM_435504}
The diatomic calculations are carried out in the prolate spheroidal
coordinate system $(\mu,\nu,\phi)$, which has been described in \citerefs{Lehtola2019_IJQC_25968}
and \citenum{Lehtola2019_IJQC_25944}. The numerical basis set is
given by
\begin{align}
\chi_{nlm}(\mu,\nu,\phi)= & B_{n}(\mu)Y_{l}^{m}(\nu,\phi),\label{eq:basis}
\end{align}
where the spherical harmonics are given by
\begin{equation}
Y_{l}^{m}(\theta,\phi)=\left(-1\right)^{m}\sqrt{\frac{2l+1}{4\pi}\frac{\left(l-m\right)!}{\left(l+m\right)!}}P_{l}^{m}(\cos\theta)e^{im\phi}.\label{eq:Ylm}
\end{equation}
For simplicity, we will assume complex AOs also in \ref{eq:AO}. 

Computing the importance profile of \ref{eq:base-importance-profile}
requires the calculation of integrals by quadrature. The volume element
in the prolate spheroidal coordinate system is given by\citep{Lehtola2019_IJQC_25944}
\begin{equation}
{\rm d}V=R_{h}^{3}\sinh\mu\sin\nu\left(\cosh^{2}\mu-\cos^{2}\nu\right){\rm d}\phi{\rm d}\nu{\rm d}\mu.\label{eq:dV}
\end{equation}
The integral over $\phi$ in the AO projection $\braket{\psi_{i}|\alpha}$
of \ref{eq:base-importance-profile}, can now be done analytically,
as can be seen from \ref{eq:Ylm}. This integral yields 
\begin{align}
\int_{0}^{2\pi}e^{-im'\phi}e^{im\phi}{\rm d}\phi= & 2\pi\delta_{mm'}.\label{eq:phi-integral}
\end{align}
The integrals over the $\mu$ and $\nu$ dimensions, in turn, are
evaluated by quadrature with the methodology discussed in \citeref{Lehtola2019_IJQC_25944}.
A similar technique was also used to implement the superposition of
atomic potentials initial guess described in \citeref{Lehtola2019_JCTC_1593}
in \textsc{HelFEM}.

Importance profiles are computed for both nuclei in the system. The
necessary relations between the $(\mu,\nu,\phi)$ coordinates of the
fully numerical calculation and the $(r,\theta,\phi)$ coordinates
needed to evaluate the AOs at the two nuclei A and B at $(0,0,-R_{h})$
and $(0,0,R_{h})$ are\citep{Yasui1982_JCP_468}
\begin{align*}
r_{A/B}= & R_{h}\left(\cosh\mu\pm\cos\nu\right),\\
z_{A/B}= & R_{h}\left(\cosh\mu\cos\nu\pm1\right),\\
\cos\theta_{A/B}= & z/r_{A/B},
\end{align*}
where $R_{h}=R/2$ is one half of the bond length $R$, and the upper
and lower signs are chosen for A and B, respectively.

We also looked into mid-bond projections, for which
\begin{align*}
r= & R_{h}\sqrt{\cosh^{2}\mu+\cos^{2}\nu-1},\\
z= & R_{h}\cosh\mu\cos\nu,\\
\cos\theta= & z/r.
\end{align*}
However, as the completeness profiles for the mid-bond projections
suggested that the employed numerical basis set is not sufficiently
complete to afford a thorough analysis on the importance of mid-bond
functions, we do not discuss mid-bond projections in this work.

\section{Results \label{sec:Results}}

We study a set of chemically diverse diatomic molecules from the database
of \citet{Weigend2005_PCCP_305}. \citet{Weigend2005_PCCP_305} employed
their database, which also contains larger molecules, to test the
TURBOMOLE default basis sets. The diatomics for the first three rows
suffice for the present study; the studied systems and the computed
CBS limit energies are shown in \cref{tab:energy}. To assess the
role of the non-orthogonality of atomic orbitals in polyatomic molecules
as well as the exactness of the NAO basis, the list of systems studied
also includes the H atom at the \ce{H2} geometry, the Li atom at
the LiH geometry, and the F atom at the HF geometry.

The numerical basis sets were determined with the proxy method of
\citeref{Lehtola2019_IJQC_25944} with the threshold $\epsilon=10^{-10}$,
which have been shown to lead to $\mu E_{h}$ level precision total
Hartree--Fock energies in \citeref{Lehtola2019_IJQC_25944}. All
calculations employed the Perdew--Burke--Ernzerhof (PBE) generalized-gradient
approximation (GGA) functional\citep{Perdew1996_PRL_3865,Perdew1997_PRL_1396}
as implemented in the Libxc library of density functional approximations.\citep{Lehtola2018_S_1}
The calculations were started from a superposition of atomic potentials
(SAP)\citep{Lehtola2019_JCTC_1593} which we have found to offer a
reasonable and easy to implement starting point for fully numerical
calculations. 

\begin{table}
\begin{tabular}{llrrrr}
System & $ M $ & R ($ a_0 $) & E ($ E_h $) & $ \Delta N_\alpha $ & $ \Delta N_\beta $ \\
\hline
\hline
\ce{H} & 2 & 1.449815 & $ -0.499990 $ & 0.002 & 0.000 \\
\ce{H2} & 1 & 1.449815 & $ -1.166566 $ & 0.038 & 0.038 \\
\ce{Li} & 2 & 3.093955 & $ -7.462178 $ & 0.001 & 0.000 \\
\ce{LiH} & 1 & 3.093955 & $ -8.047240 $ & 0.020 & 0.020 \\
\ce{Li2} & 1 & 5.281990 & $ -14.956003 $ & 0.025 & 0.025 \\
\ce{F} & 2 & 1.762862 & $ -99.667411 $ & 0.001 & 0.001 \\
\ce{HF} & 1 & 1.762862 & $ -100.402690 $ & 0.036 & 0.036 \\
\ce{LiF} & 1 & 2.973943 & $ -107.359771 $ & 0.059 & 0.059 \\
\ce{N2} & 1 & 2.101579 & $ -109.460030 $ & 0.099 & 0.099 \\
\ce{CO} & 1 & 2.158027 & $ -113.242611 $ & 0.085 & 0.085 \\
\ce{NaH} & 1 & 3.588406 & $ -162.740855 $ & 0.007 & 0.007 \\
\ce{F2} & 1 & 2.669445 & $ -199.436648 $ & 0.016 & 0.016 \\
\ce{NaF} & 1 & 3.632959 & $ -262.031125 $ & 0.048 & 0.048 \\
\ce{AlN} & 3 & 3.403664 & $ -296.880123 $ & 0.051 & 0.025 \\
\ce{MgF} & 2 & 3.358309 & $ -299.806497 $ & 0.171 & 0.047 \\
\ce{BeS} & 1 & 3.344885 & $ -412.721526 $ & 0.233 & 0.233 \\
\ce{HCl} & 1 & 2.450614 & $ -460.644416 $ & 0.044 & 0.044 \\
\ce{LiCl} & 1 & 3.876989 & $ -467.612398 $ & 0.050 & 0.050 \\
\ce{ClF} & 1 & 3.170192 & $ -559.766155 $ & 0.034 & 0.034 \\
\ce{KH} & 1 & 4.405059 & $ -600.275655 $ & 0.019 & 0.019 \\
\ce{NaCl} & 1 & 4.462511 & $ -622.297756 $ & 0.035 & 0.035 \\
\ce{P2} & 1 & 3.620777 & $ -682.425446 $ & 0.077 & 0.077 \\
\ce{KF} & 1 & 4.269240 & $ -699.579273 $ & 0.070 & 0.070 \\
\ce{S2} & 3 & 3.655904 & $ -796.089339 $ & 0.072 & 0.070 \\
\ce{ScO} & 2 & 3.143512 & $ -835.707028 $ & 0.526 & 0.414 \\
\ce{Cl2} & 1 & 3.891259 & $ -920.053496 $ & 0.039 & 0.039 \\
\ce{TiO} & 3 & 3.053330 & $ -924.438044 $ & 0.075 & 0.020 \\
\ce{VO} & 4 & 2.992120 & $ -1018.984001 $ & 0.060 & 0.016 \\
\ce{KCl} & 1 & 5.186142 & $ -1059.844176 $ & 0.050 & 0.050 \\
\ce{MnO} & 6 & 3.041319 & $ -1225.974540 $ & 0.878 & 0.046 \\
\ce{FeO} & 5 & 3.004636 & $ -1338.664046 $ & 0.675 & 0.057 \\
\ce{MnS} & 6 & 3.852394 & $ -1548.863976 $ & 0.632 & 0.073 \\
\ce{NiO} & 3 & 3.034982 & $ -1583.278125 $ & 0.347 & 0.340 \\
\ce{CuH} & 1 & 2.758533 & $ -1640.901677 $ & 0.015 & 0.015 \\
\ce{CuF} & 1 & 3.293533 & $ -1740.137149 $ & 0.050 & 0.050 \\
\ce{NiS} & 3 & 3.713630 & $ -1906.193757 $ & 0.256 & 0.234 \\
\ce{GaO} & 2 & 3.233973 & $ -1999.766136 $ & 0.048 & 0.029 \\
\ce{GaF} & 1 & 3.394640 & $ -2024.481823 $ & 0.036 & 0.036 \\
\ce{CuCl} & 1 & 3.919086 & $ -2100.413400 $ & 0.046 & 0.046 \\
\ce{GeO} & 1 & 3.113928 & $ -2151.946560 $ & 0.054 & 0.054 \\
\ce{GaCl} & 1 & 4.223980 & $ -2384.730041 $ & 0.033 & 0.033 \\
\ce{SeO} & 3 & 3.162357 & $ -2476.398637 $ & 0.067 & 0.049 \\
\ce{HBr} & 1 & 2.713347 & $ -2574.433761 $ & 0.039 & 0.039 \\
\ce{BrCl} & 1 & 4.152902 & $ -3033.858918 $ & 0.034 & 0.034 \\
\ce{KBr} & 1 & 5.511775 & $ -3173.640935 $ & 0.047 & 0.047 \\
\ce{Cu2} & 1 & 4.147742 & $ -3280.675482 $ & 0.022 & 0.022 \\
\ce{Br2} & 1 & 4.406727 & $ -5147.663465 $ & 0.031 & 0.031 \\

\end{tabular}
\caption{Systems included in the present study, including the employed spin multiplicity $ M $, bond length $ R $, as well as the PBE total energy $ E $ for the wave functions determined in $ C_{\infty h} $ (heteroatomics) or $ D_{\infty h}$ (homoatomics) symmetry that were used in the analysis. The last two columns show the number of alpha and beta electrons, $\Delta N_\alpha$ and $\Delta N_\beta$, which are not described by the minimal NAO basis. }
\label{tab:energy}
\end{table}

\subsection{Atomic calculations \label{subsec:Atomic-calculations}}

\begin{figure*}
\subfloat[GTO]{\includegraphics[width=0.45\textwidth]{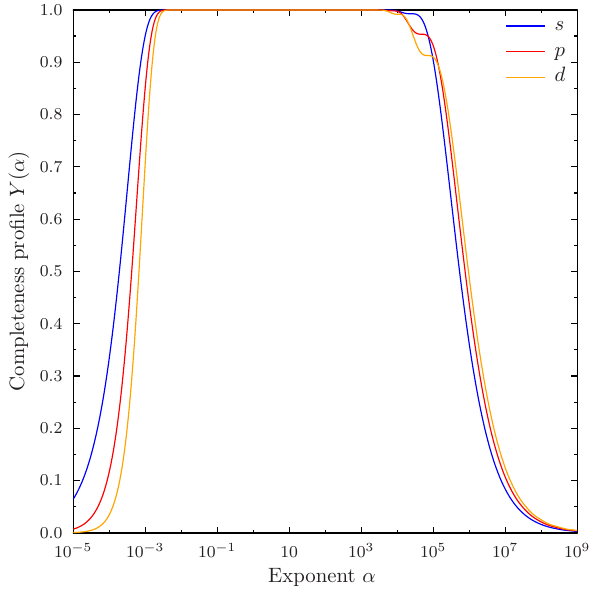}

}\subfloat[STO]{\includegraphics[width=0.45\textwidth]{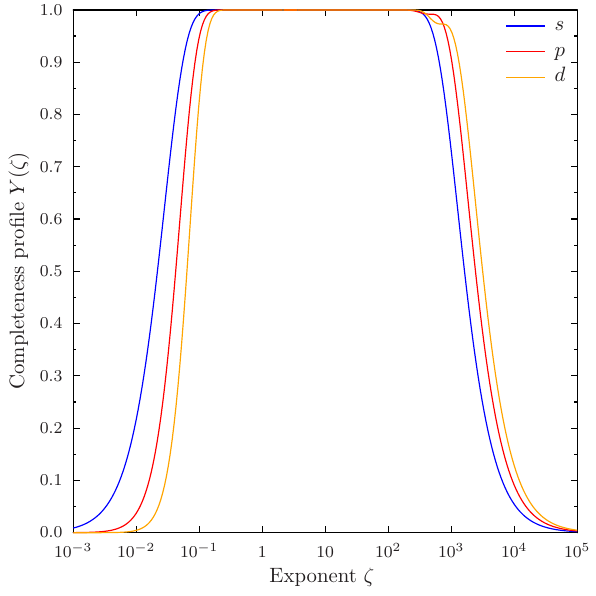}

}\caption{Completeness profile for the atomic radial finite element basis, probed
by a GTO or an STO.\label{fig:Completeness-profile}}
\end{figure*}

We begin with the analysis of the atomic calculations, which generate
the minimal basis for the molecular calculations. The NAOs were determined
with five radial elements of 15-node Lagrange interpolating polynomials
for the PBE ground state configurations given in \ref{tab:Atomic-PBE-ground}.
The atomic calculations used a practical infinity $r_{\infty}=40a_{0}$,
which is sufficient to guarantee convergence of the total energy to
sub-$\mu E_{h}$ precision to the free-atom limit. Note that typical
applications of NAOs in the literature employ confinement potentials
to enhance the locality of the atomic basis,\citep{Sankey1989_PRB_3979,Porezag1995_PRB_12947,Horsfield1997_PRB_6594,SanchezPortal1997_IJQC_453,Kenny2000_PRB_4899,Junquera2001_PRB_235111,Anglada2002_PRB_205101,Ozaki2003_PRB_155108,Ozaki2004_JCP_10879,Ozaki2004_PRB_195113,Blum2009_CPC_2175,Larsen2009_PRB_195112,Shang2010_IRPC_665,Louwerse2012_PRB_35108,Corsetti2013_JPCM_435504}
which has not been in this work as we are interested in reproducing
the free-atom limit. The completeness profile of this universal atomic
basis is shown in \ref{fig:Completeness-profile}, demonstrating that
the finite element basis can reliably describe also GTOs and STOs
with various exponents.

\begin{figure*}
\subfloat[GTO \label{fig:Br-GTO}]{\includegraphics[width=0.45\textwidth]{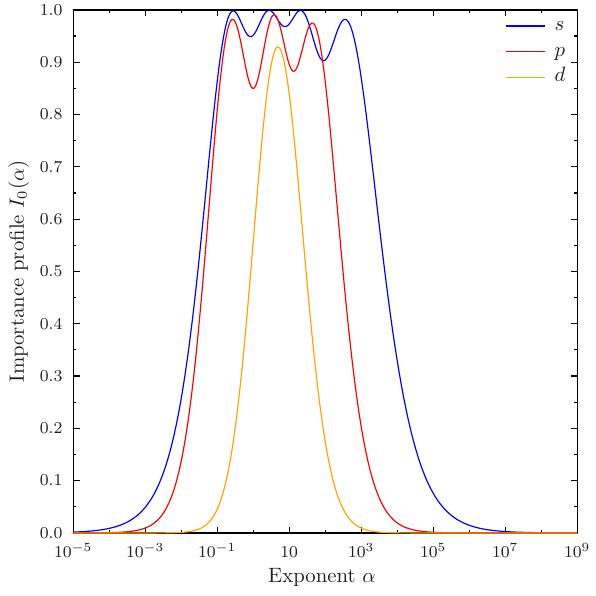}

}\subfloat[STO \label{fig:Br-STO}]{\includegraphics[width=0.45\textwidth]{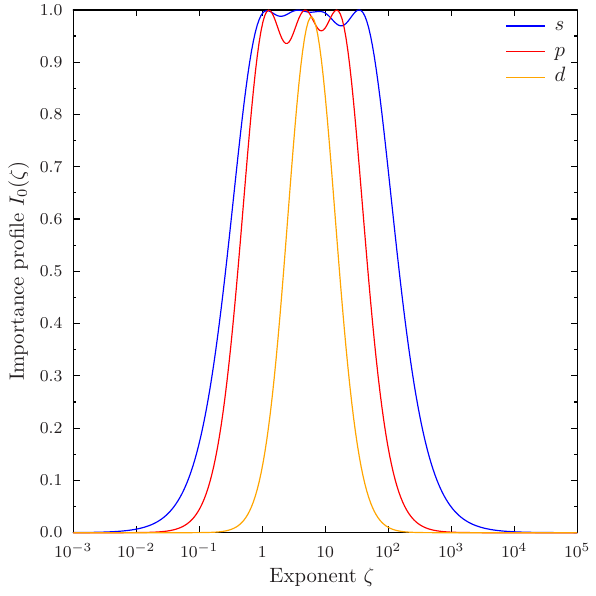}

}\caption{Importance profiles for the atomic minimal basis radial functions
of the Br atom, probed by a GTO or an STO.\label{fig:Br-importance-profile}}
\end{figure*}

The importance profile for GTOs and STOs in case of the Br atom are
shown in \ref{fig:Br-importance-profile}. Analogous plots for the
other atoms are included in the Supporting Information. In addition
to the PBE functional used for this calculation, we also ran calculations
for the atomic configurations of \ref{tab:Atomic-PBE-ground} using
the Perdew--Wang (PW92) local density approximation (LDA) \citep{Bloch1929_ZfuP_545,Dirac1930_MPCPS_376,Perdew1992_PRB_13244}
and the Tao--Perdew--Staroverov--Scuseria\citep{Tao2003_PRL_146401,Perdew2004_JCP_6898}
(TPSS) meta-GGA functional. The resulting importance profiles, which
are shown in the Supporting Information, were found to be similar
to those obtained with the PBE functional.

The atomic importance plots could be used to determine GTO and STO
basis sets that are suitable for describing the non-interacting atom.
The comparison of \ref{fig:Br-GTO,fig:Br-STO} reveals that individual
STOs do have a higher overlap with the exact numerical wave function
for the spin-restricted atom than that of individual GTOs, suggesting
that they indeed are a better basis for electronic structure calculations.
But, as we will see in the next section, few differences can be seen
in the goodness of STOs \emph{vs} GTOs for capturing polarization
effects.

\begin{figure}
\begin{centering}
\includegraphics[width=0.3\textwidth]{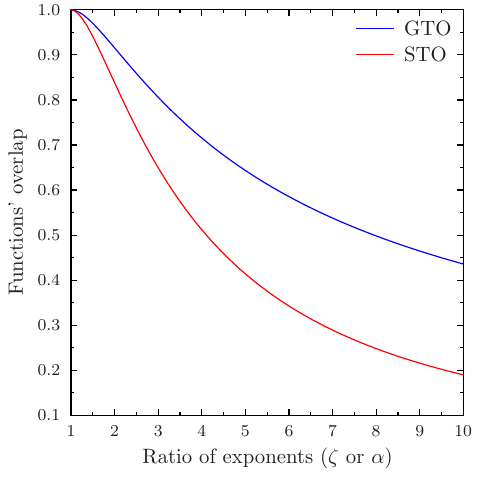}
\par\end{centering}
\caption{One-center overlap of $s$-type GTO and STO functions.\label{fig:One-center-overlap-of}}
\end{figure}

What is also noteworthy here is that the importance profiles are wider
in the GTO basis than in the STO basis. This might suggest that it
would be easier to span the AO basis set in GTOs than in STOs, since
the STO importance profile is so much more peaked. However, a fair
assessment also requires taking into account the non-orthogonality
of the STO and GTO basis functions. Taking the one-center case for
simplicity, the overlap of two $s$-type STOs is
\begin{equation}
S^{\text{STO}}(\zeta_{1},\zeta_{2})=\frac{8\sqrt{\zeta_{1}^{3}\zeta_{2}^{3}}}{\left(\zeta_{1}+\zeta_{2}\right)^{3}}=\frac{8\left(\frac{\zeta_{2}}{\zeta_{1}}\right)^{3/2}}{\left(1+\left(\frac{\zeta_{2}}{\zeta_{1}}\right)\right)^{3}}\label{eq:sto-overlap}
\end{equation}
while that of $s$-type GTOs is
\begin{equation}
S^{\text{GTO}}(\alpha_{\alpha},\zeta_{\beta})=\frac{2\sqrt{2\sqrt{\alpha_{1}^{3}\alpha_{2}^{3}}}}{\left(\alpha_{1}+\alpha_{2}\right)^{3/2}}=\frac{2\sqrt{2}\left(\frac{\alpha_{2}}{\alpha_{1}}\right)^{3/4}}{\left(1+\frac{\alpha_{2}}{\alpha_{1}}\right)^{3/2}}.\label{eq:gto-overlap}
\end{equation}
This function is shown in \ref{fig:One-center-overlap-of}: the overlap
of two STOs with different exponents decays much more quickly than
that of GTOs, which also explains why the STO importance profiles
are more peaked. This also means that a one-center expansion in terms
of STOs can use exponents that are more tightly spaced than a corresponding
Gaussian-basis one.

Note that since GTOs and STOs are not exact for the non-interacting
atom, these basis sets are more prone to basis set superposition errors
than NAO basis sets: when the atomic basis is not exact, the description
of the orbitals on the atom can be improved by ``borrowing'' the
basis functions on other nuclei.\citep{Jansen1969_CPL_140,Boys1970_MP_553}
However, as we will demonstrate later in this work, typical NAO basis
sets are also not exact for non-interacting atoms, because the NAO
basis sets are derived from spin-restricted calculations.

\begin{table}
\begin{centering}
\begin{tabular}{lll}
Atom & Configuration & Energy\tabularnewline
\hline 
\hline 
H & $1s^{1}$ & -0.458929\tabularnewline
Li & {[}He{]}$2s^{1}$ & -7.451372\tabularnewline
Be & {[}He{]}$2s^{2}$ & -14.629948\tabularnewline
C & {[}He{]}$2s^{2}2p^{2}$ & -37.748209\tabularnewline
N & {[}He{]}$2s^{2}2p^{3}$ & -54.420997\tabularnewline
O & {[}He{]}$2s^{2}2p^{4}$ & -74.945193\tabularnewline
F & {[}He{]}$2s^{2}2p^{5}$ & -99.650701\tabularnewline
Na & {[}Ne{]}$3s^{1}$ & -162.164594\tabularnewline
Mg & {[}Ne{]}$3s^{2}$ & -199.955115\tabularnewline
Al & {[}Ne{]}$3s^{2}3p^{1}$ & -242.224964\tabularnewline
P & {[}Ne{]}$3s^{2}3p^{3}$ & -341.046984\tabularnewline
S & {[}Ne{]}$3s^{2}3p^{4}$ & -397.914814\tabularnewline
Cl & {[}Ne{]}$3s^{2}3p^{5}$ & -459.962588\tabularnewline
K & {[}Ar{]}$4s^{1}$ & -599.704849\tabularnewline
Sc & {[}Ar{]}$4s^{2}4p^{1}$ & -760.251445\tabularnewline
Ti & {[}Ar{]}$4s^{2}3d^{2}$ & -849.081416\tabularnewline
V & {[}Ar{]}$4s^{2}3d^{3}$ & -943.582132\tabularnewline
Mn & {[}Ar{]}$3d^{7}$ & -1150.524945\tabularnewline
Fe & {[}Ar{]}$3d^{8}$ & -1263.294905\tabularnewline
Ni & {[}Ar{]}$3d^{10}$ & -1508.033427\tabularnewline
Cu & {[}Ar{]}$3d^{10}4s^{1}$ & -1640.290260\tabularnewline
Ga & {[}Ar{]}$3d^{10}4s^{2}4p^{1}$ & -1924.563718\tabularnewline
Ge & {[}Ar{]}$3d^{10}4s^{2}4p^{2}$ & -2076.632863\tabularnewline
Se & {[}Ar{]}$3d^{10}4s^{2}4p^{4}$ & -2401.156042\tabularnewline
Br & {[}Ar{]}$3d^{10}4s^{2}4p^{5}$ & -2573.776450\tabularnewline
\end{tabular}
\par\end{centering}
\caption{Atomic ground state configurations for spin-restricted PBE calculations,
employed as the minimal NAO basis, and the corresponding total energies.\label{tab:Atomic-PBE-ground}}
\end{table}

\subsection{Molecular calculations \label{subsec:Molecular-calculations}}

Having discussed the results of the free-atom importance profiles,
we can move on to discussing the importance profiles of the polyatomic
calculations where the minimal NAO basis is explicitly included in
the calculation of the importance profile.

No atomic basis is exact in polyatomic calculations, because atomic
symmetry is lifted in a polyatomic environment. Depending on the system,
an arbitrarily large number of higher and higher polarization shells
may be necessary. For example, reproducing the PBE atomization energy
of \ce{SF6} to 0.1 kcal/mol precision requires three polarization
shells, that is, up to $g$ functions.\citep{Lehtola2020_JCP_134108}
State-of-the-art GTO basis sets are optimized such that the error
made for free atoms is similar to the error made in neglecting higher
polarization shells in polyatomic calculations.\citep{Jensen2001_JCP_9113,Jensen2004_JCP_3463}
The purpose of this section is to analyze polarization effects.

\begin{figure*}
\subfloat[GTO projection for the $1\sigma$ orbital on the H atom in \ce{HF}.
\label{fig:H-in-HF-gto}]{\begin{centering}
\includegraphics[width=0.3\textwidth]{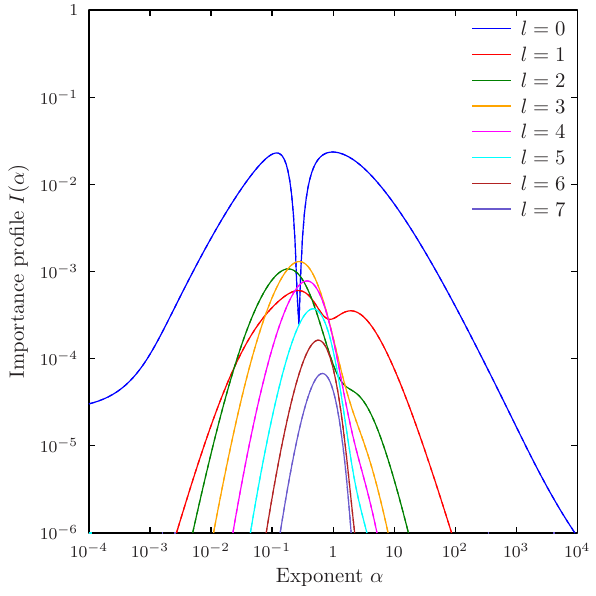}
\par\end{centering}
}\subfloat[GTO projection for the $1\sigma$ orbital on the H atom in \ce{HCl}.
\label{fig:H-in-HCl-gto}]{\begin{centering}
\includegraphics[width=0.3\textwidth]{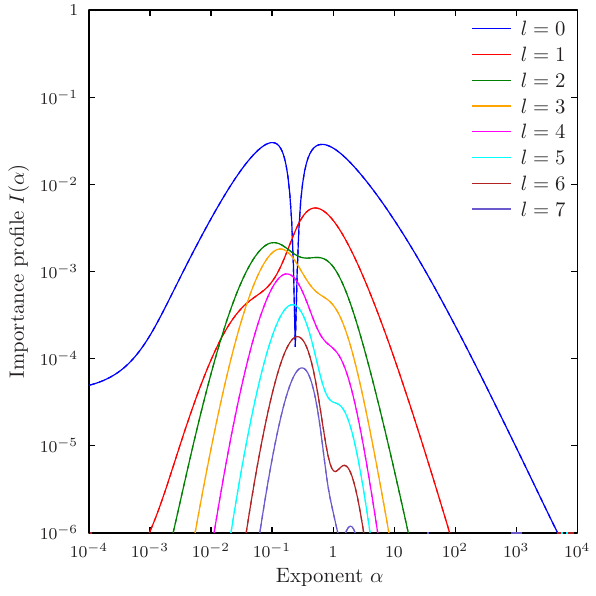}
\par\end{centering}
}\subfloat[GTO projection for the $1\sigma$ orbital on the H atom in \ce{HBr}.
\label{fig:H-in-HBr-gto}]{\begin{centering}
\includegraphics[width=0.3\textwidth]{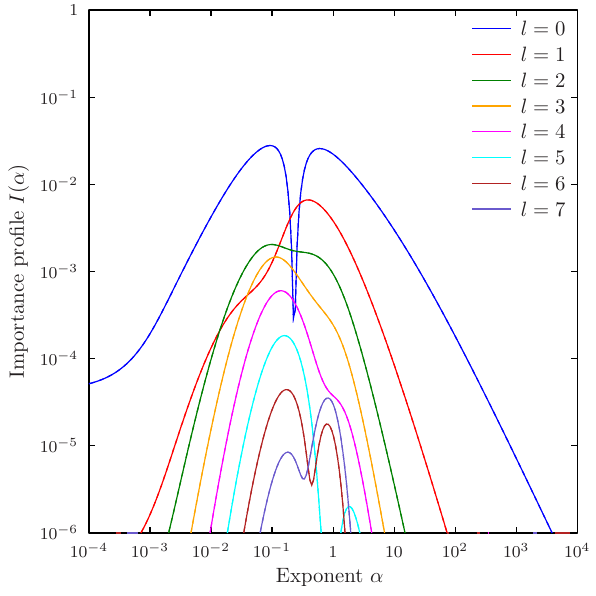}
\par\end{centering}
}

\subfloat[GTO projection for the $1\sigma_{g}$ orbital on one of the H atoms
in \ce{H2}. \label{fig:H-in-H2-gto}]{\begin{centering}
\includegraphics[width=0.3\textwidth]{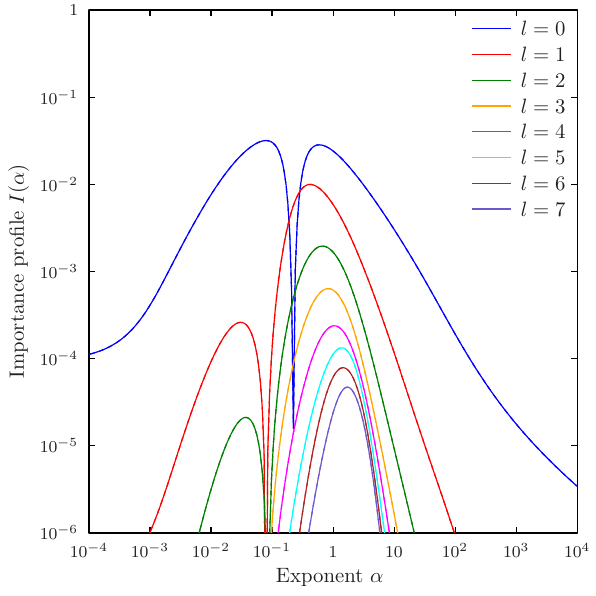}
\par\end{centering}
}\subfloat[GTO projection for the $1\sigma$ orbital on the H atom in \ce{LiH}.
\label{fig:H-in-LiH-gto}]{\begin{centering}
\includegraphics[width=0.3\textwidth]{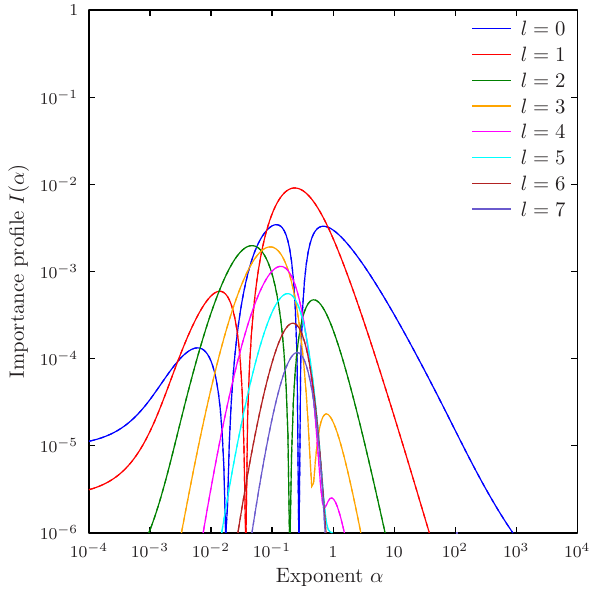}
\par\end{centering}
}\subfloat[GTO projection for the $1\sigma$ orbital on the H atom in \ce{NaH}.
\label{fig:H-in-NaH-gto}]{\begin{centering}
\includegraphics[width=0.3\textwidth]{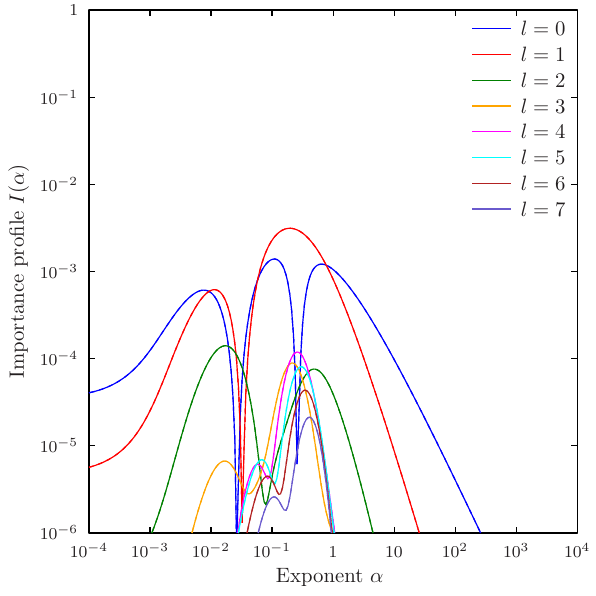}
\par\end{centering}
}

\subfloat[GTO projection for the $1\sigma$ orbital on the H atom in \ce{KH}.
\label{fig:H-in-KH-gto}]{\begin{centering}
\includegraphics[width=0.3\textwidth]{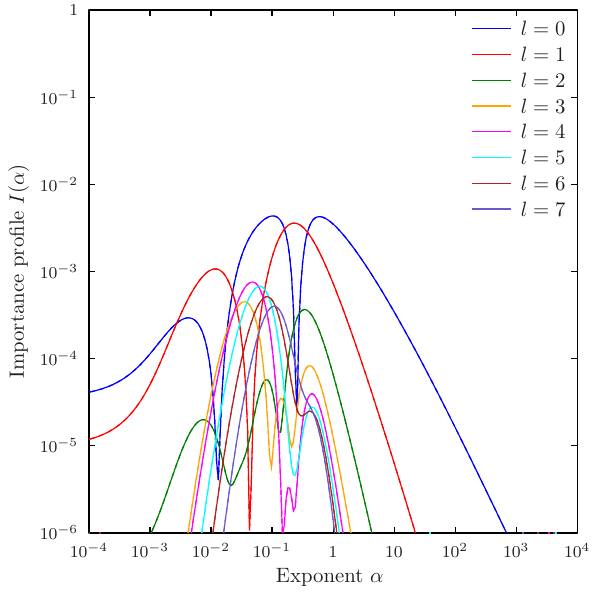}
\par\end{centering}
}\subfloat[GTO projection for the $1\sigma$ orbital on the H atom in \ce{CuH}.
\label{fig:H-in-CuH-gto}]{\begin{centering}
\includegraphics[width=0.3\textwidth]{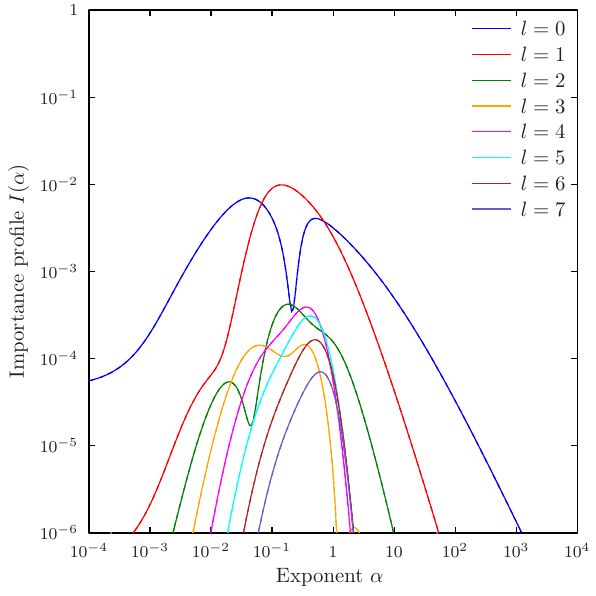}
\par\end{centering}
}\caption{Various molecular $\sigma$ orbitals' GTO projections on the hydrogen
atom. Note logarithmic scale on both axes. \label{fig:H-gto}}
\end{figure*}

Due to the large amount of computed importance profiles (618), we
will only discuss the results in detail for the hydrogen atoms in
CuH, \ce{H2}, HBr, HCl, HF, KH, LiH, and NaH. The full set of molecular
importance profiles can be found in the supporting information. We
begin the analysis with GTO projections, as this type of basis set
is most commonly used in quantum chemistry; these importance profiles
are shown in \ref{fig:H-gto}.

Starting from the top row of \ref{fig:H-gto}, it is clear that there
is a great deal of similarity between the chemistry of H in HF, HCl,
and HBr. We observe that the $\sigma$ orbitals have a large projection
onto $s$ functions on the hydrogen atom with a wide range of exponents,
indicating that the hydrogen atom requires breathing functions due
to losing electrons to the halogen atom. The importance profile for
$s$ functions experiences a sharp dip around $\alpha\approx0.28$,
possibly since the scanning function with such an exponent has a large
overlap with the NAO minimal basis.

The importance profiles for higher-angular-momentum functions for
H in HCl and for H in HBr are especially similar, and show that the
importance of the polarization shells decreases with increasing angular
momentum, as expected. H in HF, in contrast, shows somewhat unexpected
behavior, the shells arising in decreasing importance as $s$, $f$,
$d$, $g$, $p$, $h$, and so on. 

H in \ce{H2} shows a similar spectrum for breathing functions as
observed for the halide hydrides above. For the case of \ce{H2},
both the breathing and polarization curves show a double peaked structure,
as the importance curves for all angular momenta again exhibit a clear
drop at a critical value of the exponent $\alpha$, which can again
be tentatively understood by a maximal overlap with the NAO basis:
when the test function can be accurately expanded in the minimal NAO
basis, including it yields little additional flexibility to the wave
function. For \ce{H2}, the importance of the polarization shells
decreases with increasing angular momentum, as expected.

The hydrogen atoms in LiH, in NaH, and in KH likewise show great similarities
in the form of the $s$ and $p$ function importances. For the first
two molecules, the $s$ breathing function on hydrogen has smaller
importance than including a $p$ function on hydrogen, while in the
lattermost the order of the $s$ and $p$ functions is switched. There
appears to be no systematic information for the higher polarization
shells in the plots, but this can tentatively be explained by the
non-orthogonality of the atomic orbital basis functions and/or the
small minimal basis used in this work for the alkali atoms: while
the minimal basis used in this work is $2s$ for Li, $3s1p$ for Na,
and $4s2p$ for K, most minimal basis sets in quantum chemistry include
a further $p$ function on alkali atoms to account for the low-lying
$np$ excited state which is accessible to chemical bonding. The lack
of these functions in the minimal basis of the alkali atom can show
up as larger importance for higher polarization functions on hydrogen
in these molecules, and could explain the apparent lack of systematic
behavior in the series LiH$\to$NaH$\to$KH.

\subsection{Spin-polarized atomic calculations in the diatomic symmetry \label{subsec:Spin-polarized-atomic-calculatio}}

\begin{figure*}
\subfloat[GTO projection for the $1\sigma$ orbital on the H atom computed in
the \ce{H2} geometry. \label{fig:H-in-H-real}]{\begin{centering}
\includegraphics[width=0.3\textwidth]{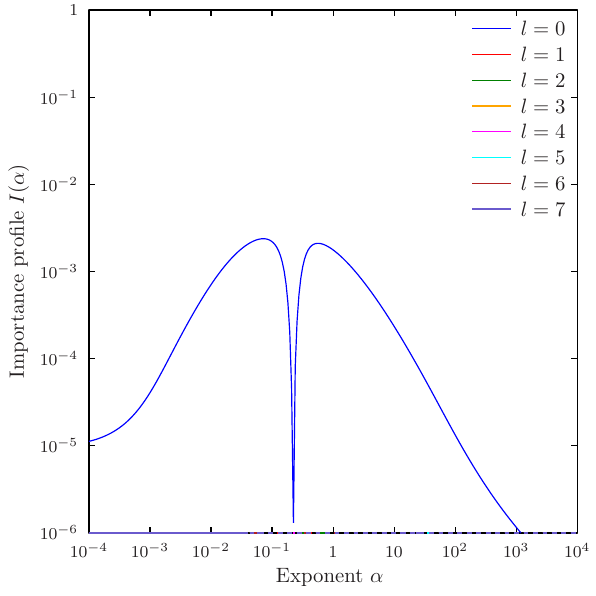}
\par\end{centering}
}\subfloat[GTO projection for the $1\sigma$ orbital on the ghost H atom for
H computed in the \ce{H2} geometry. \label{fig:H-in-H}]{\begin{centering}
\includegraphics[width=0.3\textwidth]{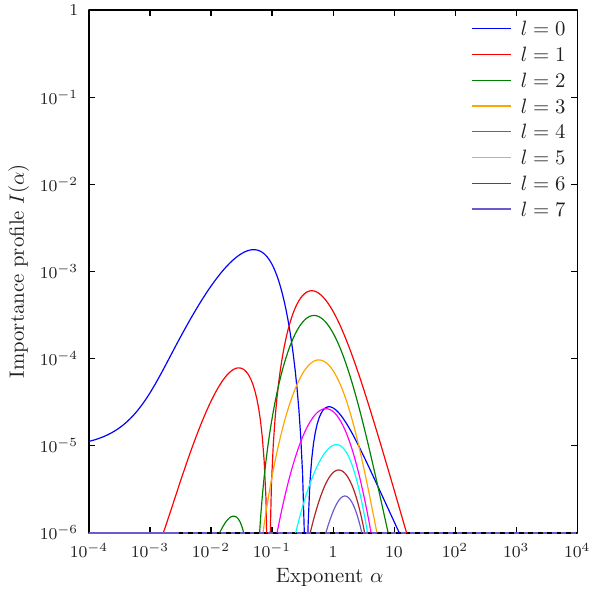}
\par\end{centering}
}

\subfloat[GTO projection for the $\sigma$ orbitals on the ghost H atom for
Li computed in the \ce{LiH} geometry. \label{fig:H-in-Li}]{\begin{centering}
\includegraphics[width=0.3\textwidth]{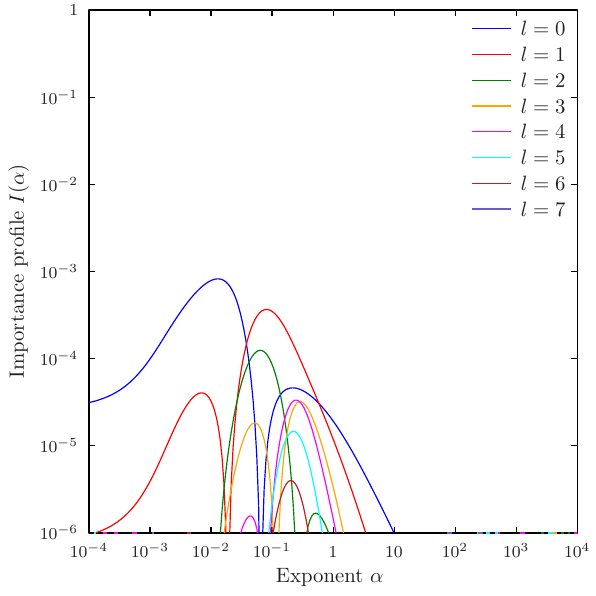}
\par\end{centering}
}\subfloat[GTO projection for the $1\sigma$ orbital on the ghost H atom for
F computed in the \ce{HF} geometry. \label{fig:H-in-F}]{\begin{centering}
\includegraphics[width=0.3\textwidth]{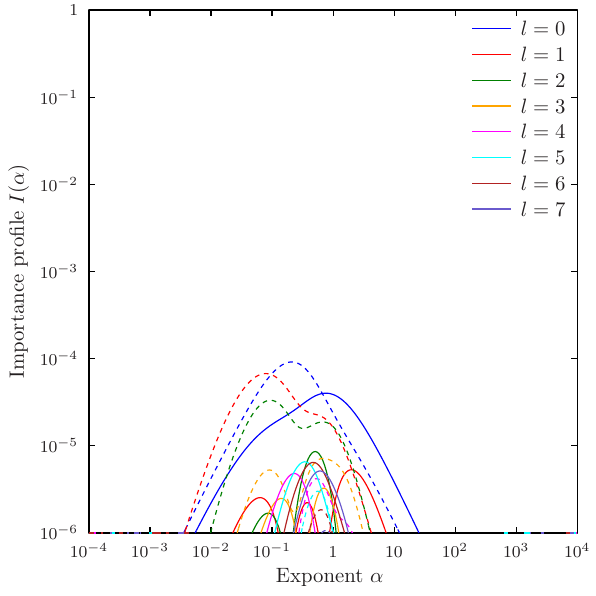}
\par\end{centering}
}

\caption{Effects of the non-orthogonality of atomic orbitals on various centers
on the importance profile. Note logarithmic scale on both axes. \label{fig:non-orthogonality}}
\end{figure*}

To assess the importance of the role of the minimal basis and the
non-orthogonality of the atomic orbital basis functions, we performed
calculations on the hydrogen atom with the diatomic numerical basis
set, in analogy to the counterpoise correction of \citet{Jansen1969_CPL_140}
and \citet{Boys1970_MP_553}. This analysis leads to the results shown
in \ref{fig:non-orthogonality}.

We begin with the calculation of the hydrogen atom in the \ce{H2}
geometry. This gives us the importance profiles shown in \cref{fig:H-in-H-real,fig:H-in-H}
for the atom itself, and the ghost atom in the molecule. Since the
exact ground state of H is well-known to be a $1s$ function, \ref{fig:H-in-H-real}
shows that the NAO does not give the exact solution: after all, this
calculation is spin-polarized, while the NAO was determined for a
spin-restricted calculation. Comparing \cref{fig:H-in-H-real,fig:H-in-H}
shows that the importance for the diffuse functions with small $\alpha$
are similar, regardless of which atom the functions are placed, while
major differences between the two centers are observed for tight functions
with large $\alpha$. Diffuse functions are indeed infamous for leading
to large interatomic overlaps and causing ill-conditioning in AO basis
sets. However, such ill conditioning can nowadays be stably and accurately
handled with the help of pivoted Cholesky decompositions to ensure
stable electronic structure calculations even in the presence of pathological
linear dependencies.\citep{Lehtola2019_JCP_241102,Lehtola2020_PRA_32504}

The polarization functions appear in \ref{fig:H-in-H} as a consequence
of the inexactness of the NAO basis on hydrogen discussed above, and
the ghost atom being off-center from the system. 

\Cref{fig:non-orthogonality} also shows two more calculations for
ghost hydrogen atoms: for Li computed with the numerical basis for
LiH, and for F computed with the numerical basis for HF, which are
shown in \ref{fig:H-in-Li,fig:H-in-F}, respectively. Like \ref{fig:H-in-H},
\ref{fig:H-in-Li} a measure goodness of the minimal NAO basis for
Li. Because the minimal NAO basis was determined with spin-restricted
orbitals, it is unable to describe the spin-polarized Li atom. A similar
story also applies to \ref{fig:H-in-F}; however, the case of fluorine
is more complicated: the hole on the $2p$ shell leads to symmetry
breaking, and one actually needs to include many polarization shells
to reproduce the energy of the fully numerical calculation in an atomic
basis set calculation.

\subsection{Comparison of STO and GTO projections \label{subsec:STO-analysis}}

\begin{figure*}
\subfloat[STO projection for the $1\sigma$ orbital on the H atom in \ce{HF}.
\label{fig:H-in-HF-sto}]{\begin{centering}
\includegraphics[width=0.3\textwidth]{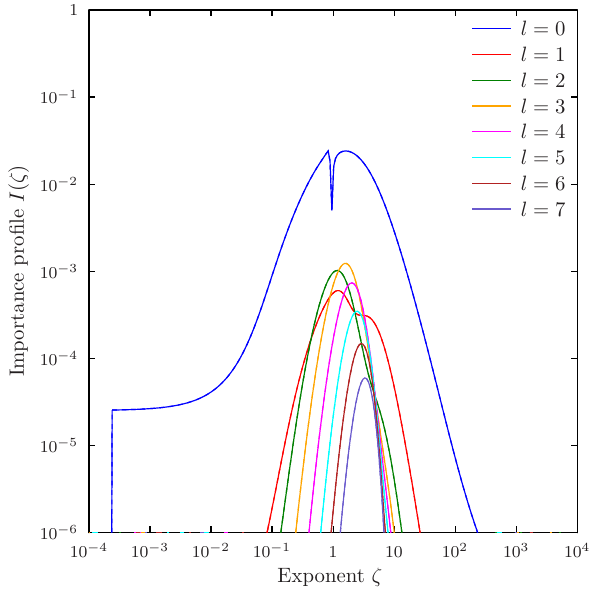}
\par\end{centering}
}\subfloat[STO projection for the $1\sigma$ orbital on the H atom in \ce{HCl}.
\label{fig:H-in-HCl-sto}]{\begin{centering}
\includegraphics[width=0.3\textwidth]{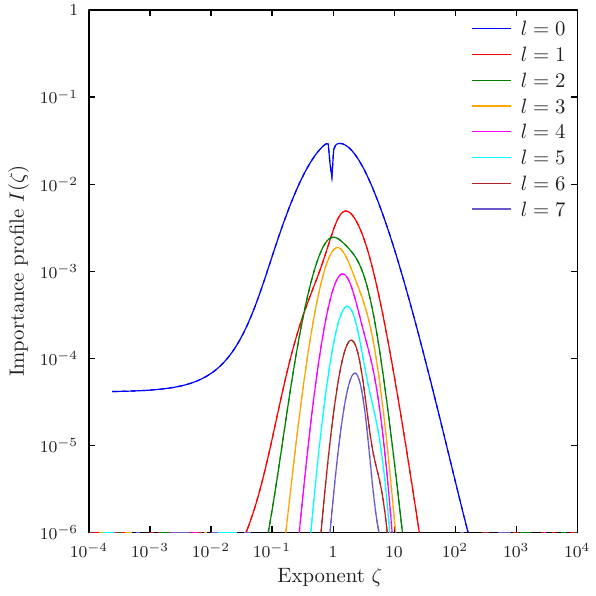}
\par\end{centering}
}\subfloat[STO projection for the $1\sigma$ orbital on the H atom in \ce{HBr}.
\label{fig:H-in-HBr-sto}]{\begin{centering}
\includegraphics[width=0.3\textwidth]{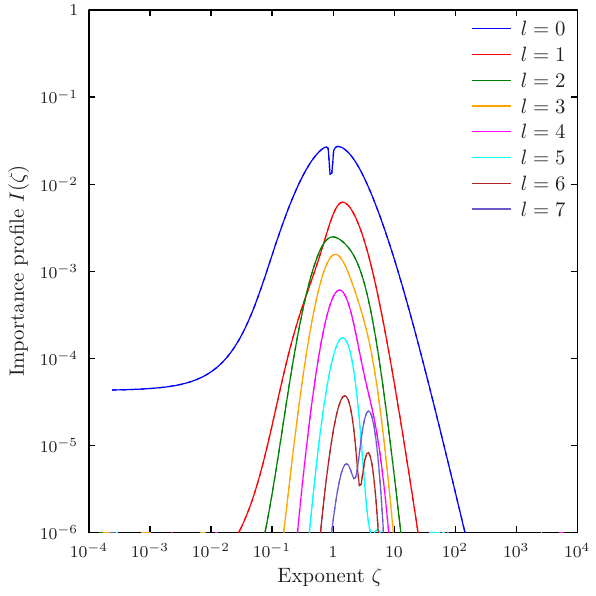}
\par\end{centering}
}

\subfloat[STO projection for the $1\sigma_{g}$ orbital on one of the H atoms
in \ce{H2}. \label{fig:H-in-H2-sto}]{\begin{centering}
\includegraphics[width=0.3\textwidth]{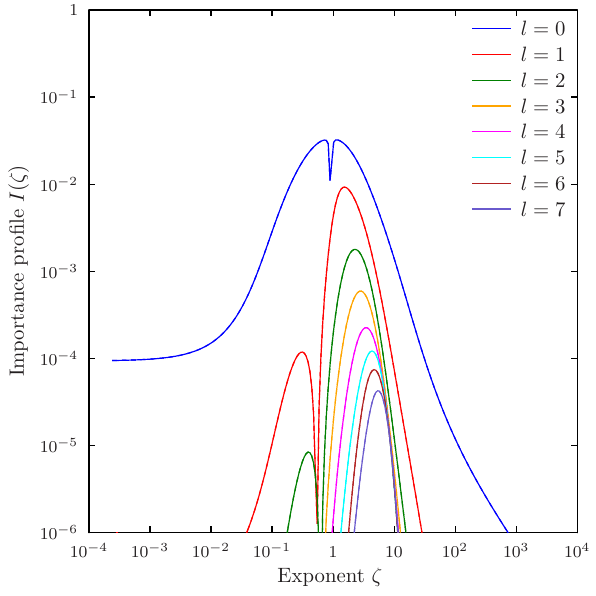}
\par\end{centering}
}\subfloat[STO projection for the $1\sigma$ orbital on the H atom in \ce{LiH}.
\label{fig:H-in-LiH-sto}]{\begin{centering}
\includegraphics[width=0.3\textwidth]{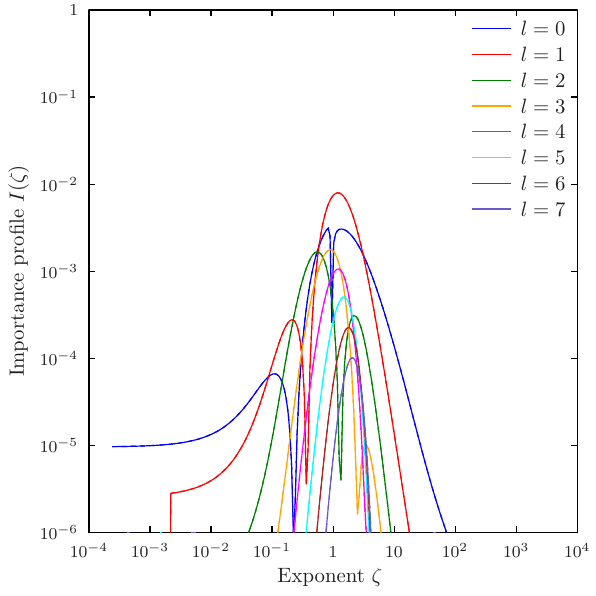}
\par\end{centering}
}\subfloat[STO projection for the $1\sigma$ orbital on the H atom in \ce{NaH}.
\label{fig:H-in-NaH-sto}]{\begin{centering}
\includegraphics[width=0.3\textwidth]{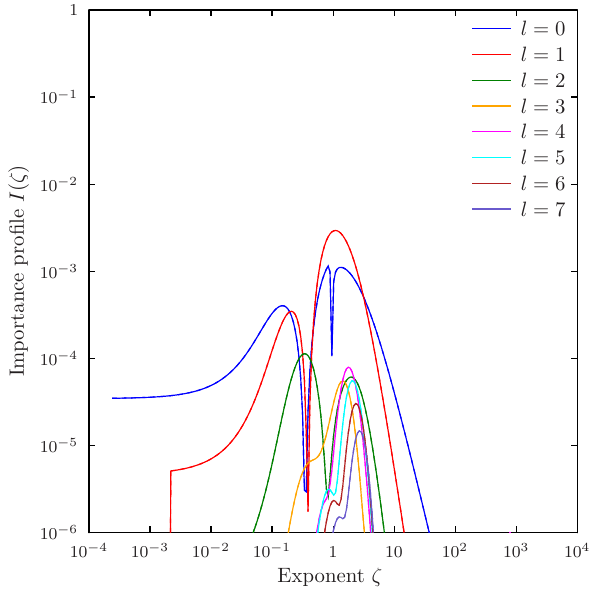}
\par\end{centering}
}

\subfloat[STO projection for the $1\sigma$ orbital on the H atom in \ce{KH}.
\label{fig:H-in-KH-sto}]{\begin{centering}
\includegraphics[width=0.3\textwidth]{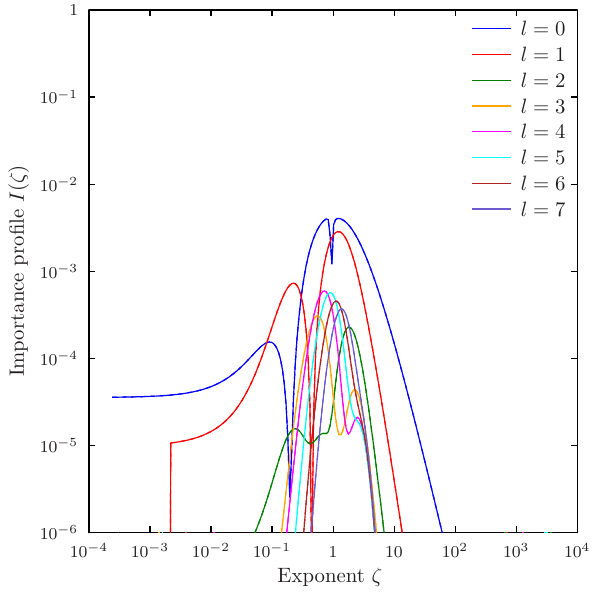}
\par\end{centering}
}\subfloat[STO projection for the $1\sigma$ orbital on the H atom in \ce{CuH}.
\label{fig:H-in-CuH-sto}]{\begin{centering}
\includegraphics[width=0.3\textwidth]{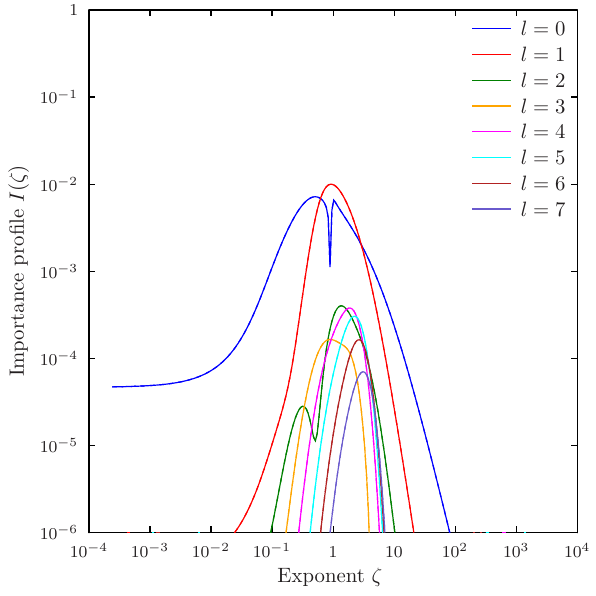}
\par\end{centering}
}\caption{Various molecular $\sigma$ orbitals' STO projections on the hydrogen
atom. Note logarithmic scale on both axes. \label{fig:H-sto}}
\end{figure*}

We now move on to the analysis of STO projections. Plots analogous
to \ref{fig:H-gto} are shown in \ref{fig:H-sto}. One-by-one visual
comparisons of \ref{fig:H-in-HF-gto,fig:H-in-HF-sto}, \ref{fig:H-in-HCl-gto,fig:H-in-HCl-sto},
\ref{fig:H-in-HBr-gto,fig:H-in-HBr-sto}, \ref{fig:H-in-H2-gto,fig:H-in-H2-sto},
\ref{fig:H-in-LiH-gto,fig:H-in-LiH-sto}, \ref{fig:H-in-NaH-gto,fig:H-in-NaH-sto},
\ref{fig:H-in-KH-gto,fig:H-in-KH-sto}, and \ref{fig:H-in-CuH-gto,fig:H-in-CuH-sto}
reveals remarkably few differences, other than the smaller scale of
the $x$ axis in the STO plots, which was discussed in \ref{subsec:Atomic-calculations}
in the context of atomic calculations.

However, the logarithmic $y$ axis scale can be misleading, and we
continued with an in-depth numerical analysis of the full database
of results, which contains 2432 GTO and STO importance profiles: profiles
for $l\in[0,\dots,7]$ for the occupied $\sigma$, $\pi$, and $\delta$
orbitals of the two centers in the 47 systems in \ref{tab:energy}. 

\begin{figure}
\begin{centering}
\includegraphics[width=0.3\textwidth]{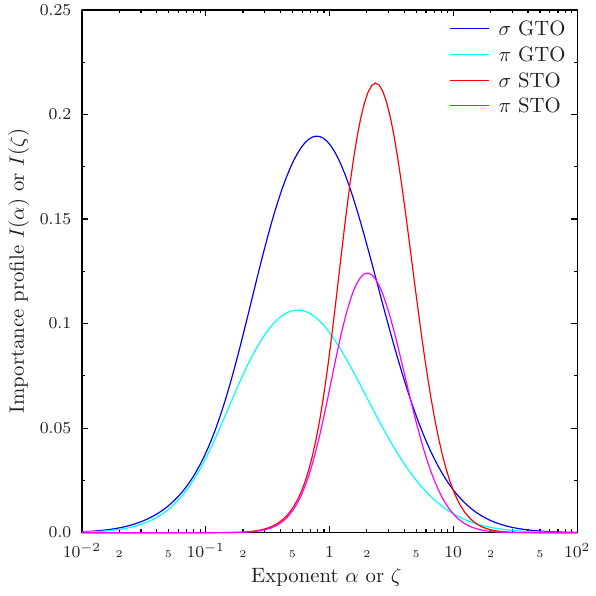}
\par\end{centering}
\caption{GTO and STO projections for the $\sigma$ and $\pi$ orbitals onto
the $d$ orbital of the Sc atom in ScO. \label{fig:ScO}}
\end{figure}

The largest difference in the database was found to arise for the
$d$ orbitals on Sc in ScO. This orbital has clear atomic character,
as indicated by the GTO and STO plots in \ref{fig:ScO}. A $d$ orbital
on Sc was not included in the minimal basis listed in \ref{tab:Atomic-PBE-ground},
because occupying the $d$ orbital raises the spin-restricted PBE
total energy by an $s\to d$ excitation energy of 1.54 eV; however,
the orbital clearly takes part in chemical bonding.

Analyzing the maximal value of the importance profile for each system,
center, and angular momentum, the remaining differences between GTOs
and STOs were found to be small, ranging between $5.4\times10^{-3}$
in favor of GTOs for a $p$ polarization function on Ni for describing
the $\sigma$ orbitals in NiS, to $6.67\times10^{-3}$ in favor of
STOs for a $p$ function on Be to describe the $\pi$ orbital polarization
in BeS. Overall, GTOs appear to win in this comparison, because the
sum of the maximal values of the GTO profiles is greater than that
of the STO profiles. However, these results do not constitute definitive
proof, for which variational optimization of the maximal overlap wave
functions---which is left to future work---is required.

\subsection{Minimal basis analysis \label{subsec:Minimal-basis-analysis}}

To continue the discussion on ScO above, we now ask if there are other
cases where the minimal NAO basis fails badly, and a more educated
choice should be made. The number of electrons that cannot be described
in the minimal NAO basis was given in \ref{tab:energy}. The analysis
of these data show that for most systems, the NAO minimal basis is
quite accurate: in all but 7 systems, the alpha and beta electron
manifolds can be described by the minimal basis to less than 0.1 electrons.
The outliers are MgF ($\Delta N_{\alpha}=0.171$), BeS ($\Delta N_{\alpha}=0.233$),
ScO ($\Delta N_{\alpha}=0.526$), MnO ($\Delta N_{\alpha}=0.878$),
FeO ($\Delta N_{\alpha}=0.675$), MnS ($\Delta N_{\alpha}=0.632$),
NiO ($\Delta N_{\alpha}=0.347$), and NiS ($\Delta N_{\alpha}=0.256$).
Examining the importance profiles for these systems leads to the following
conclusions. MgF requires an additional $p$ function on Mg and a
$s$ function on F to describe the polarization of the $\sigma$ orbitals
and the strong movement of charge from Mg to F. BeS requires a $p$
function on Be to describe the polarization of the $\pi$ orbitals
and a $d$ function on S. A $d$ function with atomic character is
missing on Sc, as was already discussed in \ref{subsec:STO-analysis}.

The discussions for FeO, MnS, NiO, and NiS are similar to the case
of MnO. MnO badly needs further $s$ functions on Mn to describe the
change in its atomic size due to its losing electrons to oxygen. Adding
an $s$ and a $p$ function on Sc appears to have around the same
importance as further $s$ and $p$ functions on oxygen. Next, one
should add a $d$ function on oxygen, followed by another $d$ function
on Sc and a $f$ function on O.

\section{Summary and discussion \label{sec:Summary-and-discussion}}

We have suggested the importance profile to extract information on
atomic basis set requirements from fully numerical calculations. First,
we discussed the free-atom importance profile, which is obtained as
a projection of the wave function at the complete basis set (CBS)
limit onto individual atomic orbital (AO) basis functions. Next, as
calculations on polyatomic systems typically include at least a minimal
basis set on each atom, we generalized the importance profile for
this case as the difference in overlap onto the CBS limit wave function
obtained with the minimal basis padded by the studied test function,
and the overlap obtained with the minimal basis, in the aim to isolate
the effects of breathing and polarization functions in polyatomic
systems.

Employing a minimal numerical atomic orbital (NAO) basis set on each
atom, we computed importance profiles for a variety of atoms in a
large database of chemically diverse diatomic molecules. Due to the
significant amount of data, importance profiles were only discussed
for hydrogen, while the full set of data is available in the supporting
information. Although only isotropic Gaussian-type orbitals (GTOs)
and Slater-type orbitals (STOs) were considered in this work, the
approach is independent of the form of the AO basis, and could also
be applied to other types of AO basis sets.

The above definition of the importance profile inherently depends
on the chosen minimal basis. Following established practice in the
NAO literature,\citep{Sankey1989_PRB_3979,Porezag1995_PRB_12947,Horsfield1997_PRB_6594,SanchezPortal1997_IJQC_453,Kenny2000_PRB_4899,Junquera2001_PRB_235111,Anglada2002_PRB_205101,Ozaki2003_PRB_155108,Ozaki2004_JCP_10879,Ozaki2004_PRB_195113,Blum2009_CPC_2175,Larsen2009_PRB_195112,Shang2010_IRPC_665,Louwerse2012_PRB_35108,Corsetti2013_JPCM_435504}
the minimal NAO basis was derived from spin-restricted atomic calculations,
and the configuration leading to the lowest total energy was used
for each atom. However, the definition of the minimal basis is not
always obvious. 

For instance, as discussed in \ref{subsec:Molecular-calculations},
many GTO basis sets include an $p$ function on alkali and alkaline
metals, as they are needed to describe the low-lying $np$ atomic
excited state. Similarly, in this study, we noticed that in the case
of ScO, a $d$ function should clearly be included in the minimal
basis of Sc, although it is an excited state according to the spin-restricted
atomic calculations. Further cases were analyzed in \ref{subsec:Minimal-basis-analysis},
where it became clear that many transition metals need breathing functions
to describe the strong changes in their electronic configuration in
chemical bonding.

The spin-restricted NAO basis is not exact for non-interacting atoms,
as was discussed for the cases of H, Li, and F in \ref{subsec:Spin-polarized-atomic-calculatio}.
The danger of such inexactness is that it invites basis set superposition
errors: if the minimal NAO basis is not exact, the atom's description
can be improved by lending basis functions from other nuclei, leading
to exaggerated binding energies.\citep{Jansen1969_CPL_140,Boys1970_MP_553} 

Importance profiles can also be computed with respect to a larger
``minimal'' atomic basis: employing a NAO basis that is sufficiently
flexible to also describe the spin-polarized atom and its low-lying
excited states would lower the estimated importances of further functions
on the atoms. Because breathing functions are essential for describing
chemistry, a double numerical basis which is also able to describe
atomic cations exactly\citep{Delley1990_JCP_508} could also be employed.
We especially believe that state-averaging\citep{Widmark1990_TCA_291,Widmark1991_TCA_419,PouAmerigo1995_TCA_149}
is an underexplored avenue in the development of NAO basis sets,\citep{Lehtola2019_IJQC_25968}
and hope to follow up with such work in the future.

The main motivation of this work was the development of new AO basis
sets starting from fully numerical methods. We believe that systematic
databases of molecular and crystalline all-electron wave functions
determined with fully numerical methods at the CBS limit would offer
an excellent starting point for developing new, compact AO basis sets
for GTO, STO, and NAO calculations. When the CBS limit wave functions
are combined with the maximal overlap method pioneered by Richardson
and coworkers,\citep{Richardson1962_JCP_1057,Richardson1963_JCP_796}
later used by many other authors\citep{Kalman1971_JCP_1841,Adamowicz1981_IJQC_545,Adamowicz1983_IJQC_19,VanLenthe2003_JCC_56,SanchezPortal1995_SSC_685,SanchezPortal1996_JPCM_3859,Chen2009_PRB_165121,Chen2010_JPCM_445501,Chen2011_JPCM_325501,Lange2011_PRB_85101,Lin2021_PRB_235131}
without attribution, different kinds of AO basis sets can be quickly
optimized for a fixed database and updated when new entries representing
different chemistries are entered in the database. An apples-to-apples
comparison of GTO, STO, and NAO basis sets would be especially interesting:
how do these three families compare in the maximal projection for
a fixed database of fully numerical reference wave functions? We hope
to follow up this study in that direction in the future.

\section*{Data Availability}

The author confirms that the data supporting the findings of this
study are available within the article and its supporting information.

\section*{Supporting Information Available}

The GTO and STO completeness and importance profiles for all atoms
and molecules considered in this study.
\begin{acknowledgments}
This work has been supported by the Academy of Finland (Suomen Akatemia)
through project numbers 350282 and 353749. Computational resources
provided by CSC -- IT Center for Science Ltd (Espoo, Finland) are
gratefully acknowledged.
\end{acknowledgments}

\bibliography{citations}

\begin{thebibliography}{118}%
\makeatletter
\providecommand \@ifxundefined [1]{%
 \@ifx{#1\undefined}
}%
\providecommand \@ifnum [1]{%
 \ifnum #1\expandafter \@firstoftwo
 \else \expandafter \@secondoftwo
 \fi
}%
\providecommand \@ifx [1]{%
 \ifx #1\expandafter \@firstoftwo
 \else \expandafter \@secondoftwo
 \fi
}%
\providecommand \natexlab [1]{#1}%
\providecommand \enquote  [1]{``#1''}%
\providecommand \bibnamefont  [1]{#1}%
\providecommand \bibfnamefont [1]{#1}%
\providecommand \citenamefont [1]{#1}%
\providecommand \href@noop [0]{\@secondoftwo}%
\providecommand \href [0]{\begingroup \@sanitize@url \@href}%
\providecommand \@href[1]{\@@startlink{#1}\@@href}%
\providecommand \@@href[1]{\endgroup#1\@@endlink}%
\providecommand \@sanitize@url [0]{\catcode `\\12\catcode `\$12\catcode
  `\&12\catcode `\#12\catcode `\^12\catcode `\_12\catcode `\%12\relax}%
\providecommand \@@startlink[1]{}%
\providecommand \@@endlink[0]{}%
\providecommand \url  [0]{\begingroup\@sanitize@url \@url }%
\providecommand \@url [1]{\endgroup\@href {#1}{\urlprefix }}%
\providecommand \urlprefix  [0]{URL }%
\providecommand \Eprint [0]{\href }%
\providecommand \doibase [0]{https://doi.org/}%
\providecommand \selectlanguage [0]{\@gobble}%
\providecommand \bibinfo  [0]{\@secondoftwo}%
\providecommand \bibfield  [0]{\@secondoftwo}%
\providecommand \translation [1]{[#1]}%
\providecommand \BibitemOpen [0]{}%
\providecommand \bibitemStop [0]{}%
\providecommand \bibitemNoStop [0]{.\EOS\space}%
\providecommand \EOS [0]{\spacefactor3000\relax}%
\providecommand \BibitemShut  [1]{\csname bibitem#1\endcsname}%
\let\auto@bib@innerbib\@empty
\bibitem [{\citenamefont
  {Lehtola}(2019{\natexlab{a}})}]{Lehtola2019_IJQC_25968}%
  \BibitemOpen
  \bibfield  {author} {\bibinfo {author} {\bibfnamefont {S.}~\bibnamefont
  {Lehtola}},\ }\bibfield  {title} {\enquote {\bibinfo {title} {A review on
  non-relativistic, fully numerical electronic structure calculations on atoms
  and diatomic molecules},}\ }\href {https://doi.org/10.1002/qua.25968}
  {\bibfield  {journal} {\bibinfo  {journal} {Int. J. Quantum Chem.}\ }\textbf
  {\bibinfo {volume} {119}},\ \bibinfo {pages} {e25968} (\bibinfo {year}
  {2019}{\natexlab{a}})},\ \Eprint {https://arxiv.org/abs/1902.01431}
  {arXiv:1902.01431} \BibitemShut {NoStop}%
\bibitem [{\citenamefont {Lehtola}, \citenamefont {Blockhuys},\ and\
  \citenamefont {{Van Alsenoy}}(2020)}]{Lehtola2020_M_1218}%
  \BibitemOpen
  \bibfield  {author} {\bibinfo {author} {\bibfnamefont {S.}~\bibnamefont
  {Lehtola}}, \bibinfo {author} {\bibfnamefont {F.}~\bibnamefont {Blockhuys}},\
  and\ \bibinfo {author} {\bibfnamefont {C.}~\bibnamefont {{Van Alsenoy}}},\
  }\bibfield  {title} {\enquote {\bibinfo {title} {An overview of
  self-consistent field calculations within finite basis sets},}\ }\href
  {https://doi.org/10.3390/molecules25051218} {\bibfield  {journal} {\bibinfo
  {journal} {Molecules}\ }\textbf {\bibinfo {volume} {25}},\ \bibinfo {pages}
  {1218} (\bibinfo {year} {2020})},\ \Eprint {https://arxiv.org/abs/1912.12029}
  {arXiv:1912.12029} \BibitemShut {NoStop}%
\bibitem [{\citenamefont {Hill}(2013)}]{Hill2013_IJQC_21}%
  \BibitemOpen
  \bibfield  {author} {\bibinfo {author} {\bibfnamefont {J.~G.}\ \bibnamefont
  {Hill}},\ }\bibfield  {title} {\enquote {\bibinfo {title} {{Gaussian basis
  sets for molecular applications}},}\ }\href
  {https://doi.org/10.1002/qua.24355} {\bibfield  {journal} {\bibinfo
  {journal} {Int. J. Quantum Chem.}\ }\textbf {\bibinfo {volume} {113}},\
  \bibinfo {pages} {21--34} (\bibinfo {year} {2013})}\BibitemShut {NoStop}%
\bibitem [{\citenamefont {Jensen}(2013)}]{Jensen2013_WIRCMS_273}%
  \BibitemOpen
  \bibfield  {author} {\bibinfo {author} {\bibfnamefont {F.}~\bibnamefont
  {Jensen}},\ }\bibfield  {title} {\enquote {\bibinfo {title} {Atomic orbital
  basis sets},}\ }\href {https://doi.org/10.1002/wcms.1123} {\bibfield
  {journal} {\bibinfo  {journal} {Wiley Interdiscip. Rev. Comput. Mol. Sci.}\
  }\textbf {\bibinfo {volume} {3}},\ \bibinfo {pages} {273--295} (\bibinfo
  {year} {2013})}\BibitemShut {NoStop}%
\bibitem [{\citenamefont
  {Lehtola}(2020{\natexlab{a}})}]{Lehtola2020_JCP_134108}%
  \BibitemOpen
  \bibfield  {author} {\bibinfo {author} {\bibfnamefont {S.}~\bibnamefont
  {Lehtola}},\ }\bibfield  {title} {\enquote {\bibinfo {title} {Polarized
  {Gaussian} basis sets from one-electron ions},}\ }\href
  {https://doi.org/10.1063/1.5144964} {\bibfield  {journal} {\bibinfo
  {journal} {J. Chem. Phys.}\ }\textbf {\bibinfo {volume} {152}},\ \bibinfo
  {pages} {134108} (\bibinfo {year} {2020}{\natexlab{a}})},\ \Eprint
  {https://arxiv.org/abs/2001.04224} {arXiv:2001.04224} \BibitemShut {NoStop}%
\bibitem [{\citenamefont {Binkley}, \citenamefont {Pople},\ and\ \citenamefont
  {Hehre}(1980)}]{Binkley1980_JACS_939}%
  \BibitemOpen
  \bibfield  {author} {\bibinfo {author} {\bibfnamefont {J.~S.}\ \bibnamefont
  {Binkley}}, \bibinfo {author} {\bibfnamefont {J.~A.}\ \bibnamefont {Pople}},\
  and\ \bibinfo {author} {\bibfnamefont {W.~J.}\ \bibnamefont {Hehre}},\
  }\bibfield  {title} {\enquote {\bibinfo {title} {{Self-consistent molecular
  orbital methods. 21. Small split-valence basis sets for first-row
  elements}},}\ }\href {https://doi.org/10.1021/ja00523a008} {\bibfield
  {journal} {\bibinfo  {journal} {J. Am. Chem. Soc.}\ }\textbf {\bibinfo
  {volume} {102}},\ \bibinfo {pages} {939} (\bibinfo {year}
  {1980})}\BibitemShut {NoStop}%
\bibitem [{\citenamefont {Hehre}, \citenamefont {Ditchfield},\ and\
  \citenamefont {Pople}(1972)}]{Hehre1972_JCP_2257}%
  \BibitemOpen
  \bibfield  {author} {\bibinfo {author} {\bibfnamefont {W.~J.}\ \bibnamefont
  {Hehre}}, \bibinfo {author} {\bibfnamefont {R.}~\bibnamefont {Ditchfield}},\
  and\ \bibinfo {author} {\bibfnamefont {J.~A.}\ \bibnamefont {Pople}},\
  }\bibfield  {title} {\enquote {\bibinfo {title} {Self-consistent molecular
  orbital methods. {XII}. {Further} extensions of {Gaussian}-type basis sets
  for use in molecular orbital studies of organic molecules},}\ }\href
  {https://doi.org/10.1063/1.1677527} {\bibfield  {journal} {\bibinfo
  {journal} {J. Chem. Phys.}\ }\textbf {\bibinfo {volume} {56}},\ \bibinfo
  {pages} {2257--2261} (\bibinfo {year} {1972})}\BibitemShut {NoStop}%
\bibitem [{\citenamefont {Krishnan}\ \emph {et~al.}(1980)\citenamefont
  {Krishnan}, \citenamefont {Binkley}, \citenamefont {Seeger},\ and\
  \citenamefont {Pople}}]{Krishnan1980_JCP_650}%
  \BibitemOpen
  \bibfield  {author} {\bibinfo {author} {\bibfnamefont {R.}~\bibnamefont
  {Krishnan}}, \bibinfo {author} {\bibfnamefont {J.~S.}\ \bibnamefont
  {Binkley}}, \bibinfo {author} {\bibfnamefont {R.}~\bibnamefont {Seeger}},\
  and\ \bibinfo {author} {\bibfnamefont {J.~A.}\ \bibnamefont {Pople}},\
  }\bibfield  {title} {\enquote {\bibinfo {title} {{Self-consistent molecular
  orbital methods. XX. A basis set for correlated wave functions}},}\ }\href
  {https://doi.org/10.1063/1.438955} {\bibfield  {journal} {\bibinfo  {journal}
  {J. Chem. Phys.}\ }\textbf {\bibinfo {volume} {72}},\ \bibinfo {pages}
  {650--654} (\bibinfo {year} {1980})}\BibitemShut {NoStop}%
\bibitem [{\citenamefont {Grev}\ and\ \citenamefont
  {Schaefer}(1989)}]{Grev1989_JCP_7305}%
  \BibitemOpen
  \bibfield  {author} {\bibinfo {author} {\bibfnamefont {R.~S.}\ \bibnamefont
  {Grev}}\ and\ \bibinfo {author} {\bibfnamefont {H.~F.}\ \bibnamefont
  {Schaefer}},\ }\bibfield  {title} {\enquote {\bibinfo {title} {6-311{G} is
  not of valence triple-zeta quality},}\ }\href
  {https://doi.org/10.1063/1.457301} {\bibfield  {journal} {\bibinfo  {journal}
  {J. Chem. Phys.}\ }\textbf {\bibinfo {volume} {91}},\ \bibinfo {pages}
  {7305--7306} (\bibinfo {year} {1989})}\BibitemShut {NoStop}%
\bibitem [{\citenamefont {Moran}\ \emph {et~al.}(2006)\citenamefont {Moran},
  \citenamefont {Simmonett}, \citenamefont {Leach}, \citenamefont {Allen},
  \citenamefont {Schleyer},\ and\ \citenamefont
  {Schaefer}}]{Moran2006_JACS_9342}%
  \BibitemOpen
  \bibfield  {author} {\bibinfo {author} {\bibfnamefont {D.}~\bibnamefont
  {Moran}}, \bibinfo {author} {\bibfnamefont {A.~C.}\ \bibnamefont
  {Simmonett}}, \bibinfo {author} {\bibfnamefont {F.~E.}\ \bibnamefont
  {Leach}}, \bibinfo {author} {\bibfnamefont {W.~D.}\ \bibnamefont {Allen}},
  \bibinfo {author} {\bibfnamefont {P.~V.~R.}\ \bibnamefont {Schleyer}},\ and\
  \bibinfo {author} {\bibfnamefont {H.~F.}\ \bibnamefont {Schaefer}},\
  }\bibfield  {title} {\enquote {\bibinfo {title} {{Popular theoretical methods
  predict benzene and arenes to be nonplanar}},}\ }\href
  {https://doi.org/10.1021/ja0630285} {\bibfield  {journal} {\bibinfo
  {journal} {J. Am. Chem. Soc.}\ }\textbf {\bibinfo {volume} {128}},\ \bibinfo
  {pages} {9342--9343} (\bibinfo {year} {2006})}\BibitemShut {NoStop}%
\bibitem [{\citenamefont {Pitman}\ \emph {et~al.}(2023)\citenamefont {Pitman},
  \citenamefont {Evans}, \citenamefont {Ireland}, \citenamefont {Lempriere},\
  and\ \citenamefont {McKemmish}}]{Pitman2023_JPCA_10295}%
  \BibitemOpen
  \bibfield  {author} {\bibinfo {author} {\bibfnamefont {S.~J.}\ \bibnamefont
  {Pitman}}, \bibinfo {author} {\bibfnamefont {A.~K.}\ \bibnamefont {Evans}},
  \bibinfo {author} {\bibfnamefont {R.~T.}\ \bibnamefont {Ireland}}, \bibinfo
  {author} {\bibfnamefont {F.}~\bibnamefont {Lempriere}},\ and\ \bibinfo
  {author} {\bibfnamefont {L.~K.}\ \bibnamefont {McKemmish}},\ }\bibfield
  {title} {\enquote {\bibinfo {title} {Benchmarking basis sets for density
  functional theory thermochemistry calculations: Why unpolarized basis sets
  and the polarized 6-311g family should be avoided},}\ }\href
  {https://doi.org/10.1021/acs.jpca.3c05573} {\bibfield  {journal} {\bibinfo
  {journal} {J. Phys. Chem. A}\ }\textbf {\bibinfo {volume} {127}},\ \bibinfo
  {pages} {10295--10306} (\bibinfo {year} {2023})}\BibitemShut {NoStop}%
\bibitem [{\citenamefont {Dunning}(1989)}]{Dunning1989_JCP_1007}%
  \BibitemOpen
  \bibfield  {author} {\bibinfo {author} {\bibfnamefont {T.~H.}\ \bibnamefont
  {Dunning}},\ }\bibfield  {title} {\enquote {\bibinfo {title} {{Gaussian basis
  sets for use in correlated molecular calculations. I. The atoms boron through
  neon and hydrogen}},}\ }\href {https://doi.org/10.1063/1.456153} {\bibfield
  {journal} {\bibinfo  {journal} {J. Chem. Phys.}\ }\textbf {\bibinfo {volume}
  {90}},\ \bibinfo {pages} {1007} (\bibinfo {year} {1989})}\BibitemShut
  {NoStop}%
\bibitem [{\citenamefont {Weigend}, \citenamefont {Furche},\ and\ \citenamefont
  {Ahlrichs}(2003)}]{Weigend2003_JCP_12753}%
  \BibitemOpen
  \bibfield  {author} {\bibinfo {author} {\bibfnamefont {F.}~\bibnamefont
  {Weigend}}, \bibinfo {author} {\bibfnamefont {F.}~\bibnamefont {Furche}},\
  and\ \bibinfo {author} {\bibfnamefont {R.}~\bibnamefont {Ahlrichs}},\
  }\bibfield  {title} {\enquote {\bibinfo {title} {{Gaussian basis sets of
  quadruple zeta valence quality for atoms H--Kr}},}\ }\href
  {https://doi.org/10.1063/1.1627293} {\bibfield  {journal} {\bibinfo
  {journal} {J. Chem. Phys.}\ }\textbf {\bibinfo {volume} {119}},\ \bibinfo
  {pages} {12753} (\bibinfo {year} {2003})}\BibitemShut {NoStop}%
\bibitem [{\citenamefont {Jensen}(2001)}]{Jensen2001_JCP_9113}%
  \BibitemOpen
  \bibfield  {author} {\bibinfo {author} {\bibfnamefont {F.}~\bibnamefont
  {Jensen}},\ }\bibfield  {title} {\enquote {\bibinfo {title} {Polarization
  consistent basis sets: Principles},}\ }\href
  {https://doi.org/10.1063/1.1413524} {\bibfield  {journal} {\bibinfo
  {journal} {J. Chem. Phys.}\ }\textbf {\bibinfo {volume} {115}},\ \bibinfo
  {pages} {9113--9125} (\bibinfo {year} {2001})}\BibitemShut {NoStop}%
\bibitem [{\citenamefont {Manninen}\ and\ \citenamefont
  {Vaara}(2006)}]{Manninen2006_JCC_434}%
  \BibitemOpen
  \bibfield  {author} {\bibinfo {author} {\bibfnamefont {P.}~\bibnamefont
  {Manninen}}\ and\ \bibinfo {author} {\bibfnamefont {J.}~\bibnamefont
  {Vaara}},\ }\bibfield  {title} {\enquote {\bibinfo {title} {{Systematic
  Gaussian basis-set limit using completeness-optimized primitive sets. A case
  for magnetic properties}},}\ }\href {https://doi.org/10.1002/jcc.20358}
  {\bibfield  {journal} {\bibinfo  {journal} {J. Comput. Chem.}\ }\textbf
  {\bibinfo {volume} {27}},\ \bibinfo {pages} {434--445} (\bibinfo {year}
  {2006})}\BibitemShut {NoStop}%
\bibitem [{\citenamefont {Lehtola}(2015)}]{Lehtola2015_JCC_335}%
  \BibitemOpen
  \bibfield  {author} {\bibinfo {author} {\bibfnamefont {S.}~\bibnamefont
  {Lehtola}},\ }\bibfield  {title} {\enquote {\bibinfo {title} {Automatic
  algorithms for completeness-optimization of {Gaussian} basis sets},}\ }\href
  {https://doi.org/10.1002/jcc.23802} {\bibfield  {journal} {\bibinfo
  {journal} {J. Comput. Chem.}\ }\textbf {\bibinfo {volume} {36}},\ \bibinfo
  {pages} {335--347} (\bibinfo {year} {2015})}\BibitemShut {NoStop}%
\bibitem [{\citenamefont {Jensen}(2006)}]{Jensen2006_JCTC_1360}%
  \BibitemOpen
  \bibfield  {author} {\bibinfo {author} {\bibfnamefont {F.}~\bibnamefont
  {Jensen}},\ }\bibfield  {title} {\enquote {\bibinfo {title} {The basis set
  convergence of spin-spin coupling constants calculated by density functional
  methods},}\ }\href {https://doi.org/10.1021/ct600166u} {\bibfield  {journal}
  {\bibinfo  {journal} {J. Chem. Theory Comput.}\ }\textbf {\bibinfo {volume}
  {2}},\ \bibinfo {pages} {1360--1369} (\bibinfo {year} {2006})}\BibitemShut
  {NoStop}%
\bibitem [{\citenamefont {Jensen}(2008)}]{Jensen2008_JCTC_719}%
  \BibitemOpen
  \bibfield  {author} {\bibinfo {author} {\bibfnamefont {F.}~\bibnamefont
  {Jensen}},\ }\bibfield  {title} {\enquote {\bibinfo {title} {Basis set
  convergence of nuclear magnetic shielding constants calculated by density
  functional methods},}\ }\href {https://doi.org/10.1021/ct800013z} {\bibfield
  {journal} {\bibinfo  {journal} {J. Chem. Theory Comput.}\ }\textbf {\bibinfo
  {volume} {4}},\ \bibinfo {pages} {719--727} (\bibinfo {year}
  {2008})}\BibitemShut {NoStop}%
\bibitem [{\citenamefont {Lehtola}\ \emph {et~al.}(2012)\citenamefont
  {Lehtola}, \citenamefont {Manninen}, \citenamefont {Hakala},\ and\
  \citenamefont {H{\"{a}}m{\"{a}}l{\"{a}}inen}}]{Lehtola2012_JCP_104105}%
  \BibitemOpen
  \bibfield  {author} {\bibinfo {author} {\bibfnamefont {J.}~\bibnamefont
  {Lehtola}}, \bibinfo {author} {\bibfnamefont {P.}~\bibnamefont {Manninen}},
  \bibinfo {author} {\bibfnamefont {M.}~\bibnamefont {Hakala}},\ and\ \bibinfo
  {author} {\bibfnamefont {K.}~\bibnamefont {H{\"{a}}m{\"{a}}l{\"{a}}inen}},\
  }\bibfield  {title} {\enquote {\bibinfo {title} {Completeness-optimized basis
  sets: Application to ground-state electron momentum densities},}\ }\href
  {https://doi.org/10.1063/1.4749272} {\bibfield  {journal} {\bibinfo
  {journal} {J. Chem. Phys.}\ }\textbf {\bibinfo {volume} {137}},\ \bibinfo
  {pages} {104105} (\bibinfo {year} {2012})}\BibitemShut {NoStop}%
\bibitem [{\citenamefont {Lehtola}\ \emph {et~al.}(2013)\citenamefont
  {Lehtola}, \citenamefont {Manninen}, \citenamefont {Hakala},\ and\
  \citenamefont {H{\"{a}}m{\"{a}}l{\"{a}}inen}}]{Lehtola2013_JCP_44109}%
  \BibitemOpen
  \bibfield  {author} {\bibinfo {author} {\bibfnamefont {S.}~\bibnamefont
  {Lehtola}}, \bibinfo {author} {\bibfnamefont {P.}~\bibnamefont {Manninen}},
  \bibinfo {author} {\bibfnamefont {M.}~\bibnamefont {Hakala}},\ and\ \bibinfo
  {author} {\bibfnamefont {K.}~\bibnamefont {H{\"{a}}m{\"{a}}l{\"{a}}inen}},\
  }\bibfield  {title} {\enquote {\bibinfo {title} {Contraction of
  completeness-optimized basis sets: application to ground-state electron
  momentum densities},}\ }\href {https://doi.org/10.1063/1.4788635} {\bibfield
  {journal} {\bibinfo  {journal} {J. Chem. Phys.}\ }\textbf {\bibinfo {volume}
  {138}},\ \bibinfo {pages} {044109} (\bibinfo {year} {2013})}\BibitemShut
  {NoStop}%
\bibitem [{\citenamefont {Lehtola}, \citenamefont {Dimitrova},\ and\
  \citenamefont {Sundholm}(2020)}]{Lehtola2020_MP_1597989}%
  \BibitemOpen
  \bibfield  {author} {\bibinfo {author} {\bibfnamefont {S.}~\bibnamefont
  {Lehtola}}, \bibinfo {author} {\bibfnamefont {M.}~\bibnamefont {Dimitrova}},\
  and\ \bibinfo {author} {\bibfnamefont {D.}~\bibnamefont {Sundholm}},\
  }\bibfield  {title} {\enquote {\bibinfo {title} {Fully numerical electronic
  structure calculations on diatomic molecules in weak to strong magnetic
  fields},}\ }\href {https://doi.org/10.1080/00268976.2019.1597989} {\bibfield
  {journal} {\bibinfo  {journal} {Mol. Phys.}\ }\textbf {\bibinfo {volume}
  {118}},\ \bibinfo {pages} {e1597989} (\bibinfo {year} {2020})},\ \Eprint
  {https://arxiv.org/abs/1812.06274} {arXiv:1812.06274} \BibitemShut {NoStop}%
\bibitem [{\citenamefont {{\AA{}}str{\"{o}}m}\ and\ \citenamefont
  {Lehtola}(2023)}]{Aastroem2023__}%
  \BibitemOpen
  \bibfield  {author} {\bibinfo {author} {\bibfnamefont {H.}~\bibnamefont
  {{\AA{}}str{\"{o}}m}}\ and\ \bibinfo {author} {\bibfnamefont
  {S.}~\bibnamefont {Lehtola}},\ }\href@noop {} {\enquote {\bibinfo {title}
  {Insight on {Gaussian} basis set truncation errors in weak to intermediate
  magnetic fields with an approximate hamiltonian},}\ } (\bibinfo {year}
  {2023}),\ \Eprint {https://arxiv.org/abs/2307.02635} {arXiv:2307.02635
  [physics.chem-ph]} \BibitemShut {NoStop}%
\bibitem [{\citenamefont {Shaw}\ and\ \citenamefont
  {Hill}(2023)}]{Shaw2023_JCP_44802}%
  \BibitemOpen
  \bibfield  {author} {\bibinfo {author} {\bibfnamefont {R.~A.}\ \bibnamefont
  {Shaw}}\ and\ \bibinfo {author} {\bibfnamefont {J.~G.}\ \bibnamefont
  {Hill}},\ }\bibfield  {title} {\enquote {\bibinfo {title} {{BasisOpt}: A
  {Python} package for quantum chemistry basis set optimization},}\ }\href
  {https://doi.org/10.1063/5.0157878} {\bibfield  {journal} {\bibinfo
  {journal} {J. Chem. Phys.}\ }\textbf {\bibinfo {volume} {159}},\ \bibinfo
  {pages} {044802} (\bibinfo {year} {2023})}\BibitemShut {NoStop}%
\bibitem [{\citenamefont {Richardson}\ \emph {et~al.}(1962)\citenamefont
  {Richardson}, \citenamefont {Nieuwpoort}, \citenamefont {Powell},\ and\
  \citenamefont {Edgell}}]{Richardson1962_JCP_1057}%
  \BibitemOpen
  \bibfield  {author} {\bibinfo {author} {\bibfnamefont {J.~W.}\ \bibnamefont
  {Richardson}}, \bibinfo {author} {\bibfnamefont {W.~C.}\ \bibnamefont
  {Nieuwpoort}}, \bibinfo {author} {\bibfnamefont {R.~R.}\ \bibnamefont
  {Powell}},\ and\ \bibinfo {author} {\bibfnamefont {W.~F.}\ \bibnamefont
  {Edgell}},\ }\bibfield  {title} {\enquote {\bibinfo {title} {Approximate
  radial functions for first-row transition-metal atoms and ions. {I}.
  {Inner}-shell, 3d and 4s atomic orbitals},}\ }\href
  {https://doi.org/10.1063/1.1732631} {\bibfield  {journal} {\bibinfo
  {journal} {J. Chem. Phys.}\ }\textbf {\bibinfo {volume} {36}},\ \bibinfo
  {pages} {1057--1061} (\bibinfo {year} {1962})}\BibitemShut {NoStop}%
\bibitem [{\citenamefont {Richardson}, \citenamefont {Powell},\ and\
  \citenamefont {Nieuwpoort}(1963)}]{Richardson1963_JCP_796}%
  \BibitemOpen
  \bibfield  {author} {\bibinfo {author} {\bibfnamefont {J.~W.}\ \bibnamefont
  {Richardson}}, \bibinfo {author} {\bibfnamefont {R.~R.}\ \bibnamefont
  {Powell}},\ and\ \bibinfo {author} {\bibfnamefont {W.~C.}\ \bibnamefont
  {Nieuwpoort}},\ }\bibfield  {title} {\enquote {\bibinfo {title} {Approximate
  radial functions for first-row transition-metal atoms and ions. {II}. 4p and
  4d atomic orbitals},}\ }\href {https://doi.org/10.1063/1.1733765} {\bibfield
  {journal} {\bibinfo  {journal} {J. Chem. Phys.}\ }\textbf {\bibinfo {volume}
  {38}},\ \bibinfo {pages} {796--801} (\bibinfo {year} {1963})}\BibitemShut
  {NoStop}%
\bibitem [{\citenamefont {Francisco}, \citenamefont {Seijo},\ and\
  \citenamefont {Pueyo}(1987)}]{Francisco1987_IJQC_279}%
  \BibitemOpen
  \bibfield  {author} {\bibinfo {author} {\bibfnamefont {E.}~\bibnamefont
  {Francisco}}, \bibinfo {author} {\bibfnamefont {L.}~\bibnamefont {Seijo}},\
  and\ \bibinfo {author} {\bibfnamefont {L.}~\bibnamefont {Pueyo}},\ }\bibfield
   {title} {\enquote {\bibinfo {title} {{Basis sets generation: Relation
  between Adamowicz's and the maximum overlap method}},}\ }\href
  {https://doi.org/10.1002/qua.560310208} {\bibfield  {journal} {\bibinfo
  {journal} {Int. J. Quantum Chem.}\ }\textbf {\bibinfo {volume} {31}},\
  \bibinfo {pages} {279--285} (\bibinfo {year} {1987})}\BibitemShut {NoStop}%
\bibitem [{\citenamefont {Kalman}(1971)}]{Kalman1971_JCP_1841}%
  \BibitemOpen
  \bibfield  {author} {\bibinfo {author} {\bibfnamefont {B.~L.}\ \bibnamefont
  {Kalman}},\ }\bibfield  {title} {\enquote {\bibinfo {title} {{MO} expansions
  in finite nonorthogonal basis sets: Conversion from an extended basis to a
  smaller basis with maximum overlap},}\ }\href
  {https://doi.org/10.1063/1.1675100} {\bibfield  {journal} {\bibinfo
  {journal} {J. Chem. Phys.}\ }\textbf {\bibinfo {volume} {54}},\ \bibinfo
  {pages} {1841--1842} (\bibinfo {year} {1971})}\BibitemShut {NoStop}%
\bibitem [{\citenamefont {Adamowicz}(1981)}]{Adamowicz1981_IJQC_545}%
  \BibitemOpen
  \bibfield  {author} {\bibinfo {author} {\bibfnamefont {L.}~\bibnamefont
  {Adamowicz}},\ }\bibfield  {title} {\enquote {\bibinfo {title} {{Basis set
  generation for the SCF calculation}},}\ }\href
  {https://doi.org/10.1002/qua.560190408} {\bibfield  {journal} {\bibinfo
  {journal} {Int. J. Quantum Chem.}\ }\textbf {\bibinfo {volume} {19}},\
  \bibinfo {pages} {545--551} (\bibinfo {year} {1981})}\BibitemShut {NoStop}%
\bibitem [{\citenamefont {Adamowicz}\ and\ \citenamefont
  {McCullough}(1983)}]{Adamowicz1983_IJQC_19}%
  \BibitemOpen
  \bibfield  {author} {\bibinfo {author} {\bibfnamefont {L.}~\bibnamefont
  {Adamowicz}}\ and\ \bibinfo {author} {\bibfnamefont {E.~A.}\ \bibnamefont
  {McCullough}},\ }\bibfield  {title} {\enquote {\bibinfo {title} {Molecular
  basis set generation: Accurate {Slater} basis sets for \ce{LiH-} ground and
  excited state and \ce{Li2-} ground state},}\ }\href
  {https://doi.org/10.1002/qua.560240103} {\bibfield  {journal} {\bibinfo
  {journal} {Int. J. Quantum Chem.}\ }\textbf {\bibinfo {volume} {24}},\
  \bibinfo {pages} {19--23} (\bibinfo {year} {1983})}\BibitemShut {NoStop}%
\bibitem [{\citenamefont {Kobus}, \citenamefont {Moncrieff},\ and\
  \citenamefont {Wilson}(1997)}]{Kobus1997_MP_1015}%
  \BibitemOpen
  \bibfield  {author} {\bibinfo {author} {\bibfnamefont {J.}~\bibnamefont
  {Kobus}}, \bibinfo {author} {\bibfnamefont {D.}~\bibnamefont {Moncrieff}},\
  and\ \bibinfo {author} {\bibfnamefont {S.}~\bibnamefont {Wilson}},\
  }\bibfield  {title} {\enquote {\bibinfo {title} {{Visualization of
  deficiencies in approximate molecular wave functions: the orbital amplitude
  difference function for the matrix Hartree--Fock description of the ground
  state of the boron fluoride molecule}},}\ }\href
  {https://doi.org/10.1080/002689797169637} {\bibfield  {journal} {\bibinfo
  {journal} {Mol. Phys.}\ }\textbf {\bibinfo {volume} {92}},\ \bibinfo {pages}
  {1015--1028} (\bibinfo {year} {1997})}\BibitemShut {NoStop}%
\bibitem [{\citenamefont {Kobus}, \citenamefont {Moncrieff},\ and\
  \citenamefont {Wilson}(2001)}]{Kobus2001_MP_315}%
  \BibitemOpen
  \bibfield  {author} {\bibinfo {author} {\bibfnamefont {J.}~\bibnamefont
  {Kobus}}, \bibinfo {author} {\bibfnamefont {D.}~\bibnamefont {Moncrieff}},\
  and\ \bibinfo {author} {\bibfnamefont {S.}~\bibnamefont {Wilson}},\
  }\bibfield  {title} {\enquote {\bibinfo {title} {{Visualization of
  deficiencies in approximate molecular wave functions: the local orbital
  energy function for the matrix Hartree--Fock model}},}\ }\href
  {https://doi.org/10.1080/00268970010011780} {\bibfield  {journal} {\bibinfo
  {journal} {Mol. Phys.}\ }\textbf {\bibinfo {volume} {99}},\ \bibinfo {pages}
  {315--326} (\bibinfo {year} {2001})}\BibitemShut {NoStop}%
\bibitem [{\citenamefont {{Van Lenthe}}\ and\ \citenamefont
  {Baerends}(2003)}]{VanLenthe2003_JCC_56}%
  \BibitemOpen
  \bibfield  {author} {\bibinfo {author} {\bibfnamefont {E.}~\bibnamefont {{Van
  Lenthe}}}\ and\ \bibinfo {author} {\bibfnamefont {E.~J.}\ \bibnamefont
  {Baerends}},\ }\bibfield  {title} {\enquote {\bibinfo {title} {{Optimized
  Slater-type basis sets for the elements 1-118.}}}\ }\href
  {https://doi.org/10.1002/jcc.10255} {\bibfield  {journal} {\bibinfo
  {journal} {J. Comput. Chem.}\ }\textbf {\bibinfo {volume} {24}},\ \bibinfo
  {pages} {1142--56} (\bibinfo {year} {2003})}\BibitemShut {NoStop}%
\bibitem [{\citenamefont {Sanchez-Portal}, \citenamefont {Artacho},\ and\
  \citenamefont {Soler}(1995)}]{SanchezPortal1995_SSC_685}%
  \BibitemOpen
  \bibfield  {author} {\bibinfo {author} {\bibfnamefont {D.}~\bibnamefont
  {Sanchez-Portal}}, \bibinfo {author} {\bibfnamefont {E.}~\bibnamefont
  {Artacho}},\ and\ \bibinfo {author} {\bibfnamefont {J.~M.}\ \bibnamefont
  {Soler}},\ }\bibfield  {title} {\enquote {\bibinfo {title} {{Projection of
  plane-wave calculations into atomic orbitals}},}\ }\href
  {https://doi.org/10.1016/0038-1098(95)00341-X} {\bibfield  {journal}
  {\bibinfo  {journal} {Solid State Commun.}\ }\textbf {\bibinfo {volume}
  {95}},\ \bibinfo {pages} {685--690} (\bibinfo {year} {1995})},\ \Eprint
  {https://arxiv.org/abs/9505075} {arXiv:9505075 [cond-mat]} \BibitemShut
  {NoStop}%
\bibitem [{\citenamefont {S{\'{a}}nchez-Portal}, \citenamefont {Artacho},\ and\
  \citenamefont {Soler}(1996)}]{SanchezPortal1996_JPCM_3859}%
  \BibitemOpen
  \bibfield  {author} {\bibinfo {author} {\bibfnamefont {D.}~\bibnamefont
  {S{\'{a}}nchez-Portal}}, \bibinfo {author} {\bibfnamefont {E.}~\bibnamefont
  {Artacho}},\ and\ \bibinfo {author} {\bibfnamefont {J.~M.}\ \bibnamefont
  {Soler}},\ }\bibfield  {title} {\enquote {\bibinfo {title} {Analysis of
  atomic orbital basis sets from the projection of plane-wave results},}\
  }\href {https://doi.org/10.1088/0953-8984/8/21/012} {\bibfield  {journal}
  {\bibinfo  {journal} {J. Phys.: Condens. Matter}\ }\textbf {\bibinfo {volume}
  {8}},\ \bibinfo {pages} {3859--3880} (\bibinfo {year} {1996})}\BibitemShut
  {NoStop}%
\bibitem [{\citenamefont {Chen}\ \emph {et~al.}(2009)\citenamefont {Chen},
  \citenamefont {Fang}, \citenamefont {Sun}, \citenamefont {Guo},\ and\
  \citenamefont {He}}]{Chen2009_PRB_165121}%
  \BibitemOpen
  \bibfield  {author} {\bibinfo {author} {\bibfnamefont {M.}~\bibnamefont
  {Chen}}, \bibinfo {author} {\bibfnamefont {W.}~\bibnamefont {Fang}}, \bibinfo
  {author} {\bibfnamefont {G.-Z.}\ \bibnamefont {Sun}}, \bibinfo {author}
  {\bibfnamefont {G.-C.}\ \bibnamefont {Guo}},\ and\ \bibinfo {author}
  {\bibfnamefont {L.}~\bibnamefont {He}},\ }\bibfield  {title} {\enquote
  {\bibinfo {title} {{Method to construct transferable minimal basis sets for
  ab initio calculations}},}\ }\href
  {https://doi.org/10.1103/PhysRevB.80.165121} {\bibfield  {journal} {\bibinfo
  {journal} {Phys. Rev. B}\ }\textbf {\bibinfo {volume} {80}},\ \bibinfo
  {pages} {165121} (\bibinfo {year} {2009})}\BibitemShut {NoStop}%
\bibitem [{\citenamefont {Chen}, \citenamefont {Guo},\ and\ \citenamefont
  {He}(2010)}]{Chen2010_JPCM_445501}%
  \BibitemOpen
  \bibfield  {author} {\bibinfo {author} {\bibfnamefont {M.}~\bibnamefont
  {Chen}}, \bibinfo {author} {\bibfnamefont {G.-C.}\ \bibnamefont {Guo}},\ and\
  \bibinfo {author} {\bibfnamefont {L.}~\bibnamefont {He}},\ }\bibfield
  {title} {\enquote {\bibinfo {title} {Systematically improvable optimized
  atomic basis sets for ab initio calculations},}\ }\href
  {https://doi.org/10.1088/0953-8984/22/44/445501} {\bibfield  {journal}
  {\bibinfo  {journal} {J. Phys.: Condens. Matter}\ }\textbf {\bibinfo {volume}
  {22}},\ \bibinfo {pages} {445501} (\bibinfo {year} {2010})}\BibitemShut
  {NoStop}%
\bibitem [{\citenamefont {Chen}, \citenamefont {Guo},\ and\ \citenamefont
  {He}(2011)}]{Chen2011_JPCM_325501}%
  \BibitemOpen
  \bibfield  {author} {\bibinfo {author} {\bibfnamefont {M.}~\bibnamefont
  {Chen}}, \bibinfo {author} {\bibfnamefont {G.-C.}\ \bibnamefont {Guo}},\ and\
  \bibinfo {author} {\bibfnamefont {L.}~\bibnamefont {He}},\ }\bibfield
  {title} {\enquote {\bibinfo {title} {Electronic structure interpolation via
  atomic orbitals},}\ }\href {https://doi.org/10.1088/0953-8984/23/32/325501}
  {\bibfield  {journal} {\bibinfo  {journal} {J. Phys.: Condens. Matter}\
  }\textbf {\bibinfo {volume} {23}},\ \bibinfo {pages} {325501} (\bibinfo
  {year} {2011})}\BibitemShut {NoStop}%
\bibitem [{\citenamefont {Lange}, \citenamefont {Freysoldt},\ and\
  \citenamefont {Neugebauer}(2011)}]{Lange2011_PRB_85101}%
  \BibitemOpen
  \bibfield  {author} {\bibinfo {author} {\bibfnamefont {B.}~\bibnamefont
  {Lange}}, \bibinfo {author} {\bibfnamefont {C.}~\bibnamefont {Freysoldt}},\
  and\ \bibinfo {author} {\bibfnamefont {J.}~\bibnamefont {Neugebauer}},\
  }\bibfield  {title} {\enquote {\bibinfo {title} {Construction and performance
  of fully numerical optimum atomic basis sets},}\ }\href
  {https://doi.org/10.1103/PhysRevB.84.085101} {\bibfield  {journal} {\bibinfo
  {journal} {Phys. Rev. B}\ }\textbf {\bibinfo {volume} {84}},\ \bibinfo
  {pages} {085101} (\bibinfo {year} {2011})}\BibitemShut {NoStop}%
\bibitem [{\citenamefont {Lin}, \citenamefont {Ren},\ and\ \citenamefont
  {He}(2021)}]{Lin2021_PRB_235131}%
  \BibitemOpen
  \bibfield  {author} {\bibinfo {author} {\bibfnamefont {P.}~\bibnamefont
  {Lin}}, \bibinfo {author} {\bibfnamefont {X.}~\bibnamefont {Ren}},\ and\
  \bibinfo {author} {\bibfnamefont {L.}~\bibnamefont {He}},\ }\bibfield
  {title} {\enquote {\bibinfo {title} {Strategy for constructing compact
  numerical atomic orbital basis sets by incorporating the gradients of
  reference wavefunctions},}\ }\href
  {https://doi.org/10.1103/physrevb.103.235131} {\bibfield  {journal} {\bibinfo
   {journal} {Phys. Rev. B}\ }\textbf {\bibinfo {volume} {103}},\ \bibinfo
  {pages} {235131} (\bibinfo {year} {2021})}\BibitemShut {NoStop}%
\bibitem [{\citenamefont {Hohenberg}\ and\ \citenamefont
  {Kohn}(1964)}]{Hohenberg1964_PR_864}%
  \BibitemOpen
  \bibfield  {author} {\bibinfo {author} {\bibfnamefont {P.}~\bibnamefont
  {Hohenberg}}\ and\ \bibinfo {author} {\bibfnamefont {W.}~\bibnamefont
  {Kohn}},\ }\bibfield  {title} {\enquote {\bibinfo {title} {Inhomogeneous
  electron gas},}\ }\href {https://doi.org/10.1103/PhysRev.136.B864} {\bibfield
   {journal} {\bibinfo  {journal} {Phys. Rev.}\ }\textbf {\bibinfo {volume}
  {136}},\ \bibinfo {pages} {B864--B871} (\bibinfo {year} {1964})}\BibitemShut
  {NoStop}%
\bibitem [{\citenamefont {Kohn}\ and\ \citenamefont
  {Sham}(1965)}]{Kohn1965_PR_1133}%
  \BibitemOpen
  \bibfield  {author} {\bibinfo {author} {\bibfnamefont {W.}~\bibnamefont
  {Kohn}}\ and\ \bibinfo {author} {\bibfnamefont {L.~J.}\ \bibnamefont
  {Sham}},\ }\bibfield  {title} {\enquote {\bibinfo {title} {Self-consistent
  equations including exchange and correlation effects},}\ }\href
  {https://doi.org/10.1103/PhysRev.140.A1133} {\bibfield  {journal} {\bibinfo
  {journal} {Phys. Rev.}\ }\textbf {\bibinfo {volume} {140}},\ \bibinfo {pages}
  {A1133--A1138} (\bibinfo {year} {1965})}\BibitemShut {NoStop}%
\bibitem [{\citenamefont {Jensen}\ \emph {et~al.}(2017)\citenamefont {Jensen},
  \citenamefont {Saha}, \citenamefont {Flores-Livas}, \citenamefont {Huhn},
  \citenamefont {Blum}, \citenamefont {Goedecker},\ and\ \citenamefont
  {Frediani}}]{Jensen2017_JPCL_1449}%
  \BibitemOpen
  \bibfield  {author} {\bibinfo {author} {\bibfnamefont {S.~R.}\ \bibnamefont
  {Jensen}}, \bibinfo {author} {\bibfnamefont {S.}~\bibnamefont {Saha}},
  \bibinfo {author} {\bibfnamefont {J.~A.}\ \bibnamefont {Flores-Livas}},
  \bibinfo {author} {\bibfnamefont {W.}~\bibnamefont {Huhn}}, \bibinfo {author}
  {\bibfnamefont {V.}~\bibnamefont {Blum}}, \bibinfo {author} {\bibfnamefont
  {S.}~\bibnamefont {Goedecker}},\ and\ \bibinfo {author} {\bibfnamefont
  {L.}~\bibnamefont {Frediani}},\ }\bibfield  {title} {\enquote {\bibinfo
  {title} {The elephant in the room of density functional theory
  calculations},}\ }\href {https://doi.org/10.1021/acs.jpclett.7b00255}
  {\bibfield  {journal} {\bibinfo  {journal} {J. Phys. Chem. Lett.}\ }\textbf
  {\bibinfo {volume} {8}},\ \bibinfo {pages} {1449--1457} (\bibinfo {year}
  {2017})},\ \Eprint {https://arxiv.org/abs/1702.00957} {arXiv:1702.00957}
  \BibitemShut {NoStop}%
\bibitem [{\citenamefont {Brakestad}\ \emph {et~al.}(2021)\citenamefont
  {Brakestad}, \citenamefont {Wind}, \citenamefont {Jensen}, \citenamefont
  {Frediani},\ and\ \citenamefont {Hopmann}}]{Brakestad2021_JCP_214302}%
  \BibitemOpen
  \bibfield  {author} {\bibinfo {author} {\bibfnamefont {A.}~\bibnamefont
  {Brakestad}}, \bibinfo {author} {\bibfnamefont {P.}~\bibnamefont {Wind}},
  \bibinfo {author} {\bibfnamefont {S.~R.}\ \bibnamefont {Jensen}}, \bibinfo
  {author} {\bibfnamefont {L.}~\bibnamefont {Frediani}},\ and\ \bibinfo
  {author} {\bibfnamefont {K.~H.}\ \bibnamefont {Hopmann}},\ }\bibfield
  {title} {\enquote {\bibinfo {title} {Multiwavelets applied to metal--ligand
  interactions: Energies free from basis set errors},}\ }\href
  {https://doi.org/10.1063/5.0046023} {\bibfield  {journal} {\bibinfo
  {journal} {J. Chem. Phys.}\ }\textbf {\bibinfo {volume} {154}},\ \bibinfo
  {pages} {214302} (\bibinfo {year} {2021})}\BibitemShut {NoStop}%
\bibitem [{\citenamefont {Vaughn}, \citenamefont {Gavini},\ and\ \citenamefont
  {Krasny}(2021)}]{Vaughn2021_JCP_110101}%
  \BibitemOpen
  \bibfield  {author} {\bibinfo {author} {\bibfnamefont {N.}~\bibnamefont
  {Vaughn}}, \bibinfo {author} {\bibfnamefont {V.}~\bibnamefont {Gavini}},\
  and\ \bibinfo {author} {\bibfnamefont {R.}~\bibnamefont {Krasny}},\
  }\bibfield  {title} {\enquote {\bibinfo {title} {Treecode-accelerated green
  iteration for kohn--sham density functional theory},}\ }\href
  {https://doi.org/10.1016/j.jcp.2020.110101} {\bibfield  {journal} {\bibinfo
  {journal} {J. Comput. Phys.}\ }\textbf {\bibinfo {volume} {430}},\ \bibinfo
  {pages} {110101} (\bibinfo {year} {2021})}\BibitemShut {NoStop}%
\bibitem [{\citenamefont {Valeev}\ \emph {et~al.}(2023)\citenamefont {Valeev},
  \citenamefont {Harrison}, \citenamefont {Holmes}, \citenamefont {Peterson},\
  and\ \citenamefont {Penchoff}}]{Valeev2023__}%
  \BibitemOpen
  \bibfield  {author} {\bibinfo {author} {\bibfnamefont {E.~F.}\ \bibnamefont
  {Valeev}}, \bibinfo {author} {\bibfnamefont {R.~J.}\ \bibnamefont
  {Harrison}}, \bibinfo {author} {\bibfnamefont {A.~A.}\ \bibnamefont
  {Holmes}}, \bibinfo {author} {\bibfnamefont {C.~C.}\ \bibnamefont
  {Peterson}},\ and\ \bibinfo {author} {\bibfnamefont {D.~A.}\ \bibnamefont
  {Penchoff}},\ }\bibfield  {title} {\enquote {\bibinfo {title} {Direct
  determination of optimal real-space orbitals for correlated electronic
  structure of molecules},}\ }\href {https://doi.org/10.1021/acs.jctc.3c00732}
  {\bibfield  {journal} {\bibinfo  {journal} {J. Chem. Theory Comput.}\ }
  (\bibinfo {year} {2023}),\ 10.1021/acs.jctc.3c00732},\ \Eprint
  {https://arxiv.org/abs/2207.10841} {2207.10841} \BibitemShut {NoStop}%
\bibitem [{\citenamefont {Gygi}(2023)}]{Gygi2023_JCTC_1300}%
  \BibitemOpen
  \bibfield  {author} {\bibinfo {author} {\bibfnamefont {F.}~\bibnamefont
  {Gygi}},\ }\bibfield  {title} {\enquote {\bibinfo {title} {All-electron
  plane-wave electronic structure calculations},}\ }\href
  {https://doi.org/10.1021/acs.jctc.2c01191} {\bibfield  {journal} {\bibinfo
  {journal} {J. Chem. Theory Comput.}\ }\textbf {\bibinfo {volume} {19}},\
  \bibinfo {pages} {1300--1309} (\bibinfo {year} {2023})}\BibitemShut {NoStop}%
\bibitem [{\citenamefont
  {Lehtola}(2023{\natexlab{a}})}]{Lehtola2023_JCTC_4033}%
  \BibitemOpen
  \bibfield  {author} {\bibinfo {author} {\bibfnamefont {S.}~\bibnamefont
  {Lehtola}},\ }\bibfield  {title} {\enquote {\bibinfo {title} {Accuracy of a
  recent regularized nuclear potential},}\ }\href
  {https://doi.org/10.1021/acs.jctc.3c00530} {\bibfield  {journal} {\bibinfo
  {journal} {J. Chem. Theory Comput.}\ }\textbf {\bibinfo {volume} {19}},\
  \bibinfo {pages} {4033--4039} (\bibinfo {year} {2023}{\natexlab{a}})},\
  \Eprint {https://arxiv.org/abs/2302.09557} {2302.09557} \BibitemShut
  {NoStop}%
\bibitem [{\citenamefont {Chong}(1995)}]{Chong1995_CJC_79}%
  \BibitemOpen
  \bibfield  {author} {\bibinfo {author} {\bibfnamefont {D.~P.}\ \bibnamefont
  {Chong}},\ }\bibfield  {title} {\enquote {\bibinfo {title} {Completeness
  profiles of one-electron basis sets},}\ }\href
  {https://doi.org/10.1139/v95-011} {\bibfield  {journal} {\bibinfo  {journal}
  {Can. J. Chem.}\ }\textbf {\bibinfo {volume} {73}},\ \bibinfo {pages}
  {79--83} (\bibinfo {year} {1995})}\BibitemShut {NoStop}%
\bibitem [{\citenamefont {G{\"{u}}ell}\ \emph {et~al.}(2008)\citenamefont
  {G{\"{u}}ell}, \citenamefont {Luis}, \citenamefont {Sol{\`{a}}},\ and\
  \citenamefont {Swart}}]{Gueell2008_JPCA_91}%
  \BibitemOpen
  \bibfield  {author} {\bibinfo {author} {\bibfnamefont {M.}~\bibnamefont
  {G{\"{u}}ell}}, \bibinfo {author} {\bibfnamefont {J.~M.}\ \bibnamefont
  {Luis}}, \bibinfo {author} {\bibfnamefont {M.}~\bibnamefont {Sol{\`{a}}}},\
  and\ \bibinfo {author} {\bibfnamefont {M.}~\bibnamefont {Swart}},\ }\bibfield
   {title} {\enquote {\bibinfo {title} {Importance of the basis set for the
  spin-state energetics of iron complexes},}\ }\href
  {https://doi.org/10.1021/jp803441m} {\bibfield  {journal} {\bibinfo
  {journal} {J. Phys. Chem. A}\ }\textbf {\bibinfo {volume} {112}},\ \bibinfo
  {pages} {6384--91} (\bibinfo {year} {2008})}\BibitemShut {NoStop}%
\bibitem [{\citenamefont {Kato}(1957)}]{Kato1957_CPAM_151}%
  \BibitemOpen
  \bibfield  {author} {\bibinfo {author} {\bibfnamefont {T.}~\bibnamefont
  {Kato}},\ }\bibfield  {title} {\enquote {\bibinfo {title} {{On the
  eigenfunctions of many-particle systems in quantum mechanics}},}\ }\href
  {https://doi.org/10.1002/cpa.3160100201} {\bibfield  {journal} {\bibinfo
  {journal} {Commun. Pure Appl. Math.}\ }\textbf {\bibinfo {volume} {10}},\
  \bibinfo {pages} {151--177} (\bibinfo {year} {1957})}\BibitemShut {NoStop}%
\bibitem [{\citenamefont {Ahlrichs}(1972)}]{Ahlrichs1972_CPL_609}%
  \BibitemOpen
  \bibfield  {author} {\bibinfo {author} {\bibfnamefont {R.}~\bibnamefont
  {Ahlrichs}},\ }\bibfield  {title} {\enquote {\bibinfo {title} {Asymptotic
  behaviour of atomic bound state wavefunctions},}\ }\href
  {https://doi.org/10.1016/0009-2614(72)80386-5} {\bibfield  {journal}
  {\bibinfo  {journal} {Chem. Phys. Lett.}\ }\textbf {\bibinfo {volume} {15}},\
  \bibinfo {pages} {609--612} (\bibinfo {year} {1972})}\BibitemShut {NoStop}%
\bibitem [{\citenamefont {Ahlrichs}(1973)}]{Ahlrichs1973_CPL_521}%
  \BibitemOpen
  \bibfield  {author} {\bibinfo {author} {\bibfnamefont {R.}~\bibnamefont
  {Ahlrichs}},\ }\bibfield  {title} {\enquote {\bibinfo {title} {Asymptotic
  behaviour of molecular bound state wavefunctions},}\ }\href
  {https://doi.org/10.1016/0009-2614(73)80455-5} {\bibfield  {journal}
  {\bibinfo  {journal} {Chem. Phys. Lett.}\ }\textbf {\bibinfo {volume} {18}},\
  \bibinfo {pages} {521--524} (\bibinfo {year} {1973})}\BibitemShut {NoStop}%
\bibitem [{\citenamefont {Katriel}\ and\ \citenamefont
  {Davidson}(1980)}]{Katriel1980_PNASUSA_4403}%
  \BibitemOpen
  \bibfield  {author} {\bibinfo {author} {\bibfnamefont {J.}~\bibnamefont
  {Katriel}}\ and\ \bibinfo {author} {\bibfnamefont {E.~R.}\ \bibnamefont
  {Davidson}},\ }\bibfield  {title} {\enquote {\bibinfo {title} {{Asymptotic
  behavior of atomic and molecular wave functions}},}\ }\href
  {https://doi.org/10.1073/pnas.77.8.4403} {\bibfield  {journal} {\bibinfo
  {journal} {Proc. Natl. Acad. Sci. U. S. A.}\ }\textbf {\bibinfo {volume}
  {77}},\ \bibinfo {pages} {4403--4406} (\bibinfo {year} {1980})}\BibitemShut
  {NoStop}%
\bibitem [{\citenamefont {Ishida}\ and\ \citenamefont
  {Ohno}(1992)}]{Ishida1992_TCA_355}%
  \BibitemOpen
  \bibfield  {author} {\bibinfo {author} {\bibfnamefont {T.}~\bibnamefont
  {Ishida}}\ and\ \bibinfo {author} {\bibfnamefont {K.}~\bibnamefont {Ohno}},\
  }\bibfield  {title} {\enquote {\bibinfo {title} {{On the asymptotic behavior
  of Hartree--Fock orbitals}},}\ }\href {https://doi.org/10.1007/BF01134860}
  {\bibfield  {journal} {\bibinfo  {journal} {Theor. Chim. Acta}\ }\textbf
  {\bibinfo {volume} {81}},\ \bibinfo {pages} {355--364} (\bibinfo {year}
  {1992})}\BibitemShut {NoStop}%
\bibitem [{\citenamefont {Lin}\ \emph {et~al.}(2023)\citenamefont {Lin},
  \citenamefont {Ren}, \citenamefont {Liu},\ and\ \citenamefont
  {He}}]{Lin2023_WCMS_1687}%
  \BibitemOpen
  \bibfield  {author} {\bibinfo {author} {\bibfnamefont {P.}~\bibnamefont
  {Lin}}, \bibinfo {author} {\bibfnamefont {X.}~\bibnamefont {Ren}}, \bibinfo
  {author} {\bibfnamefont {X.}~\bibnamefont {Liu}},\ and\ \bibinfo {author}
  {\bibfnamefont {L.}~\bibnamefont {He}},\ }\bibfield  {title} {\enquote
  {\bibinfo {title} {Ab initio electronic structure calculations based on
  numerical atomic orbitals: Basic fomalisms and recent progresses},}\ }\href
  {https://doi.org/10.1002/wcms.1687} {\bibfield  {journal} {\bibinfo
  {journal} {WIREs Comput. Mol. Sci.}\ ,\ \bibinfo {pages} {e1687}} (\bibinfo
  {year} {2023})}\BibitemShut {NoStop}%
\bibitem [{\citenamefont {Larsen}\ \emph {et~al.}(2009)\citenamefont {Larsen},
  \citenamefont {Vanin}, \citenamefont {Mortensen}, \citenamefont {Thygesen},\
  and\ \citenamefont {Jacobsen}}]{Larsen2009_PRB_195112}%
  \BibitemOpen
  \bibfield  {author} {\bibinfo {author} {\bibfnamefont {A.~H.}\ \bibnamefont
  {Larsen}}, \bibinfo {author} {\bibfnamefont {M.}~\bibnamefont {Vanin}},
  \bibinfo {author} {\bibfnamefont {J.~J.}\ \bibnamefont {Mortensen}}, \bibinfo
  {author} {\bibfnamefont {K.~S.}\ \bibnamefont {Thygesen}},\ and\ \bibinfo
  {author} {\bibfnamefont {K.~W.}\ \bibnamefont {Jacobsen}},\ }\bibfield
  {title} {\enquote {\bibinfo {title} {{Localized atomic basis set in the
  projector augmented wave method}},}\ }\href
  {https://doi.org/10.1103/PhysRevB.80.195112} {\bibfield  {journal} {\bibinfo
  {journal} {Phys. Rev. B}\ }\textbf {\bibinfo {volume} {80}},\ \bibinfo
  {pages} {195112} (\bibinfo {year} {2009})},\ \Eprint
  {https://arxiv.org/abs/arXiv:1303.0348v1} {arXiv:arXiv:1303.0348v1}
  \BibitemShut {NoStop}%
\bibitem [{\citenamefont {Blum}\ \emph {et~al.}(2009)\citenamefont {Blum},
  \citenamefont {Gehrke}, \citenamefont {Hanke}, \citenamefont {Havu},
  \citenamefont {Havu}, \citenamefont {Ren}, \citenamefont {Reuter},\ and\
  \citenamefont {Scheffler}}]{Blum2009_CPC_2175}%
  \BibitemOpen
  \bibfield  {author} {\bibinfo {author} {\bibfnamefont {V.}~\bibnamefont
  {Blum}}, \bibinfo {author} {\bibfnamefont {R.}~\bibnamefont {Gehrke}},
  \bibinfo {author} {\bibfnamefont {F.}~\bibnamefont {Hanke}}, \bibinfo
  {author} {\bibfnamefont {P.}~\bibnamefont {Havu}}, \bibinfo {author}
  {\bibfnamefont {V.}~\bibnamefont {Havu}}, \bibinfo {author} {\bibfnamefont
  {X.}~\bibnamefont {Ren}}, \bibinfo {author} {\bibfnamefont {K.}~\bibnamefont
  {Reuter}},\ and\ \bibinfo {author} {\bibfnamefont {M.}~\bibnamefont
  {Scheffler}},\ }\bibfield  {title} {\enquote {\bibinfo {title} {{Ab initio
  molecular simulations with numeric atom-centered orbitals}},}\ }\href
  {https://doi.org/10.1016/j.cpc.2009.06.022} {\bibfield  {journal} {\bibinfo
  {journal} {Comput. Phys. Commun.}\ }\textbf {\bibinfo {volume} {180}},\
  \bibinfo {pages} {2175--2196} (\bibinfo {year} {2009})}\BibitemShut {NoStop}%
\bibitem [{\citenamefont {Jensen}(2017)}]{Jensen2017_JPCA_6104}%
  \BibitemOpen
  \bibfield  {author} {\bibinfo {author} {\bibfnamefont {F.}~\bibnamefont
  {Jensen}},\ }\bibfield  {title} {\enquote {\bibinfo {title} {How large is the
  elephant in the density functional theory room?}}\ }\href
  {https://doi.org/10.1021/acs.jpca.7b04760} {\bibfield  {journal} {\bibinfo
  {journal} {J. Phys. Chem. A}\ }\textbf {\bibinfo {volume} {121}},\ \bibinfo
  {pages} {6104--6107} (\bibinfo {year} {2017})},\ \Eprint
  {https://arxiv.org/abs/1704.08832} {arXiv:1704.08832} \BibitemShut {NoStop}%
\bibitem [{\citenamefont {Feller}\ and\ \citenamefont
  {Dixon}(2018)}]{Feller2018_JPCA_2598}%
  \BibitemOpen
  \bibfield  {author} {\bibinfo {author} {\bibfnamefont {D.}~\bibnamefont
  {Feller}}\ and\ \bibinfo {author} {\bibfnamefont {D.~A.}\ \bibnamefont
  {Dixon}},\ }\bibfield  {title} {\enquote {\bibinfo {title} {{Density
  Functional Theory and the Basis Set Truncation Problem with Correlation
  Consistent Basis Sets: Elephant in the Room or Mouse in the Closet?}}}\
  }\href {https://doi.org/10.1021/acs.jpca.8b00392} {\bibfield  {journal}
  {\bibinfo  {journal} {J. Phys. Chem. A}\ }\textbf {\bibinfo {volume} {122}},\
  \bibinfo {pages} {2598--2603} (\bibinfo {year} {2018})}\BibitemShut {NoStop}%
\bibitem [{\citenamefont {Ik{\"{a}}l{\"{a}}inen}\ \emph
  {et~al.}(2008)\citenamefont {Ik{\"{a}}l{\"{a}}inen}, \citenamefont {Lantto},
  \citenamefont {Manninen},\ and\ \citenamefont
  {Vaara}}]{Ikaelaeinen2008_JCP_124102}%
  \BibitemOpen
  \bibfield  {author} {\bibinfo {author} {\bibfnamefont {S.}~\bibnamefont
  {Ik{\"{a}}l{\"{a}}inen}}, \bibinfo {author} {\bibfnamefont {P.}~\bibnamefont
  {Lantto}}, \bibinfo {author} {\bibfnamefont {P.}~\bibnamefont {Manninen}},\
  and\ \bibinfo {author} {\bibfnamefont {J.}~\bibnamefont {Vaara}},\ }\bibfield
   {title} {\enquote {\bibinfo {title} {Laser-induced nuclear magnetic
  resonance splitting in hydrocarbons},}\ }\href
  {https://doi.org/10.1063/1.2977741} {\bibfield  {journal} {\bibinfo
  {journal} {J. Chem. Phys.}\ }\textbf {\bibinfo {volume} {129}},\ \bibinfo
  {pages} {124102} (\bibinfo {year} {2008})}\BibitemShut {NoStop}%
\bibitem [{\citenamefont {Ik{\"{a}}l{\"{a}}inen}\ \emph
  {et~al.}(2009)\citenamefont {Ik{\"{a}}l{\"{a}}inen}, \citenamefont {Lantto},
  \citenamefont {Manninen},\ and\ \citenamefont
  {Vaara}}]{Ikaelaeinen2009_PCCP_14}%
  \BibitemOpen
  \bibfield  {author} {\bibinfo {author} {\bibfnamefont {S.}~\bibnamefont
  {Ik{\"{a}}l{\"{a}}inen}}, \bibinfo {author} {\bibfnamefont {P.}~\bibnamefont
  {Lantto}}, \bibinfo {author} {\bibfnamefont {P.}~\bibnamefont {Manninen}},\
  and\ \bibinfo {author} {\bibfnamefont {J.}~\bibnamefont {Vaara}},\ }\bibfield
   {title} {\enquote {\bibinfo {title} {{NMR} tensors in planar hydrocarbons of
  increasing size},}\ }\href {https://doi.org/10.1039/b919860a} {\bibfield
  {journal} {\bibinfo  {journal} {Phys. Chem. Chem. Phys.}\ }\textbf {\bibinfo
  {volume} {11}},\ \bibinfo {pages} {11404--14} (\bibinfo {year}
  {2009})}\BibitemShut {NoStop}%
\bibitem [{\citenamefont {Ik{\"{a}}l{\"{a}}inen}\ \emph
  {et~al.}(2010)\citenamefont {Ik{\"{a}}l{\"{a}}inen}, \citenamefont {Romalis},
  \citenamefont {Lantto},\ and\ \citenamefont
  {Vaara}}]{Ikaelaeinen2010_PRL_153001}%
  \BibitemOpen
  \bibfield  {author} {\bibinfo {author} {\bibfnamefont {S.}~\bibnamefont
  {Ik{\"{a}}l{\"{a}}inen}}, \bibinfo {author} {\bibfnamefont {M.}~\bibnamefont
  {Romalis}}, \bibinfo {author} {\bibfnamefont {P.}~\bibnamefont {Lantto}},\
  and\ \bibinfo {author} {\bibfnamefont {J.}~\bibnamefont {Vaara}},\ }\bibfield
   {title} {\enquote {\bibinfo {title} {Chemical distinction by nuclear spin
  optical rotation},}\ }\href {https://doi.org/10.1103/PhysRevLett.105.153001}
  {\bibfield  {journal} {\bibinfo  {journal} {Phys. Rev. Lett.}\ }\textbf
  {\bibinfo {volume} {105}},\ \bibinfo {pages} {153001} (\bibinfo {year}
  {2010})}\BibitemShut {NoStop}%
\bibitem [{\citenamefont {Ik{\"{a}}l{\"{a}}inen}, \citenamefont {Lantto},\ and\
  \citenamefont {Vaara}(2012)}]{Ikaelaeinen2012_JCTC_91}%
  \BibitemOpen
  \bibfield  {author} {\bibinfo {author} {\bibfnamefont {S.}~\bibnamefont
  {Ik{\"{a}}l{\"{a}}inen}}, \bibinfo {author} {\bibfnamefont {P.}~\bibnamefont
  {Lantto}},\ and\ \bibinfo {author} {\bibfnamefont {J.}~\bibnamefont
  {Vaara}},\ }\bibfield  {title} {\enquote {\bibinfo {title} {Fully
  relativistic calculations of {Faraday} and nuclear spin-induced optical
  rotation in xenon},}\ }\href {https://doi.org/10.1021/ct200636m} {\bibfield
  {journal} {\bibinfo  {journal} {J. Chem. Theory Comput.}\ }\textbf {\bibinfo
  {volume} {8}},\ \bibinfo {pages} {91--98} (\bibinfo {year}
  {2012})}\BibitemShut {NoStop}%
\bibitem [{\citenamefont {Lantto}\ \emph {et~al.}(2011)\citenamefont {Lantto},
  \citenamefont {Jackowski}, \citenamefont {Makulski}, \citenamefont
  {Olejniczak},\ and\ \citenamefont {Jaszu{\'{n}}ski}}]{Lantto2011_JPCA_23}%
  \BibitemOpen
  \bibfield  {author} {\bibinfo {author} {\bibfnamefont {P.}~\bibnamefont
  {Lantto}}, \bibinfo {author} {\bibfnamefont {K.}~\bibnamefont {Jackowski}},
  \bibinfo {author} {\bibfnamefont {W.}~\bibnamefont {Makulski}}, \bibinfo
  {author} {\bibfnamefont {M.}~\bibnamefont {Olejniczak}},\ and\ \bibinfo
  {author} {\bibfnamefont {M.}~\bibnamefont {Jaszu{\'{n}}ski}},\ }\bibfield
  {title} {\enquote {\bibinfo {title} {{NMR} shielding constants in \ce{PH3},
  absolute shielding scale, and the nuclear magnetic moment of $^{31}${P}},}\
  }\href {https://doi.org/10.1021/jp2052739} {\bibfield  {journal} {\bibinfo
  {journal} {J. Phys. Chem. A}\ }\textbf {\bibinfo {volume} {115}},\ \bibinfo
  {pages} {10617--23} (\bibinfo {year} {2011})}\BibitemShut {NoStop}%
\bibitem [{\citenamefont {Fu}\ and\ \citenamefont
  {Vaara}(2013)}]{Fu2013_JCP_204110}%
  \BibitemOpen
  \bibfield  {author} {\bibinfo {author} {\bibfnamefont {L.-J.}\ \bibnamefont
  {Fu}}\ and\ \bibinfo {author} {\bibfnamefont {J.}~\bibnamefont {Vaara}},\
  }\bibfield  {title} {\enquote {\bibinfo {title} {{Nuclear spin-induced
  Cotton-Mouton effect in molecules.}}}\ }\href
  {https://doi.org/10.1063/1.4807396} {\bibfield  {journal} {\bibinfo
  {journal} {J. Chem. Phys.}\ }\textbf {\bibinfo {volume} {138}},\ \bibinfo
  {pages} {204110} (\bibinfo {year} {2013})}\BibitemShut {NoStop}%
\bibitem [{\citenamefont {Vaara}, \citenamefont {Hanni},\ and\ \citenamefont
  {Jokisaari}(2013)}]{Vaara2013_JCP_104313}%
  \BibitemOpen
  \bibfield  {author} {\bibinfo {author} {\bibfnamefont {J.}~\bibnamefont
  {Vaara}}, \bibinfo {author} {\bibfnamefont {M.}~\bibnamefont {Hanni}},\ and\
  \bibinfo {author} {\bibfnamefont {J.}~\bibnamefont {Jokisaari}},\ }\bibfield
  {title} {\enquote {\bibinfo {title} {{Nuclear spin-spin coupling in a van der
  Waals-bonded system: xenon dimer.}}}\ }\href
  {https://doi.org/10.1063/1.4793745} {\bibfield  {journal} {\bibinfo
  {journal} {J. Chem. Phys.}\ }\textbf {\bibinfo {volume} {138}},\ \bibinfo
  {pages} {104313} (\bibinfo {year} {2013})}\BibitemShut {NoStop}%
\bibitem [{\citenamefont {Abuzaid}, \citenamefont {Kantola},\ and\
  \citenamefont {Vaara}(2013)}]{Abuzaid2013_MP_1390}%
  \BibitemOpen
  \bibfield  {author} {\bibinfo {author} {\bibfnamefont {N.}~\bibnamefont
  {Abuzaid}}, \bibinfo {author} {\bibfnamefont {A.~M.}\ \bibnamefont
  {Kantola}},\ and\ \bibinfo {author} {\bibfnamefont {J.}~\bibnamefont
  {Vaara}},\ }\bibfield  {title} {\enquote {\bibinfo {title} {{Magnetic
  field-induced nuclear quadrupole coupling in atomic $^{131}$Xe}},}\ }\href
  {https://doi.org/10.1080/00268976.2013.793840} {\bibfield  {journal}
  {\bibinfo  {journal} {Mol. Phys.}\ }\textbf {\bibinfo {volume} {111}},\
  \bibinfo {pages} {1390--1400} (\bibinfo {year} {2013})}\BibitemShut {NoStop}%
\bibitem [{\citenamefont {V{\"{a}}h{\"{a}}kangas}\ \emph
  {et~al.}(2013)\citenamefont {V{\"{a}}h{\"{a}}kangas}, \citenamefont
  {Ik{\"{a}}l{\"{a}}inen}, \citenamefont {Lantto},\ and\ \citenamefont
  {Vaara}}]{Vaehaekangas2013_PCCP_41}%
  \BibitemOpen
  \bibfield  {author} {\bibinfo {author} {\bibfnamefont {J.}~\bibnamefont
  {V{\"{a}}h{\"{a}}kangas}}, \bibinfo {author} {\bibfnamefont {S.}~\bibnamefont
  {Ik{\"{a}}l{\"{a}}inen}}, \bibinfo {author} {\bibfnamefont {P.}~\bibnamefont
  {Lantto}},\ and\ \bibinfo {author} {\bibfnamefont {J.}~\bibnamefont
  {Vaara}},\ }\bibfield  {title} {\enquote {\bibinfo {title} {Nuclear magnetic
  resonance predictions for graphenes: concentric finite models and
  extrapolation to large systems},}\ }\href
  {https://doi.org/10.1039/c3cp44631j} {\bibfield  {journal} {\bibinfo
  {journal} {Phys. Chem. Chem. Phys.}\ }\textbf {\bibinfo {volume} {15}},\
  \bibinfo {pages} {4634--41} (\bibinfo {year} {2013})}\BibitemShut {NoStop}%
\bibitem [{\citenamefont {V{\"{a}}h{\"{a}}kangas}, \citenamefont {Lantto},\
  and\ \citenamefont {Vaara}(2014)}]{Vaehaekangas2014_JPCC_23996}%
  \BibitemOpen
  \bibfield  {author} {\bibinfo {author} {\bibfnamefont {J.}~\bibnamefont
  {V{\"{a}}h{\"{a}}kangas}}, \bibinfo {author} {\bibfnamefont {P.}~\bibnamefont
  {Lantto}},\ and\ \bibinfo {author} {\bibfnamefont {J.}~\bibnamefont
  {Vaara}},\ }\bibfield  {title} {\enquote {\bibinfo {title} {{Faraday Rotation
  in Graphene Quantum Dots: Interplay of Size, Perimeter Type, and
  Functionalization}},}\ }\href {https://doi.org/10.1021/jp507892j} {\bibfield
  {journal} {\bibinfo  {journal} {J. Phys. Chem. C}\ }\textbf {\bibinfo
  {volume} {118}},\ \bibinfo {pages} {23996--24005} (\bibinfo {year}
  {2014})}\BibitemShut {NoStop}%
\bibitem [{\citenamefont {Rossi}\ \emph {et~al.}(2015)\citenamefont {Rossi},
  \citenamefont {Lehtola}, \citenamefont {Sakko}, \citenamefont {Puska},\ and\
  \citenamefont {Nieminen}}]{Rossi2015_JCP_94114}%
  \BibitemOpen
  \bibfield  {author} {\bibinfo {author} {\bibfnamefont {T.~P.}\ \bibnamefont
  {Rossi}}, \bibinfo {author} {\bibfnamefont {S.}~\bibnamefont {Lehtola}},
  \bibinfo {author} {\bibfnamefont {A.}~\bibnamefont {Sakko}}, \bibinfo
  {author} {\bibfnamefont {M.~J.}\ \bibnamefont {Puska}},\ and\ \bibinfo
  {author} {\bibfnamefont {R.~M.}\ \bibnamefont {Nieminen}},\ }\bibfield
  {title} {\enquote {\bibinfo {title} {Nanoplasmonics simulations at the basis
  set limit through completeness-optimized, local numerical basis sets},}\
  }\href {https://doi.org/10.1063/1.4913739} {\bibfield  {journal} {\bibinfo
  {journal} {J. Chem. Phys.}\ }\textbf {\bibinfo {volume} {142}},\ \bibinfo
  {pages} {094114} (\bibinfo {year} {2015})}\BibitemShut {NoStop}%
\bibitem [{\citenamefont {Hanni}\ \emph {et~al.}(2017)\citenamefont {Hanni},
  \citenamefont {Lantto}, \citenamefont {Repisk{\'{y}}}, \citenamefont
  {Mare{\v{s}}}, \citenamefont {Saam},\ and\ \citenamefont
  {Vaara}}]{Hanni2017_PRA_32509}%
  \BibitemOpen
  \bibfield  {author} {\bibinfo {author} {\bibfnamefont {M.}~\bibnamefont
  {Hanni}}, \bibinfo {author} {\bibfnamefont {P.}~\bibnamefont {Lantto}},
  \bibinfo {author} {\bibfnamefont {M.}~\bibnamefont {Repisk{\'{y}}}}, \bibinfo
  {author} {\bibfnamefont {J.}~\bibnamefont {Mare{\v{s}}}}, \bibinfo {author}
  {\bibfnamefont {B.}~\bibnamefont {Saam}},\ and\ \bibinfo {author}
  {\bibfnamefont {J.}~\bibnamefont {Vaara}},\ }\bibfield  {title} {\enquote
  {\bibinfo {title} {{Electron and nuclear spin polarization in Rb-Xe
  spin-exchange optical hyperpolarization}},}\ }\href
  {https://doi.org/10.1103/PhysRevA.95.032509} {\bibfield  {journal} {\bibinfo
  {journal} {Phys. Rev. A}\ }\textbf {\bibinfo {volume} {95}},\ \bibinfo
  {pages} {032509} (\bibinfo {year} {2017})}\BibitemShut {NoStop}%
\bibitem [{\citenamefont {Auer}, \citenamefont {Helgaker},\ and\ \citenamefont
  {Klopper}(2002)}]{Auer2002_JCC_5}%
  \BibitemOpen
  \bibfield  {author} {\bibinfo {author} {\bibfnamefont {A.~A.}\ \bibnamefont
  {Auer}}, \bibinfo {author} {\bibfnamefont {T.}~\bibnamefont {Helgaker}},\
  and\ \bibinfo {author} {\bibfnamefont {W.}~\bibnamefont {Klopper}},\
  }\bibfield  {title} {\enquote {\bibinfo {title} {{Software news and updates.
  Basis-set completeness profiles in two dimensions.}}}\ }\href
  {https://doi.org/10.1002/jcc.1169} {\bibfield  {journal} {\bibinfo  {journal}
  {J. Comput. Chem.}\ }\textbf {\bibinfo {volume} {23}},\ \bibinfo {pages}
  {420--5} (\bibinfo {year} {2002})}\BibitemShut {NoStop}%
\bibitem [{\citenamefont {Lu}\ \emph {et~al.}(2004{\natexlab{a}})\citenamefont
  {Lu}, \citenamefont {Wang}, \citenamefont {Schmidt}, \citenamefont
  {Bytautas}, \citenamefont {Ho},\ and\ \citenamefont
  {Ruedenberg}}]{Lu2004_JCP_37}%
  \BibitemOpen
  \bibfield  {author} {\bibinfo {author} {\bibfnamefont {W.~C.}\ \bibnamefont
  {Lu}}, \bibinfo {author} {\bibfnamefont {C.~Z.}\ \bibnamefont {Wang}},
  \bibinfo {author} {\bibfnamefont {M.~W.}\ \bibnamefont {Schmidt}}, \bibinfo
  {author} {\bibfnamefont {L.}~\bibnamefont {Bytautas}}, \bibinfo {author}
  {\bibfnamefont {K.~M.}\ \bibnamefont {Ho}},\ and\ \bibinfo {author}
  {\bibfnamefont {K.}~\bibnamefont {Ruedenberg}},\ }\bibfield  {title}
  {\enquote {\bibinfo {title} {{Molecule intrinsic minimal basis sets. I. Exact
  resolution of ab initio optimized molecular orbitals in terms of deformed
  atomic minimal-basis orbitals.}}}\ }\href {https://doi.org/10.1063/1.1638731}
  {\bibfield  {journal} {\bibinfo  {journal} {J. Chem. Phys.}\ }\textbf
  {\bibinfo {volume} {120}},\ \bibinfo {pages} {2629--37} (\bibinfo {year}
  {2004}{\natexlab{a}})}\BibitemShut {NoStop}%
\bibitem [{\citenamefont {Lu}\ \emph {et~al.}(2004{\natexlab{b}})\citenamefont
  {Lu}, \citenamefont {Wang}, \citenamefont {Schmidt}, \citenamefont
  {Bytautas}, \citenamefont {Ho},\ and\ \citenamefont
  {Ruedenberg}}]{Lu2004_JCP_2638}%
  \BibitemOpen
  \bibfield  {author} {\bibinfo {author} {\bibfnamefont {W.~C.}\ \bibnamefont
  {Lu}}, \bibinfo {author} {\bibfnamefont {C.~Z.}\ \bibnamefont {Wang}},
  \bibinfo {author} {\bibfnamefont {M.~W.}\ \bibnamefont {Schmidt}}, \bibinfo
  {author} {\bibfnamefont {L.}~\bibnamefont {Bytautas}}, \bibinfo {author}
  {\bibfnamefont {K.~M.}\ \bibnamefont {Ho}},\ and\ \bibinfo {author}
  {\bibfnamefont {K.}~\bibnamefont {Ruedenberg}},\ }\bibfield  {title}
  {\enquote {\bibinfo {title} {Molecule intrinsic minimal basis sets. {II}.
  bonding analyses for \ce{Si4H6} and \ce{Si2} to \ce{Si10}},}\ }\href
  {https://doi.org/10.1063/1.1638732} {\bibfield  {journal} {\bibinfo
  {journal} {J. Chem. Phys.}\ }\textbf {\bibinfo {volume} {120}},\ \bibinfo
  {pages} {2638--2651} (\bibinfo {year} {2004}{\natexlab{b}})}\BibitemShut
  {NoStop}%
\bibitem [{\citenamefont {Lu}\ \emph {et~al.}(2004{\natexlab{c}})\citenamefont
  {Lu}, \citenamefont {Wang}, \citenamefont {Chan}, \citenamefont
  {Ruedenberg},\ and\ \citenamefont {Ho}}]{Lu2004_PRB_41101}%
  \BibitemOpen
  \bibfield  {author} {\bibinfo {author} {\bibfnamefont {W.~C.}\ \bibnamefont
  {Lu}}, \bibinfo {author} {\bibfnamefont {C.~Z.}\ \bibnamefont {Wang}},
  \bibinfo {author} {\bibfnamefont {T.~L.}\ \bibnamefont {Chan}}, \bibinfo
  {author} {\bibfnamefont {K.}~\bibnamefont {Ruedenberg}},\ and\ \bibinfo
  {author} {\bibfnamefont {K.~M.}\ \bibnamefont {Ho}},\ }\bibfield  {title}
  {\enquote {\bibinfo {title} {{Representation of electronic structures in
  crystals in terms of highly localized quasiatomic minimal basis orbitals}},}\
  }\href {https://doi.org/10.1103/PhysRevB.70.041101} {\bibfield  {journal}
  {\bibinfo  {journal} {Phys. Rev. B}\ }\textbf {\bibinfo {volume} {70}},\
  \bibinfo {pages} {041101} (\bibinfo {year} {2004}{\natexlab{c}})}\BibitemShut
  {NoStop}%
\bibitem [{\citenamefont {Knizia}(2013)}]{Knizia2013_JCTC_4834}%
  \BibitemOpen
  \bibfield  {author} {\bibinfo {author} {\bibfnamefont {G.}~\bibnamefont
  {Knizia}},\ }\bibfield  {title} {\enquote {\bibinfo {title} {{Intrinsic
  Atomic Orbitals: An Unbiased Bridge between Quantum Theory and Chemical
  Concepts}},}\ }\href {https://doi.org/10.1021/ct400687b} {\bibfield
  {journal} {\bibinfo  {journal} {J. Chem. Theory Comput.}\ }\textbf {\bibinfo
  {volume} {9}},\ \bibinfo {pages} {4834--4843} (\bibinfo {year}
  {2013})}\BibitemShut {NoStop}%
\bibitem [{\citenamefont {Knizia}\ and\ \citenamefont
  {Klein}(2015)}]{Knizia2015_ACIE_5518}%
  \BibitemOpen
  \bibfield  {author} {\bibinfo {author} {\bibfnamefont {G.}~\bibnamefont
  {Knizia}}\ and\ \bibinfo {author} {\bibfnamefont {J.~E. M.~N.}\ \bibnamefont
  {Klein}},\ }\bibfield  {title} {\enquote {\bibinfo {title} {{Electron Flow in
  Reaction Mechanisms-Revealed from First Principles}},}\ }\href
  {https://doi.org/10.1002/anie.201410637} {\bibfield  {journal} {\bibinfo
  {journal} {Angew. Chemie Int. Ed.}\ }\textbf {\bibinfo {volume} {54}},\
  \bibinfo {pages} {5518--5522} (\bibinfo {year} {2015})}\BibitemShut {NoStop}%
\bibitem [{\citenamefont {Janowski}(2014)}]{Janowski2014_JCTC_3085}%
  \BibitemOpen
  \bibfield  {author} {\bibinfo {author} {\bibfnamefont {T.}~\bibnamefont
  {Janowski}},\ }\bibfield  {title} {\enquote {\bibinfo {title} {{Near
  Equivalence of Intrinsic Atomic Orbitals and Quasiatomic Orbitals}},}\ }\href
  {https://doi.org/10.1021/ct500245f} {\bibfield  {journal} {\bibinfo
  {journal} {J. Chem. Theory Comput.}\ }\textbf {\bibinfo {volume} {10}},\
  \bibinfo {pages} {3085--3091} (\bibinfo {year} {2014})}\BibitemShut {NoStop}%
\bibitem [{\citenamefont {Clement}, \citenamefont {Wang},\ and\ \citenamefont
  {Valeev}(2021)}]{Clement2021_JCTC_7406}%
  \BibitemOpen
  \bibfield  {author} {\bibinfo {author} {\bibfnamefont {M.~C.}\ \bibnamefont
  {Clement}}, \bibinfo {author} {\bibfnamefont {X.}~\bibnamefont {Wang}},\ and\
  \bibinfo {author} {\bibfnamefont {E.~F.}\ \bibnamefont {Valeev}},\ }\bibfield
   {title} {\enquote {\bibinfo {title} {Robust {Pipek}--{Mezey} orbital
  localization in periodic solids},}\ }\href
  {https://doi.org/10.1021/acs.jctc.1c00238} {\bibfield  {journal} {\bibinfo
  {journal} {J. Chem. Theory Comput.}\ }\textbf {\bibinfo {volume} {17}},\
  \bibinfo {pages} {7406--7415} (\bibinfo {year} {2021})}\BibitemShut {NoStop}%
\bibitem [{\citenamefont
  {Lehtola}(2019{\natexlab{b}})}]{Lehtola2019_IJQC_25945}%
  \BibitemOpen
  \bibfield  {author} {\bibinfo {author} {\bibfnamefont {S.}~\bibnamefont
  {Lehtola}},\ }\bibfield  {title} {\enquote {\bibinfo {title} {Fully numerical
  {Hartree}--{Fock} and density functional calculations. {I}. {Atoms}},}\
  }\href {https://doi.org/10.1002/qua.25945} {\bibfield  {journal} {\bibinfo
  {journal} {Int. J. Quantum Chem.}\ }\textbf {\bibinfo {volume} {119}},\
  \bibinfo {pages} {e25945} (\bibinfo {year} {2019}{\natexlab{b}})},\ \Eprint
  {https://arxiv.org/abs/1810.11651} {arXiv:1810.11651} \BibitemShut {NoStop}%
\bibitem [{\citenamefont
  {Lehtola}(2020{\natexlab{b}})}]{Lehtola2020_PRA_12516}%
  \BibitemOpen
  \bibfield  {author} {\bibinfo {author} {\bibfnamefont {S.}~\bibnamefont
  {Lehtola}},\ }\bibfield  {title} {\enquote {\bibinfo {title} {Fully numerical
  calculations on atoms with fractional occupations and range-separated
  exchange functionals},}\ }\href {https://doi.org/10.1103/PhysRevA.101.012516}
  {\bibfield  {journal} {\bibinfo  {journal} {Phys. Rev. A}\ }\textbf {\bibinfo
  {volume} {101}},\ \bibinfo {pages} {012516} (\bibinfo {year}
  {2020}{\natexlab{b}})},\ \Eprint {https://arxiv.org/abs/1908.02528}
  {arXiv:1908.02528} \BibitemShut {NoStop}%
\bibitem [{\citenamefont
  {Lehtola}(2023{\natexlab{b}})}]{Lehtola2023_JCTC_2502}%
  \BibitemOpen
  \bibfield  {author} {\bibinfo {author} {\bibfnamefont {S.}~\bibnamefont
  {Lehtola}},\ }\bibfield  {title} {\enquote {\bibinfo {title} {Meta-{GGA}
  density functional calculations on atoms with spherically symmetric densities
  in the finite element formalism},}\ }\href
  {https://doi.org/10.1021/acs.jctc.3c00183} {\bibfield  {journal} {\bibinfo
  {journal} {J. Chem. Theory Comput.}\ }\textbf {\bibinfo {volume} {19}},\
  \bibinfo {pages} {2502--2517} (\bibinfo {year} {2023}{\natexlab{b}})},\
  \Eprint {https://arxiv.org/abs/2302.06284} {2302.06284} \BibitemShut
  {NoStop}%
\bibitem [{\citenamefont
  {Lehtola}(2023{\natexlab{c}})}]{Lehtola2023_JPCA_4180}%
  \BibitemOpen
  \bibfield  {author} {\bibinfo {author} {\bibfnamefont {S.}~\bibnamefont
  {Lehtola}},\ }\bibfield  {title} {\enquote {\bibinfo {title} {Atomic
  electronic structure calculations with {Hermite} interpolating
  polynomials},}\ }\href {https://doi.org/10.1021/acs.jpca.3c00729} {\bibfield
  {journal} {\bibinfo  {journal} {J. Phys. Chem. A}\ }\textbf {\bibinfo
  {volume} {127}},\ \bibinfo {pages} {4180--4193} (\bibinfo {year}
  {2023}{\natexlab{c}})},\ \Eprint {https://arxiv.org/abs/2302.00440}
  {2302.00440} \BibitemShut {NoStop}%
\bibitem [{\citenamefont
  {Lehtola}(2019{\natexlab{c}})}]{Lehtola2019_IJQC_25944}%
  \BibitemOpen
  \bibfield  {author} {\bibinfo {author} {\bibfnamefont {S.}~\bibnamefont
  {Lehtola}},\ }\bibfield  {title} {\enquote {\bibinfo {title} {Fully numerical
  {Hartree}--{Fock} and density functional calculations. {II}. {Diatomic}
  molecules},}\ }\href {https://doi.org/10.1002/qua.25944} {\bibfield
  {journal} {\bibinfo  {journal} {Int. J. Quantum Chem.}\ }\textbf {\bibinfo
  {volume} {119}},\ \bibinfo {pages} {e25944} (\bibinfo {year}
  {2019}{\natexlab{c}})},\ \Eprint {https://arxiv.org/abs/1810.11653}
  {arXiv:1810.11653} \BibitemShut {NoStop}%
\bibitem [{\citenamefont {Lehtola}(2023{\natexlab{d}})}]{Lehtola2018__}%
  \BibitemOpen
  \bibfield  {author} {\bibinfo {author} {\bibfnamefont {S.}~\bibnamefont
  {Lehtola}},\ }\href {http://github.com/susilehtola/HelFEM} {\enquote
  {\bibinfo {title} {{HelFEM -- Finite element methods for electronic structure
  calculations on small systems}},}\ } (\bibinfo {year} {2023}{\natexlab{d}}),\
  \bibinfo {note} {accessed 26 March 2023.}\BibitemShut {Stop}%
\bibitem [{\citenamefont {Sankey}\ and\ \citenamefont
  {Niklewski}(1989)}]{Sankey1989_PRB_3979}%
  \BibitemOpen
  \bibfield  {author} {\bibinfo {author} {\bibfnamefont {O.}~\bibnamefont
  {Sankey}}\ and\ \bibinfo {author} {\bibfnamefont {D.}~\bibnamefont
  {Niklewski}},\ }\bibfield  {title} {\enquote {\bibinfo {title} {{Ab initio
  multicenter tight-binding model for molecular-dynamics simulations and other
  applications in covalent systems}},}\ }\href
  {https://doi.org/10.1103/PhysRevB.40.3979} {\bibfield  {journal} {\bibinfo
  {journal} {Phys. Rev. B}\ }\textbf {\bibinfo {volume} {40}},\ \bibinfo
  {pages} {3979--3995} (\bibinfo {year} {1989})}\BibitemShut {NoStop}%
\bibitem [{\citenamefont {Porezag}\ \emph {et~al.}(1995)\citenamefont
  {Porezag}, \citenamefont {Frauenheim}, \citenamefont {K{\"{o}}hler},
  \citenamefont {Seifert},\ and\ \citenamefont
  {Kaschner}}]{Porezag1995_PRB_12947}%
  \BibitemOpen
  \bibfield  {author} {\bibinfo {author} {\bibfnamefont {D.}~\bibnamefont
  {Porezag}}, \bibinfo {author} {\bibfnamefont {T.}~\bibnamefont {Frauenheim}},
  \bibinfo {author} {\bibfnamefont {T.}~\bibnamefont {K{\"{o}}hler}}, \bibinfo
  {author} {\bibfnamefont {G.}~\bibnamefont {Seifert}},\ and\ \bibinfo {author}
  {\bibfnamefont {R.}~\bibnamefont {Kaschner}},\ }\bibfield  {title} {\enquote
  {\bibinfo {title} {Construction of tight-binding-like potentials on the basis
  of density-functional theory: Application to carbon},}\ }\href
  {https://doi.org/10.1103/PhysRevB.51.12947} {\bibfield  {journal} {\bibinfo
  {journal} {Phys. Rev. B}\ }\textbf {\bibinfo {volume} {51}},\ \bibinfo
  {pages} {12947--12957} (\bibinfo {year} {1995})}\BibitemShut {NoStop}%
\bibitem [{\citenamefont {Horsfield}(1997)}]{Horsfield1997_PRB_6594}%
  \BibitemOpen
  \bibfield  {author} {\bibinfo {author} {\bibfnamefont {A.~P.}\ \bibnamefont
  {Horsfield}},\ }\bibfield  {title} {\enquote {\bibinfo {title} {{Efficient ab
  initio tight binding}},}\ }\href {https://doi.org/10.1103/PhysRevB.56.6594}
  {\bibfield  {journal} {\bibinfo  {journal} {Phys. Rev. B}\ }\textbf {\bibinfo
  {volume} {56}},\ \bibinfo {pages} {6594--6602} (\bibinfo {year}
  {1997})}\BibitemShut {NoStop}%
\bibitem [{\citenamefont {S{\'{a}}nchez-Portal}\ \emph
  {et~al.}(1997)\citenamefont {S{\'{a}}nchez-Portal}, \citenamefont
  {Ordej{\'{o}}n}, \citenamefont {Artacho},\ and\ \citenamefont
  {Soler}}]{SanchezPortal1997_IJQC_453}%
  \BibitemOpen
  \bibfield  {author} {\bibinfo {author} {\bibfnamefont {D.}~\bibnamefont
  {S{\'{a}}nchez-Portal}}, \bibinfo {author} {\bibfnamefont {P.}~\bibnamefont
  {Ordej{\'{o}}n}}, \bibinfo {author} {\bibfnamefont {E.}~\bibnamefont
  {Artacho}},\ and\ \bibinfo {author} {\bibfnamefont {J.~M.}\ \bibnamefont
  {Soler}},\ }\bibfield  {title} {\enquote {\bibinfo {title}
  {{Density-functional method for very large systems with LCAO basis sets}},}\
  }\href
  {https://doi.org/10.1002/(SICI)1097-461X(1997)65:5<453::AID-QUA9>3.0.CO;2-V}
  {\bibfield  {journal} {\bibinfo  {journal} {Int. J. Quantum Chem.}\ }\textbf
  {\bibinfo {volume} {65}},\ \bibinfo {pages} {453--461} (\bibinfo {year}
  {1997})}\BibitemShut {NoStop}%
\bibitem [{\citenamefont {Kenny}, \citenamefont {Horsfield},\ and\
  \citenamefont {Fujitani}(2000)}]{Kenny2000_PRB_4899}%
  \BibitemOpen
  \bibfield  {author} {\bibinfo {author} {\bibfnamefont {S.}~\bibnamefont
  {Kenny}}, \bibinfo {author} {\bibfnamefont {A.}~\bibnamefont {Horsfield}},\
  and\ \bibinfo {author} {\bibfnamefont {H.}~\bibnamefont {Fujitani}},\
  }\bibfield  {title} {\enquote {\bibinfo {title} {{Transferable atomic-type
  orbital basis sets for solids}},}\ }\href
  {https://doi.org/10.1103/PhysRevB.62.4899} {\bibfield  {journal} {\bibinfo
  {journal} {Phys. Rev. B}\ }\textbf {\bibinfo {volume} {62}},\ \bibinfo
  {pages} {4899--4905} (\bibinfo {year} {2000})}\BibitemShut {NoStop}%
\bibitem [{\citenamefont {Junquera}\ \emph {et~al.}(2001)\citenamefont
  {Junquera}, \citenamefont {Paz}, \citenamefont {S{\'{a}}nchez-Portal},\ and\
  \citenamefont {Artacho}}]{Junquera2001_PRB_235111}%
  \BibitemOpen
  \bibfield  {author} {\bibinfo {author} {\bibfnamefont {J.}~\bibnamefont
  {Junquera}}, \bibinfo {author} {\bibfnamefont {{\'{O}}.}~\bibnamefont {Paz}},
  \bibinfo {author} {\bibfnamefont {D.}~\bibnamefont {S{\'{a}}nchez-Portal}},\
  and\ \bibinfo {author} {\bibfnamefont {E.}~\bibnamefont {Artacho}},\
  }\bibfield  {title} {\enquote {\bibinfo {title} {{Numerical atomic orbitals
  for linear-scaling calculations}},}\ }\href
  {https://doi.org/10.1103/PhysRevB.64.235111} {\bibfield  {journal} {\bibinfo
  {journal} {Phys. Rev. B}\ }\textbf {\bibinfo {volume} {64}},\ \bibinfo
  {pages} {235111} (\bibinfo {year} {2001})}\BibitemShut {NoStop}%
\bibitem [{\citenamefont {Anglada}\ \emph {et~al.}(2002)\citenamefont
  {Anglada}, \citenamefont {{M. Soler}}, \citenamefont {Junquera},\ and\
  \citenamefont {Artacho}}]{Anglada2002_PRB_205101}%
  \BibitemOpen
  \bibfield  {author} {\bibinfo {author} {\bibfnamefont {E.}~\bibnamefont
  {Anglada}}, \bibinfo {author} {\bibfnamefont {J.}~\bibnamefont {{M. Soler}}},
  \bibinfo {author} {\bibfnamefont {J.}~\bibnamefont {Junquera}},\ and\
  \bibinfo {author} {\bibfnamefont {E.}~\bibnamefont {Artacho}},\ }\bibfield
  {title} {\enquote {\bibinfo {title} {{Systematic generation of finite-range
  atomic basis sets for linear-scaling calculations}},}\ }\href
  {https://doi.org/10.1103/PhysRevB.66.205101} {\bibfield  {journal} {\bibinfo
  {journal} {Phys. Rev. B}\ }\textbf {\bibinfo {volume} {66}},\ \bibinfo
  {pages} {205101} (\bibinfo {year} {2002})}\BibitemShut {NoStop}%
\bibitem [{\citenamefont {Ozaki}(2003)}]{Ozaki2003_PRB_155108}%
  \BibitemOpen
  \bibfield  {author} {\bibinfo {author} {\bibfnamefont {T.}~\bibnamefont
  {Ozaki}},\ }\bibfield  {title} {\enquote {\bibinfo {title} {{Variationally
  optimized atomic orbitals for large-scale electronic structures}},}\ }\href
  {https://doi.org/10.1103/PhysRevB.67.155108} {\bibfield  {journal} {\bibinfo
  {journal} {Phys. Rev. B}\ }\textbf {\bibinfo {volume} {67}},\ \bibinfo
  {pages} {155108} (\bibinfo {year} {2003})}\BibitemShut {NoStop}%
\bibitem [{\citenamefont {Ozaki}\ and\ \citenamefont
  {Kino}(2004{\natexlab{a}})}]{Ozaki2004_JCP_10879}%
  \BibitemOpen
  \bibfield  {author} {\bibinfo {author} {\bibfnamefont {T.}~\bibnamefont
  {Ozaki}}\ and\ \bibinfo {author} {\bibfnamefont {H.}~\bibnamefont {Kino}},\
  }\bibfield  {title} {\enquote {\bibinfo {title} {{Variationally optimized
  basis orbitals for biological molecules}},}\ }\href
  {https://doi.org/10.1063/1.1794591} {\bibfield  {journal} {\bibinfo
  {journal} {J. Chem. Phys.}\ }\textbf {\bibinfo {volume} {121}},\ \bibinfo
  {pages} {10879} (\bibinfo {year} {2004}{\natexlab{a}})}\BibitemShut {NoStop}%
\bibitem [{\citenamefont {Ozaki}\ and\ \citenamefont
  {Kino}(2004{\natexlab{b}})}]{Ozaki2004_PRB_195113}%
  \BibitemOpen
  \bibfield  {author} {\bibinfo {author} {\bibfnamefont {T.}~\bibnamefont
  {Ozaki}}\ and\ \bibinfo {author} {\bibfnamefont {H.}~\bibnamefont {Kino}},\
  }\bibfield  {title} {\enquote {\bibinfo {title} {{Numerical atomic basis
  orbitals from H to Kr}},}\ }\href
  {https://doi.org/10.1103/PhysRevB.69.195113} {\bibfield  {journal} {\bibinfo
  {journal} {Phys. Rev. B}\ }\textbf {\bibinfo {volume} {69}},\ \bibinfo
  {pages} {195113} (\bibinfo {year} {2004}{\natexlab{b}})}\BibitemShut
  {NoStop}%
\bibitem [{\citenamefont {Shang}\ \emph {et~al.}(2010)\citenamefont {Shang},
  \citenamefont {Xiang}, \citenamefont {Li},\ and\ \citenamefont
  {Yang}}]{Shang2010_IRPC_665}%
  \BibitemOpen
  \bibfield  {author} {\bibinfo {author} {\bibfnamefont {H.}~\bibnamefont
  {Shang}}, \bibinfo {author} {\bibfnamefont {H.}~\bibnamefont {Xiang}},
  \bibinfo {author} {\bibfnamefont {Z.}~\bibnamefont {Li}},\ and\ \bibinfo
  {author} {\bibfnamefont {J.}~\bibnamefont {Yang}},\ }\bibfield  {title}
  {\enquote {\bibinfo {title} {Linear scaling electronic structure calculations
  with numerical atomic basis set},}\ }\href
  {https://doi.org/10.1080/0144235X.2010.520454} {\bibfield  {journal}
  {\bibinfo  {journal} {Int. Rev. Phys. Chem.}\ }\textbf {\bibinfo {volume}
  {29}},\ \bibinfo {pages} {665--691} (\bibinfo {year} {2010})}\BibitemShut
  {NoStop}%
\bibitem [{\citenamefont {Louwerse}\ and\ \citenamefont
  {Rothenberg}(2012)}]{Louwerse2012_PRB_35108}%
  \BibitemOpen
  \bibfield  {author} {\bibinfo {author} {\bibfnamefont {M.~J.}\ \bibnamefont
  {Louwerse}}\ and\ \bibinfo {author} {\bibfnamefont {G.}~\bibnamefont
  {Rothenberg}},\ }\bibfield  {title} {\enquote {\bibinfo {title}
  {{Transferable basis sets of numerical atomic orbitals}},}\ }\href
  {https://doi.org/10.1103/PhysRevB.85.035108} {\bibfield  {journal} {\bibinfo
  {journal} {Phys. Rev. B}\ }\textbf {\bibinfo {volume} {85}},\ \bibinfo
  {pages} {035108} (\bibinfo {year} {2012})}\BibitemShut {NoStop}%
\bibitem [{\citenamefont {Corsetti}\ \emph {et~al.}(2013)\citenamefont
  {Corsetti}, \citenamefont {Fern{\'{a}}ndez-Serra}, \citenamefont {Soler},\
  and\ \citenamefont {Artacho}}]{Corsetti2013_JPCM_435504}%
  \BibitemOpen
  \bibfield  {author} {\bibinfo {author} {\bibfnamefont {F.}~\bibnamefont
  {Corsetti}}, \bibinfo {author} {\bibfnamefont {M.-V.}\ \bibnamefont
  {Fern{\'{a}}ndez-Serra}}, \bibinfo {author} {\bibfnamefont {J.~M.}\
  \bibnamefont {Soler}},\ and\ \bibinfo {author} {\bibfnamefont
  {E.}~\bibnamefont {Artacho}},\ }\bibfield  {title} {\enquote {\bibinfo
  {title} {{Optimal finite-range atomic basis sets for liquid water and
  ice}},}\ }\href {https://doi.org/10.1088/0953-8984/25/43/435504} {\bibfield
  {journal} {\bibinfo  {journal} {J. Phys. Condens. Matter}\ }\textbf {\bibinfo
  {volume} {25}},\ \bibinfo {pages} {435504} (\bibinfo {year}
  {2013})}\BibitemShut {NoStop}%
\bibitem [{\citenamefont
  {Lehtola}(2019{\natexlab{d}})}]{Lehtola2019_JCTC_1593}%
  \BibitemOpen
  \bibfield  {author} {\bibinfo {author} {\bibfnamefont {S.}~\bibnamefont
  {Lehtola}},\ }\bibfield  {title} {\enquote {\bibinfo {title} {Assessment of
  initial guesses for self-consistent field calculations. superposition of
  atomic potentials: Simple yet efficient},}\ }\href
  {https://doi.org/10.1021/acs.jctc.8b01089} {\bibfield  {journal} {\bibinfo
  {journal} {J. Chem. Theory Comput.}\ }\textbf {\bibinfo {volume} {15}},\
  \bibinfo {pages} {1593--1604} (\bibinfo {year} {2019}{\natexlab{d}})},\
  \Eprint {https://arxiv.org/abs/1810.11659} {arXiv:1810.11659} \BibitemShut
  {NoStop}%
\bibitem [{\citenamefont {Yasui}\ and\ \citenamefont
  {Saika}(1982)}]{Yasui1982_JCP_468}%
  \BibitemOpen
  \bibfield  {author} {\bibinfo {author} {\bibfnamefont {J.}~\bibnamefont
  {Yasui}}\ and\ \bibinfo {author} {\bibfnamefont {A.}~\bibnamefont {Saika}},\
  }\bibfield  {title} {\enquote {\bibinfo {title} {Unified analytical
  evaluation of two-center, two-electron integrals over {Slater}-type
  orbitals},}\ }\href {https://doi.org/10.1063/1.442745} {\bibfield  {journal}
  {\bibinfo  {journal} {J. Chem. Phys.}\ }\textbf {\bibinfo {volume} {76}},\
  \bibinfo {pages} {468--472} (\bibinfo {year} {1982})}\BibitemShut {NoStop}%
\bibitem [{\citenamefont {Weigend}\ and\ \citenamefont
  {Ahlrichs}(2005)}]{Weigend2005_PCCP_305}%
  \BibitemOpen
  \bibfield  {author} {\bibinfo {author} {\bibfnamefont {F.}~\bibnamefont
  {Weigend}}\ and\ \bibinfo {author} {\bibfnamefont {R.}~\bibnamefont
  {Ahlrichs}},\ }\bibfield  {title} {\enquote {\bibinfo {title} {{Balanced
  basis sets of split valence, triple zeta valence and quadruple zeta valence
  quality for H to Rn: Design and assessment of accuracy.}}}\ }\href
  {https://doi.org/10.1039/b508541a} {\bibfield  {journal} {\bibinfo  {journal}
  {Phys. Chem. Chem. Phys.}\ }\textbf {\bibinfo {volume} {7}},\ \bibinfo
  {pages} {3297--305} (\bibinfo {year} {2005})}\BibitemShut {NoStop}%
\bibitem [{\citenamefont {Perdew}, \citenamefont {Burke},\ and\ \citenamefont
  {Ernzerhof}(1996)}]{Perdew1996_PRL_3865}%
  \BibitemOpen
  \bibfield  {author} {\bibinfo {author} {\bibfnamefont {J.~P.}\ \bibnamefont
  {Perdew}}, \bibinfo {author} {\bibfnamefont {K.}~\bibnamefont {Burke}},\ and\
  \bibinfo {author} {\bibfnamefont {M.}~\bibnamefont {Ernzerhof}},\ }\bibfield
  {title} {\enquote {\bibinfo {title} {Generalized gradient approximation made
  simple},}\ }\href {https://doi.org/10.1103/PhysRevLett.77.3865} {\bibfield
  {journal} {\bibinfo  {journal} {Phys. Rev. Lett.}\ }\textbf {\bibinfo
  {volume} {77}},\ \bibinfo {pages} {3865--3868} (\bibinfo {year}
  {1996})}\BibitemShut {NoStop}%
\bibitem [{\citenamefont {Perdew}, \citenamefont {Burke},\ and\ \citenamefont
  {Ernzerhof}(1997)}]{Perdew1997_PRL_1396}%
  \BibitemOpen
  \bibfield  {author} {\bibinfo {author} {\bibfnamefont {J.~P.}\ \bibnamefont
  {Perdew}}, \bibinfo {author} {\bibfnamefont {K.}~\bibnamefont {Burke}},\ and\
  \bibinfo {author} {\bibfnamefont {M.}~\bibnamefont {Ernzerhof}},\ }\bibfield
  {title} {\enquote {\bibinfo {title} {Generalized gradient approximation made
  simple [{Phys}. {Rev}. {Lett}. 77, 3865 (1996)]},}\ }\href
  {https://doi.org/10.1103/PhysRevLett.78.1396} {\bibfield  {journal} {\bibinfo
   {journal} {Phys. Rev. Lett.}\ }\textbf {\bibinfo {volume} {78}},\ \bibinfo
  {pages} {1396--1396} (\bibinfo {year} {1997})}\BibitemShut {NoStop}%
\bibitem [{\citenamefont {Lehtola}\ \emph {et~al.}(2018)\citenamefont
  {Lehtola}, \citenamefont {Steigemann}, \citenamefont {Oliveira},\ and\
  \citenamefont {Marques}}]{Lehtola2018_S_1}%
  \BibitemOpen
  \bibfield  {author} {\bibinfo {author} {\bibfnamefont {S.}~\bibnamefont
  {Lehtola}}, \bibinfo {author} {\bibfnamefont {C.}~\bibnamefont {Steigemann}},
  \bibinfo {author} {\bibfnamefont {M.~J.~T.}\ \bibnamefont {Oliveira}},\ and\
  \bibinfo {author} {\bibfnamefont {M.~A.~L.}\ \bibnamefont {Marques}},\
  }\bibfield  {title} {\enquote {\bibinfo {title} {Recent developments in
  {LIBXC}---a comprehensive library of functionals for density functional
  theory},}\ }\href {https://doi.org/10.1016/j.softx.2017.11.002} {\bibfield
  {journal} {\bibinfo  {journal} {SoftwareX}\ }\textbf {\bibinfo {volume}
  {7}},\ \bibinfo {pages} {1--5} (\bibinfo {year} {2018})}\BibitemShut
  {NoStop}%
\bibitem [{\citenamefont {Bloch}(1929)}]{Bloch1929_ZfuP_545}%
  \BibitemOpen
  \bibfield  {author} {\bibinfo {author} {\bibfnamefont {F.}~\bibnamefont
  {Bloch}},\ }\bibfield  {title} {\enquote {\bibinfo {title} {{Bemerkung zur
  Elektronentheorie des Ferromagnetismus und der elektrischen
  Leitf{\"{a}}higkeit}},}\ }\href {https://doi.org/10.1007/BF01340281}
  {\bibfield  {journal} {\bibinfo  {journal} {Z. Phys.}\ }\textbf {\bibinfo
  {volume} {57}},\ \bibinfo {pages} {545--555} (\bibinfo {year}
  {1929})}\BibitemShut {NoStop}%
\bibitem [{\citenamefont {Dirac}(1930)}]{Dirac1930_MPCPS_376}%
  \BibitemOpen
  \bibfield  {author} {\bibinfo {author} {\bibfnamefont {P.~A.~M.}\
  \bibnamefont {Dirac}},\ }\bibfield  {title} {\enquote {\bibinfo {title} {Note
  on exchange phenomena in the {Thomas} atom},}\ }\href
  {https://doi.org/10.1017/S0305004100016108} {\bibfield  {journal} {\bibinfo
  {journal} {Math. Proc. Cambridge Philos. Soc.}\ }\textbf {\bibinfo {volume}
  {26}},\ \bibinfo {pages} {376--385} (\bibinfo {year} {1930})}\BibitemShut
  {NoStop}%
\bibitem [{\citenamefont {Perdew}\ and\ \citenamefont
  {Wang}(1992)}]{Perdew1992_PRB_13244}%
  \BibitemOpen
  \bibfield  {author} {\bibinfo {author} {\bibfnamefont {J.~P.}\ \bibnamefont
  {Perdew}}\ and\ \bibinfo {author} {\bibfnamefont {Y.}~\bibnamefont {Wang}},\
  }\bibfield  {title} {\enquote {\bibinfo {title} {{Accurate and simple
  analytic representation of the electron-gas correlation energy}},}\ }\href
  {https://doi.org/10.1103/PhysRevB.45.13244} {\bibfield  {journal} {\bibinfo
  {journal} {Phys. Rev. B}\ }\textbf {\bibinfo {volume} {45}},\ \bibinfo
  {pages} {13244--13249} (\bibinfo {year} {1992})}\BibitemShut {NoStop}%
\bibitem [{\citenamefont {Tao}\ \emph {et~al.}(2003)\citenamefont {Tao},
  \citenamefont {Perdew}, \citenamefont {Staroverov},\ and\ \citenamefont
  {Scuseria}}]{Tao2003_PRL_146401}%
  \BibitemOpen
  \bibfield  {author} {\bibinfo {author} {\bibfnamefont {J.}~\bibnamefont
  {Tao}}, \bibinfo {author} {\bibfnamefont {J.~P.}\ \bibnamefont {Perdew}},
  \bibinfo {author} {\bibfnamefont {V.~N.}\ \bibnamefont {Staroverov}},\ and\
  \bibinfo {author} {\bibfnamefont {G.~E.}\ \bibnamefont {Scuseria}},\
  }\bibfield  {title} {\enquote {\bibinfo {title} {Climbing the density
  functional ladder: Nonempirical meta-generalized gradient approximation
  designed for molecules and solids},}\ }\href
  {https://doi.org/10.1103/PhysRevLett.91.146401} {\bibfield  {journal}
  {\bibinfo  {journal} {Phys. Rev. Lett.}\ }\textbf {\bibinfo {volume} {91}},\
  \bibinfo {pages} {146401} (\bibinfo {year} {2003})}\BibitemShut {NoStop}%
\bibitem [{\citenamefont {Perdew}\ \emph {et~al.}(2004)\citenamefont {Perdew},
  \citenamefont {Tao}, \citenamefont {Staroverov},\ and\ \citenamefont
  {Scuseria}}]{Perdew2004_JCP_6898}%
  \BibitemOpen
  \bibfield  {author} {\bibinfo {author} {\bibfnamefont {J.~P.}\ \bibnamefont
  {Perdew}}, \bibinfo {author} {\bibfnamefont {J.}~\bibnamefont {Tao}},
  \bibinfo {author} {\bibfnamefont {V.~N.}\ \bibnamefont {Staroverov}},\ and\
  \bibinfo {author} {\bibfnamefont {G.~E.}\ \bibnamefont {Scuseria}},\
  }\bibfield  {title} {\enquote {\bibinfo {title} {{Meta-generalized gradient
  approximation: Explanation of a realistic nonempirical density
  functional}},}\ }\href {https://doi.org/10.1063/1.1665298} {\bibfield
  {journal} {\bibinfo  {journal} {J. Chem. Phys.}\ }\textbf {\bibinfo {volume}
  {120}},\ \bibinfo {pages} {6898--6911} (\bibinfo {year} {2004})}\BibitemShut
  {NoStop}%
\bibitem [{\citenamefont {Jansen}\ and\ \citenamefont
  {Ros}(1969)}]{Jansen1969_CPL_140}%
  \BibitemOpen
  \bibfield  {author} {\bibinfo {author} {\bibfnamefont {H.~B.}\ \bibnamefont
  {Jansen}}\ and\ \bibinfo {author} {\bibfnamefont {P.}~\bibnamefont {Ros}},\
  }\bibfield  {title} {\enquote {\bibinfo {title} {{Non-empirical molecular
  orbital calculations on the protonation of carbon monoxide}},}\ }\href
  {https://doi.org/10.1016/0009-2614(69)80118-1} {\bibfield  {journal}
  {\bibinfo  {journal} {Chem. Phys. Lett.}\ }\textbf {\bibinfo {volume} {3}},\
  \bibinfo {pages} {140--143} (\bibinfo {year} {1969})}\BibitemShut {NoStop}%
\bibitem [{\citenamefont {Boys}\ and\ \citenamefont
  {Bernardi}(1970)}]{Boys1970_MP_553}%
  \BibitemOpen
  \bibfield  {author} {\bibinfo {author} {\bibfnamefont {S.~F.}\ \bibnamefont
  {Boys}}\ and\ \bibinfo {author} {\bibfnamefont {F.}~\bibnamefont
  {Bernardi}},\ }\bibfield  {title} {\enquote {\bibinfo {title} {{The
  calculation of small molecular interactions by the differences of separate
  total energies. Some procedures with reduced errors}},}\ }\href
  {https://doi.org/10.1080/00268977000101561} {\bibfield  {journal} {\bibinfo
  {journal} {Mol. Phys.}\ }\textbf {\bibinfo {volume} {19}},\ \bibinfo {pages}
  {553--566} (\bibinfo {year} {1970})}\BibitemShut {NoStop}%
\bibitem [{\citenamefont {Jensen}\ and\ \citenamefont
  {Helgaker}(2004)}]{Jensen2004_JCP_3463}%
  \BibitemOpen
  \bibfield  {author} {\bibinfo {author} {\bibfnamefont {F.}~\bibnamefont
  {Jensen}}\ and\ \bibinfo {author} {\bibfnamefont {T.}~\bibnamefont
  {Helgaker}},\ }\bibfield  {title} {\enquote {\bibinfo {title} {Polarization
  consistent basis sets. {V}. the elements {Si}--{Cl}},}\ }\href
  {https://doi.org/10.1063/1.1756866} {\bibfield  {journal} {\bibinfo
  {journal} {J. Chem. Phys.}\ }\textbf {\bibinfo {volume} {121}},\ \bibinfo
  {pages} {3463--3470} (\bibinfo {year} {2004})}\BibitemShut {NoStop}%
\bibitem [{\citenamefont
  {Lehtola}(2019{\natexlab{e}})}]{Lehtola2019_JCP_241102}%
  \BibitemOpen
  \bibfield  {author} {\bibinfo {author} {\bibfnamefont {S.}~\bibnamefont
  {Lehtola}},\ }\bibfield  {title} {\enquote {\bibinfo {title} {Curing basis
  set overcompleteness with pivoted {Cholesky} decompositions},}\ }\href
  {https://doi.org/10.1063/1.5139948} {\bibfield  {journal} {\bibinfo
  {journal} {J. Chem. Phys.}\ }\textbf {\bibinfo {volume} {151}},\ \bibinfo
  {pages} {241102} (\bibinfo {year} {2019}{\natexlab{e}})},\ \Eprint
  {https://arxiv.org/abs/1911.10372} {arXiv:1911.10372} \BibitemShut {NoStop}%
\bibitem [{\citenamefont
  {Lehtola}(2020{\natexlab{c}})}]{Lehtola2020_PRA_32504}%
  \BibitemOpen
  \bibfield  {author} {\bibinfo {author} {\bibfnamefont {S.}~\bibnamefont
  {Lehtola}},\ }\bibfield  {title} {\enquote {\bibinfo {title} {Accurate
  reproduction of strongly repulsive interatomic potentials},}\ }\href
  {https://doi.org/10.1103/PhysRevA.101.032504} {\bibfield  {journal} {\bibinfo
   {journal} {Phys. Rev. A}\ }\textbf {\bibinfo {volume} {101}},\ \bibinfo
  {pages} {032504} (\bibinfo {year} {2020}{\natexlab{c}})},\ \Eprint
  {https://arxiv.org/abs/1912.12624} {arXiv:1912.12624} \BibitemShut {NoStop}%
\bibitem [{\citenamefont {Delley}(1990)}]{Delley1990_JCP_508}%
  \BibitemOpen
  \bibfield  {author} {\bibinfo {author} {\bibfnamefont {B.}~\bibnamefont
  {Delley}},\ }\bibfield  {title} {\enquote {\bibinfo {title} {{An all-electron
  numerical method for solving the local density functional for polyatomic
  molecules}},}\ }\href {https://doi.org/10.1063/1.458452} {\bibfield
  {journal} {\bibinfo  {journal} {J. Chem. Phys.}\ }\textbf {\bibinfo {volume}
  {92}},\ \bibinfo {pages} {508} (\bibinfo {year} {1990})}\BibitemShut
  {NoStop}%
\bibitem [{\citenamefont {Widmark}, \citenamefont {Malmqvist},\ and\
  \citenamefont {Roos}(1990)}]{Widmark1990_TCA_291}%
  \BibitemOpen
  \bibfield  {author} {\bibinfo {author} {\bibfnamefont {P.-O.}\ \bibnamefont
  {Widmark}}, \bibinfo {author} {\bibfnamefont {P.-{\AA}.}\ \bibnamefont
  {Malmqvist}},\ and\ \bibinfo {author} {\bibfnamefont {B.~O.}\ \bibnamefont
  {Roos}},\ }\bibfield  {title} {\enquote {\bibinfo {title} {{Density matrix
  averaged atomic natural orbital (ANO) basis sets for correlated molecular
  wave functions}},}\ }\href {https://doi.org/10.1007/BF01120130} {\bibfield
  {journal} {\bibinfo  {journal} {Theor. Chim. Acta}\ }\textbf {\bibinfo
  {volume} {77}},\ \bibinfo {pages} {291--306} (\bibinfo {year}
  {1990})}\BibitemShut {NoStop}%
\bibitem [{\citenamefont {Widmark}, \citenamefont {Persson},\ and\
  \citenamefont {Roos}(1991)}]{Widmark1991_TCA_419}%
  \BibitemOpen
  \bibfield  {author} {\bibinfo {author} {\bibfnamefont {P.-O.}\ \bibnamefont
  {Widmark}}, \bibinfo {author} {\bibfnamefont {B.~J.}\ \bibnamefont
  {Persson}},\ and\ \bibinfo {author} {\bibfnamefont {B.~O.}\ \bibnamefont
  {Roos}},\ }\bibfield  {title} {\enquote {\bibinfo {title} {{Density matrix
  averaged atomic natural orbital (ANO) basis sets for correlated molecular
  wave functions}},}\ }\href {https://doi.org/10.1007/BF01112569} {\bibfield
  {journal} {\bibinfo  {journal} {Theor. Chim. Acta}\ }\textbf {\bibinfo
  {volume} {79}},\ \bibinfo {pages} {419--432} (\bibinfo {year}
  {1991})}\BibitemShut {NoStop}%
\bibitem [{\citenamefont {Pou-Am{\'{e}}rigo}\ \emph {et~al.}(1995)\citenamefont
  {Pou-Am{\'{e}}rigo}, \citenamefont {Merch{\'{a}}n}, \citenamefont
  {Nebot-Gil}, \citenamefont {Widmark},\ and\ \citenamefont
  {Roos}}]{PouAmerigo1995_TCA_149}%
  \BibitemOpen
  \bibfield  {author} {\bibinfo {author} {\bibfnamefont {R.}~\bibnamefont
  {Pou-Am{\'{e}}rigo}}, \bibinfo {author} {\bibfnamefont {M.}~\bibnamefont
  {Merch{\'{a}}n}}, \bibinfo {author} {\bibfnamefont {I.}~\bibnamefont
  {Nebot-Gil}}, \bibinfo {author} {\bibfnamefont {P.-O.}\ \bibnamefont
  {Widmark}},\ and\ \bibinfo {author} {\bibfnamefont {B.~O.}\ \bibnamefont
  {Roos}},\ }\bibfield  {title} {\enquote {\bibinfo {title} {Density matrix
  averaged atomic natural orbital ({ANO}) basis sets for correlated molecular
  wave functions},}\ }\href {https://doi.org/10.1007/BF01114922} {\bibfield
  {journal} {\bibinfo  {journal} {Theor. Chim. Acta}\ }\textbf {\bibinfo
  {volume} {92}},\ \bibinfo {pages} {149--181} (\bibinfo {year}
  {1995})}\BibitemShut {NoStop}%
\end{thebibliography}%

\end{document}